\documentclass[11pt]{article}
\pdfoutput=1
\usepackage[utf8]{inputenc}
\usepackage{cite}
\usepackage{epigraph}
\usepackage[dvipsnames]{xcolor}
\usepackage{amsmath,amssymb,amsbsy,amstext,amsthm,simplewick,amsfonts,mathrsfs}
\usepackage{graphicx}
\usepackage{wrapfig}
\usepackage{upgreek}
\usepackage{cancel}
\usepackage{bm} 
\usepackage{framed}
\usepackage{bbm}
\usepackage{textcomp}
\usepackage{tikz}
\usepackage{pifont}
\usetikzlibrary{matrix,shapes,fit,tikzmark,calc}
\usepackage{tikzlines}
\usepackage{tkz-euclide}
\usepackage{tikz-cd}
\usepackage[labelfont=bf]{caption}
\usepackage{adjustbox}
\usepackage{makecell}
\usepackage{bbold}
\usepackage{tcolorbox}
\usepackage{physics}
\usepackage{wrapfig}
\usepackage{empheq}
\usepackage[normalem]{ulem}
\usepackage{enumitem}
\usepackage{braket} 
\usepackage{array}
\usepackage{dsfont}
\usepackage{ulem}
\usetikzlibrary {decorations.markings,fpu}
\usepackage{enumitem}
\usepackage{slashed}
\usepackage{titlesec}
\titleformat{\section}{\normalfont\fontsize{12}{16}\bfseries}{\thesection}{1em}{}
\numberwithin{equation}{section}

\def\be{\begin{equation}}
\def\ee{\end{equation}}

\def\d{\delta}

\def\ba{\begin{eqnarray}}
\def\ea{\end{eqnarray}}

\def\L{\mathcal{L}}

\def\aa{\mathsf{a}}
\def\bb{\mathsf{b}}

\def\x{\xi}
\def\vpi{\varphi}

\usepackage{caption}
\usepackage{subcaption}

\definecolor{blue3}{RGB}{31,119,180}
\definecolor{red3}{RGB}{214,39,40}
\definecolor{orange3}{RGB}{255,127,14}
\definecolor{green3}{RGB}{44,160,44}
\definecolor{mmaGrey}{RGB}{127, 127, 127}
\definecolor{mmaBlue}{RGB}{93, 129, 180}
\definecolor{mmaYellow}{RGB}{224, 155, 36}
\definecolor{figure-purple}{RGB}{203,194,244}
\definecolor{figure-green}{RGB}{47,199,190}

\newmuskip\pFqmuskip
\def\pFqcomma{\mathchar"002C\mskip\pFqmuskip}
\newcommand*\pFq[6][8]{%
	\begingroup 
	\pFqmuskip=#1mu\relax
	\mathcode`\,=\string"8000
	\begingroup\lccode`\~=`\,
	\lowercase{\endgroup\let~}\pFqcomma
	{}_{#2}{\rm{F}}_{#3}{\left[\genfrac..{0pt}{}{#4}{#5};#6\right]}%
	\endgroup
}

\newcommand*\pregFq[6][8]{%
	\begingroup 
	\pFqmuskip=#1mu\relax
	\mathcode`\,=\string"8000
	\begingroup\lccode`\~=`\,
	\lowercase{\endgroup\let~}\pFqcomma
	{}_{#2}\tilde{{\rm{F}}}_{#3}{\left[\genfrac..{0pt}{}{#4}{#5};#6\right]}%
	\endgroup
}

\usepackage{colortbl}
\definecolor{lightgreen}{cmyk}{0.2, 0, 0.2, 0.2}
\definecolor{lightgray}{cmyk}{0.1,0.2,0,0.1}
\definecolor{lightgray2}{cmyk}{0.1,0.1,0,0.1}
\definecolor{pygreen}{RGB}{30, 100, 64}

\definecolor{myblue}{RGB}{0, 100, 200}

\usepackage{hyperref}
\hypersetup{colorlinks=true,linkcolor=blue3,linktocpage=true,citecolor=red3,urlcolor=green3,pdfencoding=auto}

\makeatletter
\newlength{\apb@width}
\newcommand{\autoparbox}[2][c]{\settowidth{\apb@width}{#2}\parbox[#1]{\apb@width}{#2}}

\makeatother


\def\d{{\rm d}}

\def\x{{\bm x}}

\def\Mpl{M_{\text{P}}}

\def\B{{\cal B}}

\def\Mp{M_{\rm P}}

\def\nn{\nonumber}

\def\beq{\begin{equation}}
\def\eeq{\end{equation}}

\def\bgm{\begin{matrix}}
\def\edm{\end{matrix}}

\newmdenv[backgroundcolor=gray!10,%
skipabove=5pt,%
skipbelow=5pt,%
leftmargin=2pt,%
rightmargin=2pt,%
innertopmargin=-6pt,%
innerbottommargin=5pt,%
innerleftmargin=5pt,%
innerrightmargin=5pt,%
splittopskip=0pt,%
splitbottomskip=0pt,%
linewidth=0pt,%
nobreak=true]%
{keyeqn}

\graphicspath{{Fig/}}

\allowdisplaybreaks[1]
\begin{document}

\begin{titlepage}
\setcounter{page}{1} \baselineskip=15.5pt

\thispagestyle{empty}

\renewcommand*{\thefootnote}{\fnsymbol{footnote}}

\begin{center}
{\fontsize
{15}{15} \bf Every Wrinkle Carries A Memory:}\\ 
{\textit{ \Large
An Integro-differential Bootstrap for Features in Cosmological Correlators }}
\end{center}

\vskip 18pt
\begin{center}
\noindent
{\fontsize{12}{18}\selectfont 
Sadra Jazayeri\footnote{\tt s.jazayeri@imperial.ac.uk}$^{,a}$, Xi Tong\footnote{\tt xt246@cam.ac.uk}$^{,b}$
and Yuhang Zhu\footnote{\tt yhzhu@ibs.re.kr}$^{,c}$
}
\end{center}

\begin{center}
\vskip 12pt
\textit{$^{a}$Abdus Salam Centre for Theoretical Physics, Imperial College, London, SW7 2AZ, UK}
\\ [5pt]
\textit{$^{b}$Department of Applied Mathematics and Theoretical Physics, University of Cambridge, Wilberforce Road, Cambridge, CB3 0WA, UK}
\\ [5pt]
\textit{$^{c}$Cosmology, Gravity and Astroparticle Physics Group,\\Center for Theoretical Physics of the Universe,\\
Institute for Basic Science, Daejeon 34126, Korea}
\end{center}

\vskip 20pt

	
\noindent\rule{\textwidth}{0.4pt}
\noindent \textbf{Abstract}

\noindent Motivated by cosmological observations, we push the cosmological bootstrap program beyond the de Sitter invariance lamppost by considering correlators that explicitly break scale invariance, thereby exhibiting \textit{primordial features}. For exchange processes involving heavy fields with time-dependent masses and sound speeds, we demonstrate that \textit{locality} in the bulk implies a set of \textit{integro-differential equations} for correlators on the boundary. These scale-breaking boundary equations come with a built-in memory kernel in momentum-kinematic space encapsulating the universe's evolution during inflation. Specialising to heavy fields with sinusoidal masses such as those found in axion monodromy scenarios, we show that a powerful synthesis of \textit{microcausality} and \textit{analyticity} allows an analytical solution of these equations at leading order in the amplitude of mass oscillations. Meanwhile, we also unveil non-perturbative information in the integro-differential equations by resumming apparent infrared divergences as parametric resonances. In addition, we provide a first-of-its-kind example of \textit{numerical bootstrap} that directly maps out the solution space of such boundary equations. Finally, we compute the bispectrum and uncover, in the squeezed limit, a scale-breaking cosmological collider signal, whose amplitude can be exponentially enhanced (with respect to the Boltzmann suppression) due to particle production triggered by high-frequency mass oscillations.


\end{titlepage}

\setcounter{tocdepth}{2}
{
\tableofcontents
}

\renewcommand*{\thefootnote}{\arabic{footnote}}
\setcounter{footnote}{0}

\newpage
\section{Introduction}

The statistical patterns present in the late-time distribution of matter and radiation in our universe are the fossil records of its quantum history before the hot Big Bang, during an accelerated phase of expansion known as inflation. Through the measurement of cosmological correlation functions (or correlators in short) at late times, modern cosmology is partly concerned with reconstructing this history in a manner that adheres to well-established physical principles such as unitarity, locality and causality. These principles are usually made explicit by writing down e.g. local quantum field theories with associated unitary time evolutions to describe inflation. The boundary correlators that emerge as consequences of these bulk evolutions, are therefore tightly constrained by the fundamental principles therein. However, such a bulk perspective is not directly reachable for us observers in the late-time universe. Instead, we are only granted access to the boundary data at the end of inflation, where the fundamental principles appear completely obscured. For instance, it is not immediately apparent what consistency requirements determine if a given correlator, even in perturbation theory, can result from a healthy time evolution during inflation, and what diagnosis decides otherwise. Indeed, much recent effort in the cosmological bootstrap program\cite{Arkani-Hamed:2018kmz,Baumann:2020dch,Hogervorst:2021uvp, Baumann:2022jpr,deRham:2025mjh} has been devoted to finding such consistency conditions, some of which have inspired new ways to ``bootstrap'' perturbative diagrams without explicitly evaluating their notoriously difficult, nested time integrals. See \cite{Bzowski:2013sza, Sleight:2019hfp,Sleight:2019mgd,Pajer:2020wnj, Qin:2022lva, Qin:2022fbv,Wang:2022eop,Salcedo:2022aal, Jazayeri:2022kjy,Pimentel:2022fsc,Melville:2023kgd, Armstrong:2023phb, Qin:2023nhv, Qin:2023bjk, Qin:2023ejc,Chowdhury:2023arc, Fan:2024iek,  Aoki:2024uyi,Goodhew:2024eup, Werth:2024mjg, Melville:2024ove, Qin:2025xct,Wang:2025qww, Cespedes:2025dnq,Chowdhury:2025ohm,Fan:2025scu,CarrilloGonzalez:2025qjk} for an incomplete list of references. 

\vskip 7pt

\begin{figure}
    \centering
    \includegraphics[scale=0.8]{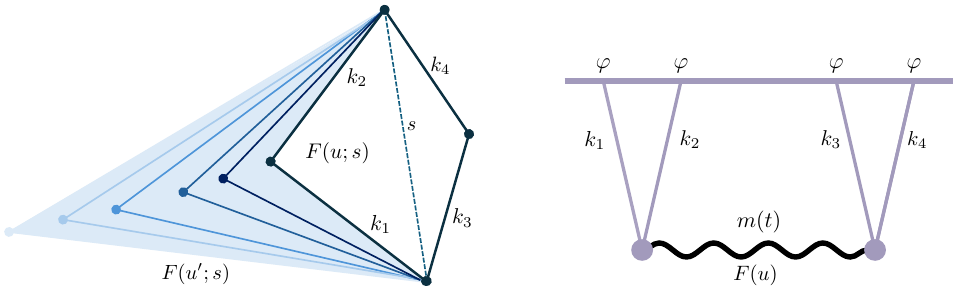}
    \caption{\textit{Right}: The single-exchange diagram of our interest, characterised by four external conformally coupled fields $\vpi$ and a heavy intermediate scalar $\sigma$, endowed with a time-dependent mass $m(t)$. For constant masses, the diagram satisfies the ordinary differential equation \eqref{bootstrapODE}. \textit{Left}: For time-dependent masses, the integro-differential equation \eqref{IDEschematic} takes over, relating the exchange diagram in one momentum configuration to its values at more squeezed configurations with $u>u'$.}
    \label{fig:IDEfig}
\end{figure}

A prototypical example is a massive single-exchange diagram in exact de Sitter space, characterised by four external legs of conformally coupled scalars (with the associated 3-momenta $\bm{k}_{i}$ along with their magnitudes $k_i\equiv |\bm{k}_i|$, $i=1,\dots,4$) and vertices connected by the propagator of a massive scalar field $\sigma$ in the $s$-channel (with the associated internal momentum $\bm{s}=\bm{k}_1+\bm{k}_2$); see Figure \ref{fig:IDEfig}. 
This diagram has been bootstrapped in \cite{Arkani-Hamed:2018kmz} using a set of \textit{differential} equations of the schematic form
\begin{align}
\label{bootstrapODE}
    ({\Delta}_{u}+m_0^2)F(u,v)=\frac{u v}{u+v}\,,
\end{align}
where $u\equiv |\bm {s}|/(k_1+k_2)$, $v\equiv |\bm{s}|/(k_3+k_4)$, $m_0$ denotes the mass of the intermediate field in Hubble units, and $\Delta_u=u^2(1-u^2)\partial^2_u-2u^3\partial_u-2$ is a differential operator whose form follows from the Klein-Gordon operator in the bulk. It is straightforward to show that the four-point diagram $F(u,v)$ directly inherits its bootstrap equation \eqref{bootstrapODE} from the local equation of motion satisfied by the bulk-bulk propagator. In other words, \eqref{bootstrapODE} is a boundary manifestation of \textit{locality} in the bulk\footnote{See also \cite{Jazayeri:2021fvk} for a \textit{manifest locality test} applicable to Witten diagrams with external massless fields and manifestly local interactions in the bulk.}. See \cite{Arkani-Hamed:2023kig,Grimm:2024tbg,Chen:2024glu,Liu:2024str} for recent developments on generalised differential equations for correlators. 

\vskip 5pt

The above diagram lives in exact de Sitter space and therefore is invariant under its full $SO(4,1)$ isometry group, which notably includes three de Sitter boosts and one dilation (a.k.a. scale invariance). However, realistic inflation turns out to be less symmetric and is thus crucially distinct from exact de Sitter. More specifically, the three de Sitter boosts are generically strongly broken during inflation since the rolling inflaton background picks a preferred rest frame, while the scale invariance is softly broken by the tilt of the inflaton potential. In fact, from established frameworks such as the EFT of inflation and its generalisations \cite{Cheung:2007st,Senatore:2010wk,Noumi:2012vr,Cabass:2022jda,Salcedo:2024smn,Pinol:2024arz}, we have learned that rich phenomenology, associated with sizeable non-Gaussianities, comes with a strong breaking of de Sitter isometries \cite{EnricoDan}. 
For instance, in cases where the inflaton kinetic energy driven by $\dot{\bar\phi}^{1/2}\sim 58H$ is pumped into the massive sector, the particle production rate can be exponentially enhanced. As a direct consequence of copious particle production, the \textit{cosmological collider} signals, which encode the particle spectrum during inflation in the form of squeezed-limit non-analytic dependences on the momentum ratio \cite{Noumi:2012vr, Chen:2009zp,Baumann:2011nk,Arkani-Hamed:2015bza,Lee:2016vti}, are dramatically amplified, alleviating their Boltzmann suppression from a de Sitter-invariant universe \cite{Flauger:2016idt,Chen:2018xck,Hook:2019zxa,Wang:2020ioa,Bodas:2020yho,Sou:2021juh,Tong:2022cdz,Chen:2022vzh,Qin:2022fbv,Qin:2022lva,Chen:2023txq,Stefanyszyn:2023qov, Bodas:2024hih,An:2025mdb, Werth:2023pfl, Pinol:2023oux}. Motivated by these observations, significant recent progress has been made in bootstrapping boostless correlators using non-symmetry-based tools. These include, for example, modified versions of the bootstrap differential equation \eqref{bootstrapODE} (e.g. in the presence of non-trivial sound speeds\cite{Jazayeri:2022kjy,Jazayeri:2023xcj,Pimentel:2022fsc,Qin:2025xct}), cutting rules based on unitarity \cite{Meltzer:2020qbr, Goodhew:2020hob,  Melville:2021lst,Goodhew:2021oqg,Stefanyszyn:2024msm,Stefanyszyn:2025yhq} and causality \cite{Tong:2021wai,AguiSalcedo:2023nds,Ema:2024hkj,Liu:2024xyi}
, as well as  analyticity \& recursion relations \cite{Stefanyszyn:2020kay,Pajer:2020wxk, Meltzer:2021zin,Meltzer:2021bmb, Cabass:2021fnw}.

\vskip 5pt

The breaking of scale invariance, on the other hand, has been much less studied in the context of cosmological correlators and their bootstraps. There are several reasons for exploring setups that exhibit strong breaking of scale invariance:
\begin{itemize}
    \item[i.] On the observational side, while scale invariance is a reasonable assumption according to the measurements of the scalar tilt, percent-level departures from a power-law two-point function---also known as \textit{primordial features}---are still compatible with the existing Cosmic Microwave Background (CMB) and Large Scale Structure (LSS) data \cite{Achucarro:2022qrl, Braglia:2021ckn,Braglia:2022ftm}. In fact, within the next few years, data from Galaxy surveys such as EUCLID and DESI is projected to significantly improve the present bounds on primordial features, making their theoretical study especially timely \cite{Chen:2016vvw, Slosar:2019gvt, Beutler:2019ojk,Euclid:2023shr}\footnote{Primordial features are particularly fruitful targets for LSS observations because they are partially protected from gravitational non-linear evolutions at late times \cite{Beutler:2019ojk}. This protection is not generic, for instance, it does not extend to (scale-invariant) equilateral-type non-Gaussianities\cite{Achucarro:2022qrl}.}. 
    
    \item[ii.] On the phenomenological side, breaking scale invariance during inflation provides new opportunities to probe ultraviolet physics well beyond the Hubble scale. For example, via non-shift symmetric couplings to the inflaton field, particles parametrically heavier than the expansion rate can be non-adiabatically created during inflation. The subsequent decay of these species into the inflaton fluctuations would leave a variety of scale-breaking signatures in inflationary correlators, providing an unprecedented observational window into the heavy sector during inflation \cite{Chen:2011zf,Chen:2012ja,  Flauger:2016idt, Abolhasani:2019wri, Abolhasani:2019lwu,Abolhasani:2020xcg,Chen:2022vzh}. 
    
    \item[iii.] On the theoretical side, scale invariance is simultaneously a blessing that brings immense simplification to the understanding of inflationary physics, and a curse for our bias towards the analytical lamppost. Existing studies that venture beyond scale-invariant setups mostly rely on simplifying assumptions, such as breaking to a discrete subgroup \cite{Behbahani:2011it,DuasoPueyo:2023kyh}, restricting to breaking in perturbative vertices \cite{Werth:2023pfl,Pinol:2023oux,Chen:2022vzh,Qin:2023ejc,Pajer:2024ckd,Wang:2025qww} and special approximations of the breaking parameter \cite{Aoki:2023wdc}.
\end{itemize}

\vskip 7pt

In this work, we shall leap away from the scale-invariance lamppost by proposing a bootstrap of general scale-breaking tree correlators.
When de Sitter dilatation is broken, a local time evolution no longer translates into an ordinary local differential equation for exchange correlators. Instead, from the correspondence between conformal time $\eta$ and comoving scale $k$ (i.e. $k=-1/\eta$, at horizon crossing\footnote{Note that, despite the breaking of scale invariance in the bulk-bulk propagator, the conformal time-comoving scale correspondence is upheld by the scale-invariant bulk-boundary propagator $\propto \exp(ik\eta)$.}), one should expect non-local relations to arise between widely separated configurations in  kinematic space. Such boundary non-localities resonate with the notion that correlators at the end of inflation encode an integrated history over the bulk, whose details cannot be extracted from any one momentum configuration. Indeed, we will find that in cases where scale breaking solely originates from the massive sector (via explicit time dependence in either the free theory dynamics or interactions), the resulting four-point diagram satisfies an \textit{integro-differential equation} of the schematic form\footnote{More precisely, it is the Schwinger-Keldysh components $F_{\pm\pm}$ that satisfy integro differential equations like \eqref{IDEschematic}. See \eqref{IDEintro1}.}
\begin{keyeqn}
    \begin{align}
        \label{IDEschematic}
        ({\Delta}_{u}+m_0^2)\,F(u;s)=\frac{u v}{u+v}-\int_0^u \dfrac{\d u'}{u'}\left(1-\frac{u'}{u}\right)^{-1} \,K\left[\dfrac{s(u-u')}{uu'}\right]\, F(u';s)\,,
    \end{align}
\end{keyeqn}
where we have omitted the functional dependence on $v$ for simplicity. The effect of scale breaking is carried by the \textit{memory kernel} $K(x)$, whose detailed form relies on specific scale-breaking models, such as the heavy field's mass or sound-speed time-dependence. But in all cases (under the Bunch-Davies initial condition assumption) we will show that $K(x)$ must identically vanish for $x<0$. As illustrated in Figure \ref{fig:IDEfig}, this imposes a retardation in kinematic space for \eqref{IDEschematic}, where configurations with $u>u'$ (less squeezed than the left hand side) are precluded from the right hand side.  
Also note that, given the explicit dependence of the kernel $K$ on the comoving momentum $s$, neither the integro-differential equation nor its solutions $F(u;s)$ are symmetric under the rescaling $\bm{k}_i\to \lambda\,\bm{k}_i$. 

Such integro-differential equations are notoriously difficult to solve in general, even perturbatively. However, we demonstrate that their solutions are highly constrained by the principles of \textit{analyticity} and \textit{microcausality}. In more detail, consistent with previous works\cite{Lee:2016vti, Tong:2021wai,Qin:2023bjk,Ema:2024hkj}, we will show that microcausality---the vanishing of the heavy field commutator outside the light-cone---enforces a powerful factorisation property upon the exchange diagram, even in the absence of scale invariance. According to this property, the exchange diagram must \textit{factorise} into a specific product of its three-point function sub-diagrams, up to analytic terms in the exchanged momentum $\bm{s}$. This factorisation, together with the regularity of the four-point function at physical configurations (as required by the Bunch-Davies initial condition), will provide sufficient boundary conditions for solving our bootstrap integro-differential equations.

For concreteness, we shall narrow down our focus on a heavy field with sinusoidal mass oscillations as $m^2(t)=m_0^2+g^2 m_0^2 \cos(\omega t)$. Such a setup is motivated e.g. by axion monodromy inflation coupled to matter, where despite the breaking of the inflaton continuous shift symmetry a discrete subgroup is approximately preserved. 
In this simple scenario, carefully harnessing the above bootstrap constraints allows us to perturbatively solve \eqref{IDEschematic} up to the leading order in $g^2$ from a pure boundary perspective. Despite the special monochromatic time dependence assumed for the intermediate propagator, the final exchange diagram, by weighted integrations over the frequency $\omega$, can be generalised to any time-dependent intermediate mass at linear order in $\Delta m^2$. 

Moreover, expanding near the squeezed limit $u\ll 1$, we find \textit{resonant} cosmological collider signals of the schematic type $u^{\pm i m_0}s^{\mp i\omega}$ (see \cite{Wang:2025qww, Pajer:2024ckd, Chen:2022vzh, Werth:2023pfl, Pinol:2023oux} for alternative setups leading to similar non-Gaussian signatures). These signals can be enhanced by a transient resonance in the UV, where the heavy field momentum crosses the frequency scale at $s/a(t)=\omega$, and a persistent parametric resonance in the IR (for the special case with $\omega\approx 2m_0$), thereby overcoming the Boltzmann suppression and becoming relevant for future cosmological observations. To extract more non-perturbative information, we also perform a direct numerical bootstrap of \eqref{IDEschematic} using the finite-difference method, where we find a perfect agreement with the analytical prediction of the scaling exponents near the IR resonances. We note in passing that this serves as a first example of numerical bootstrap for correlators in cosmology.

Note that unlike traditional bulk methods such as in-in/Schwinger-Keldysh formalisms, where one needs to solve the mode functions (often numerically when scale invariance is broken) before performing layers of nested time integrals, the bootstrap equation \eqref{IDEschematic} constrains the observables in one go by non-perturbatively resumming scale-invariance breaking effects in the memory kernel $K$. Therefore, this boundary perspective does appear more efficient at least formally. More importantly, its integrated nature vividly shows how the information of local and causal evolution histories in a higher dimensional bulk is \textit{smeared} over the kinematic space in a lower dimensional boundary, and how this smearing ultimately stem from scale breaking.

\vskip 7pt

The rest of this paper is structured as follows. We begin by providing an outline of the bootstrap roadmap in Section \ref{RoadmapSubsection} for readers who wish to grasp an overall picture of the bootstrap without diving into technicalities. Then in Section \ref{Setup}, we explain our model setup and introduce the seed four-point function with general time-dependent parameters. We then switch to the boundary perspective and show that the seed function does satisfy an integro-differential equation as advertised above in Section \ref{BdyIDEg2BootstrapSection}. Using constraints of locality, microcausality and analyticity, we solve this integro-differential bootstrap equation up to the first non-trivial order in Section \ref{Sec: PertSolution} and discuss the implication for cosmological observables such as the curvature power spectrum and bispectrum. Section \ref{ParaResSection} is an interlude where we further explore the effect of parametric resonances in our model. We then perform a direct numerical bootstrap of the boundary equation in Section \ref{numericalBootstrapSection} to acquire non-perturbative information of the system. We conclude and give outlooks in Section \ref{conclusionSection}.

\paragraph{Conventions and notations} We use the following coordinates to chart the Poincaré patch of a $(3+1)$-dimensional de Sitter space: 
\begin{align}
    \d s^2=\dfrac{1}{H^2\eta^2}(-\d\eta^2+\d\bm{x}^2)\,,
\end{align}
where $-\infty<\eta<0$ is the conformal time. We occasionally use the FLRW coordinates, $\d s^2=-\d t^2+a^2(t)\d \bm{x}^2$ with $a(t)=e^{H t}$. 
To avoid clutter, we mostly set the Hubble rate to unity ($H=1$) unless otherwise stated. In the main body of the paper, we will consider a heavy bulk field with mass $m_0$ that is also characterised by the dimensionless mass-index $\mu=(m_0^2/H^2-9/4)^{1/2}>0$. The case with a light bulk field is discussed separately in Appendix \ref{LightFieldSection}. We introduce the following shorthand involving factors of the Euler gamma function:
\begin{align}
\label{gammasymbol}
    \Gamma\left[\bgm \alpha_1\dots \alpha_n\\
        \beta_1\dots \beta_m\edm\right]=\dfrac{\Gamma(\alpha_1)\dots \Gamma(\alpha_n)}{\Gamma(\beta_1)\dots \Gamma(\beta_m)}\,.
\end{align}
We make frequent use of hypergeometric functions ${}_p \rm{F}_q$ with different weights in the analytical calculations. Their regularised form is defined as follows:
\begin{align}
    \pregFq{p}{q}{a_1,\cdots,a_p}{b_1,\cdots,b_q}{u}\equiv\frac{1}{
    \Gamma\left[ b_1,\cdots,b_q\right]
    }\,\pFq{p}{q}{a_1,\cdots,a_p}{b_1,\cdots,b_q}{u}\,. \label{eq: reg_pFq}
\end{align}
We will refer to the conformally coupled field in de Sitter by $\vpi$ (with $m^2=2H^2$ in $3+1$ dimensions) and to the massless Goldstone boson in the EFT of inflation by $\pi$. The seed four-point function will be denoted by $F$ and characterised by the external spatial momenta $\bm{k}_i\,(i=1,\dots 4)$. We denote the exchanged momentum in the $s$-channel by $\bm{s}=\bm{k}_1+\bm{k}_2$. We use regular letters to denote the magnitudes of spatial momenta, i.e. $k_i=|\bm{k}_i|,\, s=|\bm{s}|$. Sums of these momenta magnitudes are shortened as $k_{ij}\equiv k_{i}+k_j$. The total-energy variable is denoted by $k_T\equiv k_{12}+k_{34}$. We also define    
\begin{align}
    u\equiv \dfrac{s}{k_{12}}~,\qquad v\equiv \dfrac{s}{k_{34}}~,
\end{align}
and
\begin{align}
    U\equiv \dfrac{2u}{1+u}~, \qquad V\equiv\frac{2v}{1+v}\,,
\end{align}
as convenient bootstrap variables. $\sigma$ will denote a heavy field whose mass oscillates in time as $m^2(t)=m_0^2+g^2m_0^2\cos[\omega (t-t_0)]$, where $0\leq g<1$. We use the dimensionless parameter $x_0\equiv -s\eta_0$, in which $\eta_0=-\exp(-Ht_0)$, as a proxy for the oscillation phase (note that $\eta_0$ is not to be confused with the boundary time $\eta_{\rm end}$ at the end of inflation). The power spectrum of the curvature perturbation is defined as 
\begin{align}
    \langle \zeta_{\bm{k}_1} \zeta_{\bm{k}_2}\rangle\equiv(2\pi)^3\delta^3(\bm{k}_1+\bm{k}_2)\,P(k)\equiv(2\pi)^3\delta^3(\bm{k}_1+\bm{k}_2)\,\frac{2\pi^2}{k^3}\Delta_{\zeta}^2\,,
\end{align}
and $\Delta^2_{\zeta}$ denotes the dimensionless power spectrum, whose observed amplitude is $\Delta_\zeta^2\simeq2\times10^{-9}$. 

\subsection{Bootstrap roadmap}\label{RoadmapSubsection}

Here we present a condensed summary for readers seeking a quick overview of our bootstrap method, bypassing technical details.

\paragraph{Boundary equations} The prototypical scale-breaking model is a scalar field with a time-dependent mass $m^2(t)$ that can be Fourier expanded as
\begin{equation}
   m^2(t)=m^2_0+\frac{1}{2}m_0^2\int {\rm{d}}\omega\,\rho_\omega \left(\frac{\eta}{\eta_0}\right)^{i\omega}\,. 
\end{equation}
The exchange of such a massive field sources a four-point function (see Figure \ref{fig:IDEfig}) whose Schwinger-Keldysh integrals $F_{\sf a b}$ ($\sf a,b=\pm$) satisfy an integro-differential bootstrap equation
\begin{align}
    (\hat{\Delta}_{12}+m_0^2)\,F_{\sf a b}(k_{12},k_{34},s)&=\dfrac{\delta_{\sf a b}}{k_{T}} + \int_{0}^\infty {\rm{d}}q \,K_{\sf a}(q)\,F_{\sf a b}(k_{12}+q,k_{34},s)\,, 
\label{IDEintro1}
\end{align}
where
\begin{align}
   \hat{{\Delta}}_{12}=(k_{12}^2-s^2)\partial_{k_{12}}^2+2k_{12}\partial_{k_{12}}-2\,,
\end{align}
is the usual scale-invariant Klein-Gordon operator represented in boundary kinematics and
\begin{align}
    K_{\sf a}(q)=-\frac{1}{2}m_0^2 \int{\rm{d}}\omega\, \rho_\omega  \frac{e^{-{\sf a}\pi\omega/2}}{\Gamma(i\omega)} \frac{(-q \eta_0)^{i\omega}}{q}\,,
\end{align}
is a scale-breaking memory kernel reflecting the time dependence of the mass. As an aside, note that the full correlator, $F=\sum_{\aa,\bb=\pm 1}F_{\aa\bb}$, does not satisfy on its own any integro-differential equation like Eq.~\eqref{IDEintro1}. The reason is the use of different memory kernels for individual Schwinger-Keldysh components. More details on this point will be provided in Section \ref{BdyIDEg2BootstrapSection}.

In the simplest case where the mass is a monochromatic oscillatory function of time, $\rho_{\omega'}=g^2[\delta(\omega'-\omega)+\delta(\omega'+\omega)]$, the memory kernel reduces by essentially dropping the frequency integral. Such oscillatory masses are motivated by models of axion monodromy inflation, where the approximate shift symmetry of the inflaton field $\phi$ is broken down to the discrete subgroup $\phi \to \phi+2\pi n f$, where $f$ is the axion decay constant. This allows incorporating new operators into the kinetic term of the heavy field such as $g^2 m_0^2\cos(\phi/f)\sigma^2$, which produces an oscillatory mass correction with $\omega=\dot{\bar\phi}/f$ upon setting the inflaton to its background $\phi=\bar\phi(t)$.

\paragraph{Constraints from fundamental principles} 
As we saw, \textit{locality} in the bulk manifests as integro-differential equations for correlators on the boundary. Let us now review the consequences of other fundamental principle for our exchange diagram (Figure \ref{fig:IDEfig}):
\begin{enumerate}
    \item[(i)] {\it Bunch-Davies vacuum}: The structure of the integro-differential equations \eqref{IDEintro1} is also constrained by the Bunch-Davies initial condition
    of the massive field. Such asymptotic vacuum is well-defined under the mild assumption that mass grows slower than the exponential $e^{-Ht}$ in the far past. 
    We will show that, in \eqref{IDEintro1}, the domain of the momentum integral, due to this vacuum choice, is restricted to $q>0$, which, loosely speaking, corresponds to four-point kinematics that are more squeezed than the left hand side configuration (see Figure \ref{fig:IDEfig} for an illustration). 

    \item[(ii)] {\it Analyticity}: The Bunch-Davies initial condition also has far reaching consequences for the analytic properties of correlators, even those violating scale invariance. In particular, it requires the exchange diagram $F$ to be an analytic function in the complex plane of its external energies ($k_{12}$ and $k_{34}$) as well as its internal energy ($s$) except for: the total-energy singularity at $k_T=k_{12}+k_{34}=0$ and a corresponding cut across $k_T<0$; the partial-energy singularities at $k_{12}+s=0$ and $k_{34}+s=0$, along with corresponding branch cuts across $k_{12}+s<0$ and $k_{34}+s<0$, respectively; and finally, a branch point at $s=0$ (associated with particle production) with a corresponding cut along $s^2<0$. 
    Imposing this analytic structure, in particular the regularity of the diagram in the folded limits ($k_{12}=s$ and $k_{34}=s$), provides necessary boundary conditions for solving the integro-differential equations. 

    \item[(iii)] {\it Microcausality}: In any consistent quantum field theory (QFT), local operators should commute outside the light cone, a property commonly referred to as microcausality. When applied to the massive-field operator $\sigma$ in our setup, microcausality implies that the Fourier transform of the retarded propagator 
    \begin{align}
    G_R(\bm{s},\eta,\eta')&=\int {\rm{d}}^3\bm{x}\,e^{-i\bm{s}\cdot\bm{x}}\,\langle\Omega|\left[\sigma(\eta',0),\sigma(\eta,\bm{x})\right]|\Omega\rangle\,\theta(\eta'-\eta)\,,
    \end{align}
    at fixed conformal times $\eta$ and $\eta'$, must be analytic in the 3-momentum $\bm{s}$ \cite{Lee:2016vti, Tong:2021wai, Hui:2025aja}. This imposes a powerful constraint on the functional form of the exchange diagram. That is, up to analytic terms in the exchanged momentum $\bm{s}$, the diagram must \textit{factorise} as: 
    \begin{align}
    \nn
    F(k_{12},k_{34},s)=&-[f(-k_{12}-i\epsilon,s)]^*\times f(k_{34},s)+f(k_{12},s)f^*(k_{34},s)+\text{c.c.}\\ \label{factorisationintro}
    &+\text{analytic in}\,\bm{s}^2\,,
    \end{align}
    in which $f$ is the three-point integral
    \begin{align}
    f(k_{12},s)=+i\int_{-\infty(1-i\epsilon)} \dfrac{{\rm{d}}\eta}{\eta^2}\,e^{ik_{12}\eta}\sigma_-(s,\eta)\,,
    \end{align}
    where the energy variables ($k_{12},k_{34}$ and $s$) are all assumed to be positive, and $\sigma_-$ is the negative frequency mode function of the heavy field. The factorised contributions in \eqref{IDEintro1} generically have branch points at $\bm{s}=0$. Conversely, the contribution from replacing the Feynman propagator $G_{\pm\pm}$ in $F_{\pm\pm}$ with the retarded propagator $G_R$ is analytic in the 3-momentum $\bm{s}$. For similar discussions on the implications of causality see \cite{Tong:2021wai,Qin:2023bjk,Ema:2024hkj} for in-in correlators; \cite{AguiSalcedo:2023nds,Salcedo:2022aal} for the wavefunction of the universe; and \cite{deRham:2020zyh, Creminelli:2022onn, CarrilloGonzalez:2022fwg,CarrilloGonzalez:2023emp, Creminelli:2024lhd, CarrilloGonzalez:2025fqq} for EFTs around Lorentz violating backgrounds.
\end{enumerate}

\paragraph{Bootstrap strategy} Working at leading order in the mass oscillations amplitude $g^2$, our strategy will be first to solve the bootstrap equation for the three-point building block $f$, imposing as boundary conditions: (1) analyticity in the folded limit, and (2) consistent behaviour near the total energy singularity. From $f$, the non-time-ordered components $F_{\pm\mp}$ immediately follow. We will then go ahead and solve the bootstrap equation for the ordered components $F_{\pm\pm}$, this time requesting the factorisation property \eqref{factorisationintro} as an additional boundary condition around $s=0$. To simplify the task, we will only consider the soft limit ($\bm{k}_4\to 0$) of the four-point diagram, the knowledge of which will be sufficient for extracting the bispectrum of curvature perturbations $B(k_1,k_2,k_3)$ in Section \ref{sec: bispectrum} (see Figure \ref{fig:seedweight}).  

\paragraph{Phenomenology and parametric resonance} We explore the phenomenology of our model by retreating to the simplest exchange diagram governed by the following interactions:
\begin{align}
    S_{\rm int}=\int\dfrac{\mathrm{d}\eta\,\mathrm{d}^3\bm{x}}{\eta^4}\,\left(\rho\, \eta\,\pi'_c\sigma-\dfrac{1}{\Lambda}\eta^2\pi_c'^2\sigma-\dfrac{1}{\Lambda'}\eta^2(\partial_i\pi_c)^2\sigma\right)\,.
\end{align}
Solving the bootstrap equation to $\mathcal{O}(g^2)$ order yields an analytical expression for the cosmological collider signals in the squeezed limit of the curvature bispectrum of the form
\begin{align}
\nonumber \lim_{k_3\ll k_{1,2}}B(k_1,k_2,k_3) &\supset f_{\text{NL}}(\mu,\omega)\,\,P(k_1)\,P(k_3)\left(\dfrac{k_{12}}{k_3}\right)^{-3/2}\cos\left[\mu\log(\frac{k_{12}}{k_{3}})+\omega \log(-k_3\eta_0)\right]\\
&\quad +(\mu\to -\mu)\,,
\end{align}
where the signal strength is amplified by a UV resonance between the sub-horizon oscillations of the massive field and its time-dependent mass, thereby losing the Boltzmann suppression in the regime $\omega\gtrsim \mu$. Meanwhile, the running of the bispectrum (through the phase factor $\log(-k_3\eta_0)$, where $\eta_0$ is a fiducial conformal time) is uniquely determined by an unbroken discrete subgroup of the dilation symmetry, whereas its characteristic oscillations and scaling behaviour as a function of the momentum ratio $k_{12}/k_3$ reflect the standard super-horizon dilution and oscillations of the (unmodulated) heavy field in de Sitter.

We will also show that $f_{\rm{NL}}(\mu,\omega)$ at order $g^2$ has a singularity at the characteristic frequency $\omega=2\mu$. Using boundary eigenfrequency analysis, this apparent divergence can be shown to resum into an anomalous scaling exponent of cosmological collider signals,
\begin{align}
    B\sim \left(\frac{k_{12}}{k_3}\right)^{-\frac{3}{2}+\lambda_1 \pm i\mu}~,\quad \lambda_1=\frac{g^2}{4\mu}\left(\mu ^2+\frac{9}{4}\right)~.
\end{align}
This can be attributed to parametric resonance effects in the IR which also happens for $\omega_n=2\mu/n$, $n=1,2,3,\cdots$.

\paragraph{Numerical bootstrap} We also attempt a direct numerical bootstrap using finite differences, where the integro-differential equation reduces to a simple linear algebra equation
\begin{align}
    \mathcal{D} \mathbf{F}=\bm{\mathcal{S}}+ \mathcal{Q} \mathbf{F}~,
\end{align}
where $\bm{\mathcal{S}}\sim 1/k_T$ is the source and $\mathcal{D}$, $\mathcal{Q}$ denote the discretised differential operator $\hat{\Delta}_{12}+m_0^2$ and the memory kernel $K(q)$, respectively. Solving this matrix equation with appropriate boundary conditions and regularisation schemes, we find consistency with the aforementioned analysis both in the analytical solution in the perturbative regime and the scaling exponents in the non-perturbative regime.

\section{Model setup}\label{Setup}

Let us begin by sketching the general background of our model and highlighting some of the salient features we shall be concerned with. We shall work within the extended framework of the EFT of inflation that incorporates a massive degree of freedom in addition to the Goldstone.

\paragraph{Effective Field Theory of Inflation} As was explained in the introduction, the backdrop of our study is a quasi-single field \cite{Chen:2009zp} scenario in which the approximate shift symmetry of the Goldstone boson $\pi$ typically assumed in the EFT of inflation is explicitly broken\cite{Cheung:2007st, Flauger:2016idt, Behbahani:2011it, Senatore:2010wk}. 
For concreteness, we assume that the inflaton field is a canonical scalar interacting with an additional scalar field $\sigma$. The inflaton is assumed to be approximately massless while the scalar $\sigma$ is massive. In the unitary gauge (where $\pi=0$), the action of the system at leading order in derivatives and up to quadratic order in $\sigma$ and metric perturbations is given by
\begin{align}
\label{EFToI}
    S=\int \mathrm{d}^4x\,\sqrt{-g}\,&\Bigg[ \,\dfrac{1}{2}\Mpl^2 R+\Mpl^2\dot{H}(t)g^{00}-\Mpl^2(3H^2(t)+\dot{H})+\\ \nn
    & -\dfrac{1}{2}\bar{M}_1^2(t)(g^{\mu\nu}\partial_\mu\sigma\partial_\nu\sigma)+\dfrac{1}{2}\bar{M}_2^2(t)(g^{0\mu}\partial_\mu\sigma)^2-\dfrac{1}{2}\bar{M}_3^2(t)\sigma^2\\ \nn
    &\qquad\qquad \qquad-\tilde{M}_1(t)\delta g^{00}\sigma-\tilde{M}_2(t)(\delta g^{00})^2\sigma-\tilde{M}_3(t)\delta g^{00}\sigma^2
   \Bigg]\,,
\end{align}
where tadpole terms are excluded.

Performing the Stückelberg trick, $t\to t+\pi(t,\bm{x})$, and taking the decoupling limit, the second line yields the quadratic action for $\sigma$,  
\begin{align}
\label{sigmaquadratic}
    S^{(2)}_\sigma=\int \mathrm{d}^4x\,\sqrt{-g}\,\Big[\dfrac{1}{2}\dot{\tilde{\sigma}}^2-\dfrac{1}{2a^2}c_s^2(t)(\partial_i\tilde{\sigma})^2-\dfrac{1}{2}m^2(t)\tilde{\sigma}^2\Big]\,,
\end{align}
where $\tilde{\sigma}=(\bar{M}_1^2(t)+\bar{M}_2^2(t))^{1/2}\sigma$ is the canonically normalised massive field, while $c_s(t)$ and $m(t)$ are its effective sound speed and mass, respectively. These quantities can be determined from the unitary gauge Wilson coefficients $\bar{M}_i^2(t)$. Hereafter, we drop the tilde from $\sigma$ to avoid clutter. 
For convenience, we set the heavy field's sound speed to unity ($c_s=1$), while maintaining a generic time-dependence for mass.\footnote{More precisely, to have a well-defined Bunch-Davies initial condition for the heavy field, its mass $m(t)$ should grow slower than exponential in the infinite past, if at all.} Nevertheless, our integro-differential equations will be easily adaptable to time-dependent sound speeds $c_s(t)$; see Section \ref{generalverticesandmasses} for a brief discussion. 

Moreover, the positive- and negative-frequency mode functions of the heavy field will be denoted by $\sigma_+$ and $\sigma_-$, respectively, which we assume satisfy the Bunch-Davies initial condition, i.e. 
\begin{align}
    \lim_{\eta\to -\infty}\sigma_\mp(s,\eta)=-\dfrac{H \eta}{\sqrt{2k}}\exp(\pm ik\eta)\,.
\end{align}

There are several mixing terms between $\pi$ and $\sigma$ which emerge after setting 
\begin{align}
    g^{00}\to -1-2\dot{\pi}+(\partial_\mu\pi)^2\,,\qquad g^{0\mu}\partial_\mu\sigma\to -\dot{\sigma}+g^{\mu\nu}\partial_\nu\pi\partial_\mu \sigma\,,
\end{align}
and 
\begin{align}
    &\bar{M}_i^2(t)\to \bar{M}_i^2(t+\pi)\,,\qquad \tilde{M}_1(t)\to \tilde{M}_1(t+\pi)\,.
\end{align}
We will focus on tree-level contributions from the massive field to the power spectrum and the bispectrum of curvature perturbations. Consequently, we only need to keep track of mixing terms linear in $\sigma$ and at most quadratic in $\pi$. These terms are captured by the first two building blocks in the last line of \eqref{EFToI}, which after restoring $\pi$ yield
\begin{align}
\label{Smixing}
    S_{\text{mixing}}=\int \mathrm{d}^4x\,\sqrt{-g}\Big[\tilde{M}_1(t)(-2\dot{\pi}-\dfrac{1}{a^2}(\partial_i\pi)^2)-2\partial_t\tilde{M}_1(t)\,\pi\dot{\pi}+ \tilde{c}(t)\dot{\pi}^2\Big]\sigma\,,
\end{align}
where $\tilde{c}\equiv -4\tilde{M}_2+\tilde{M}_1$. Let us highlight the non-shift-symmetric cubic term $\pi\dot{\pi}\sigma$ which arises after expanding $\bar{M}_1(t)$ around $\pi=0$. As noted in \cite{Pajer:2024ckd, Werth:2023pfl, Pinol:2023oux}, this term dominates over the shift-symmetric vertices $\dot{\pi}^2\sigma$ and $(\partial_i\pi)^2\sigma$, around highly oscillatory backgrounds with $\partial_t \bar{M}_1(t)\gg H\,\bar{M}_1(t)$.

Now we come back to the first line of \eqref{EFToI}, which generates the following action for $\pi$: 
\begin{align}
    S_\pi=\int \mathrm{d}^4x\,\sqrt{-g}\Mpl^2 |\dot{H}|(\partial_\mu \pi)^2+\dots\,,
\end{align}
with ellipses standing for higher-order terms in $\pi$, originating from the expansion of $H(t+\pi)$ and $\dot{H}(t+\pi)$ around $\pi=0$. These terms are extensively studied in the context of axion monodromy inflation, where they give rise to characteristic resonant features in the two- and higher-point functions\cite{Chen:2008wn,Flauger:2010ja,Leblond:2010yq} (see \cite{Creminelli:2024cge, Creminelli:2025tae, DuasoPueyo:2023kyh} for recent discussions). In this work, however, we will neglect these self-interactions and focus on correlators induced by the heavy field exchange. To further simplify our analysis, we also take $\dot{H}(t)$ to be nearly constant, thereby preserving the shift symmetry of $\pi$ except through its coupling to the heavy sector.  As a result, the mode function of the (canonically normalised) Goldstone $\pi_c=(2|\dot{H}|)^{1/2}\Mp\pi$ in our setup will be well captured by that of a massless scalar in de Sitter, i.e. 
\begin{align}
    \pi_c^\pm(k,\eta)=\dfrac{H}{\sqrt{2k^3}}(1\pm ik\eta)e^{\mp ik\eta}\,.
\end{align}

\begin{figure}[ht]
    \centering
    \includegraphics[scale=0.62]{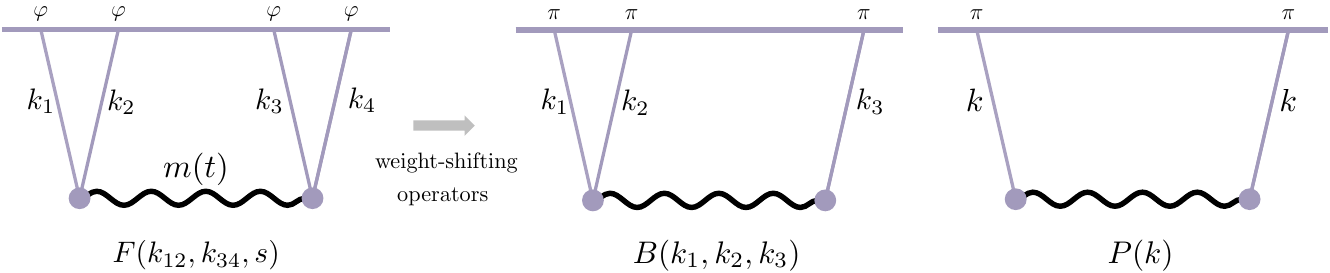}
    \caption{The seed four-point function vs. the single-exchange diagrams for the bispectrum and the power spectrum of $\zeta$, related by the action of weight-shifting operators, see Section \ref{sec: bispectrum}} 
    \label{fig:seedweight}
\end{figure}

\paragraph{Seed four-point functions} By combining the quadratic and cubic vertices in \eqref{Smixing}, one can construct single-exchange diagrams, with two or three massless external legs (see Figure \ref{fig:seedweight}). Let us consider vertices of the forms  $\lambda_1(t)\dot{\pi}\sigma$, $\lambda_2(t)\pi\dot{\pi}\sigma$, $\lambda_3(t)\dot{\pi}^2\sigma$ or $\lambda_4(t)(\partial_i\pi)^2\sigma$, where the time-dependent coupling $\lambda_i(t)$ is specified in each case by $\tilde{M}_1(t)$ and $\tilde{c}(t)$. 
Diagrams with such vertices can be efficiently extracted from a set of \textit{seed four-point functions}, defined as single-exchange diagrams with four external conformally coupled fields $\vpi$ and cubic vertices of the form $\lambda_i(t) \vpi^2 \sigma$. 
Thanks to the simpler bulk-boundary propagator of the conformally coupled field, these exchange diagrams are easier to handle analytically than their massless counterparts. Most notably, as will be shown in the next section, these diagrams satisfy a set of \textit{integro-differential equations} (IDEs), which are the generalisations of the bootstrap differential equations for de Sitter seed diagrams (with constant intermediate masses)\cite{Arkani-Hamed:2018kmz}.   
Once these IDEs are solved, the resulting seed four-point functions can be mapped onto the power spectrum and the bispectrum of curvature perturbations. We will achieve this in Section \ref{sec: bispectrum} through a set of \textit{weight-shifting operators} and by taking appropriate soft limits to reduce from the four-point to the three- and the two-point kinematics, along the lines of \cite{Baumann:2019oyu, Jazayeri:2022kjy, Pimentel:2022fsc}. 

We begin the analysis of the seed four-point function by writing it as
\begin{align}
    \langle \vpi(\bm{k}_1)\vpi(\bm{k}_2)\vpi(\bm{k}_3)\vpi(\bm{k}_4)\rangle=\dfrac{\eta_{\rm end}^4}{4k_1k_2k_3k_4}F(k_{12},k_{34},s)+t,u\,\,\text{channels}\,,
\end{align}
where $k_{ij}=k_i+k_j$, and $\eta_{\rm end}$ is the conformal time at the end of inflation. 
Using the Schwinger-Keldysh formalism we obtain
\begin{align}\label{fourpointall}
    F(\{k\},s)=F_{++}+F_{--}+F_{+-}+F_{-+}\,,
\end{align}
where
\begin{keyeqn}
\begin{align}
\label{genericfourpoint}
    F_{\aa \bb}(\{k\},s)=-\aa \bb\int_{-\infty(1- i\aa\epsilon)}^{0} {\rm{d}}\eta\frac{\lambda_L(\eta)}{\eta^2}\int_{-\infty(1- i\bb\epsilon)}^{0}{\rm{d}}\eta'\frac{\lambda_R(\eta')}{\eta'^2}\, e^{ i\aa k_{12}\eta} \,e^{i\bb k_{34}\eta'} G_{\aa \bb}(s, \eta, \eta')\,.
\end{align}
\end{keyeqn}
Here $\aa,\bb=\pm$ denote the Schwinger-Keldysh indices, $\lambda_{L}$ ($\lambda_R$) are the left (right) couplings in the diagram that could explicitly depend on time, and $G_{\aa\bb}$ are the bulk-bulk propagators in the Schwinger-Keldysh formalism satisfying
\begin{align}
    \left[\partial_\eta^2-\dfrac{2}{\eta}\partial_\eta+k^2+\dfrac{m^2(\eta)}{\eta^2 }\right]G_{\pm\pm}(s,\eta,\eta')&=\mp i\,\eta^{\prime 2}\delta(\eta-\eta')\,,\\ \nn
    \left[\partial_\eta^2-\dfrac{2}{\eta}\partial_\eta+k^2+\dfrac{m^2(\eta)}{\eta^2 }\right]G_{\pm\mp}(s,\eta,\eta')&=0\,.
\end{align}
In terms of the mode functions, 
\begin{align}
    G_{\pm\pm}(s,\eta,\eta')&=\sigma_\pm(s,\eta)\sigma_\mp(s,\eta')\theta(\eta-\eta')+(\eta\leftrightarrow\eta')\,,\\ \nn
    G_{\pm\mp}(s,\eta,\eta')&=\sigma_\mp(s,\eta)\sigma_\pm(s,\eta')\,.
\end{align}
Notice that $G_{--}=G^*_{++}$ and $G_{-+}=G^*_{+-}$ and, consequently, $F_{--}=F^*_{++}$ and $F_{-+}=F^*_{+-}$. These relations ensure that $F(\{k\},s)$ is a real quantity. For future convenience, we also decompose the mass into a constant and a time-dependent piece by writing $m^2(\eta)=m_0^2+\Delta m^2(\eta)$.

It is useful to decompose our four-point $F(\{k\},s)$ into a superposition of exchange diagrams with monochromatic vertices, oscillating as $e^{\pm i\Omega t}$. So we Fourier transform the couplings as 
\begin{align}
    \lambda_{L,R}=\int \dfrac{{\rm{d}}\Omega}{2\pi}\tilde{\lambda}_{L,R}(\Omega)\,e^{i\Omega t}=\int \dfrac{{\rm{d}}\Omega}{2\pi}\tilde{\lambda}_{L,R}(\Omega)\,(\eta/\eta_0)^{-i\Omega}\,,
\end{align}
where $\eta_0$ is an arbitrary, fiducial conformal time. Plugging these transformations back into the Schwinger-Keldysh time integral, we get 
\begin{align}
    F(k_{12},k_{34},s)=\int_{-\infty}^{+\infty} \dfrac{{\rm{d}}\Omega_L}{2\pi}\dfrac{{\rm{d}}\Omega_R}{2\pi}\tilde{\lambda}_{L}(\Omega_L)\,\tilde{\lambda}_{R}(\Omega_R)\,F^{\Omega_L,\Omega_R}(k_{12},k_{34},s)\,,
\end{align}
in which the correlator basis $F^{\Omega_L,\Omega_R}$ is defined by 
\begin{align}
\nn
    F^{\Omega_L,\Omega_R}=\sum_{\aa,\bb=\pm } F^{\Omega_L,\Omega_R}_{\aa\bb} 
    &=\sum_{\aa,\bb}(-\aa\bb)\int_{-\infty(1- i\aa \epsilon)}^{0}\int_{-\infty(1- i\bb \epsilon)}^{0}\dfrac{{\rm{d}}\eta}{\eta^2}\dfrac{{\rm{d}}\eta'}{\eta'^2}\,\left(\dfrac{\eta}{\eta_0}\right)^{-i\Omega_L}\,\left(\dfrac{\eta'}{\eta_0}\right)^{-i\Omega_R}\\ \nn
    &\,\times e^{i\aa k_{12}\eta} \,e^{i\bb k_{34}\eta'} G_{\aa\bb}(s, \eta, \eta')\,.
\end{align}
Note that, for a time-independent intermediate mass, $F^{\Omega_L,\Omega_R}$ satisfies a pair of differential equations governing its behaviour as a function of the external kinematic variables $k_{12}$ and $k_{34}$ \cite{Qin:2023ejc}. These equations can be analytically solved with suitable boundary conditions imposed e.g. by analyticity and consistent factorisation around certain poles. We will see in the next section that a new set of \textit{integro-differential equations} takes over when the intermediate mass evolves with time. Specialising to monochromatic masses ($\Delta m^2\propto \cos(\omega t)$) and time-independent vertices, we will solve these integro-differential equations using additional constraints from \textit{microcausality} and the \textit{Bunch-Davies} initial state, and to leading order in the oscillation amplitude. 

\section{Boundary integro-differential equations}\label{BdyIDEg2BootstrapSection}

In this section, we switch gear to boundary techniques and translate the Schwinger-Keldysh time integrals into constraint equations in kinematics space. We will start off with general scale-breaking scenarios, showcasing the full capability of our method, before specialising down to a monochromatic model and working towards solving it.

\subsection{General vertices, sound speeds and masses}\label{generalverticesandmasses}

To declutter our notation, we suppress the superscripts of $F_{\pm\pm}^{\Omega_L,\Omega_R}$. Utilising the bulk differential equation for the massive propagator, one can show after successive integration by parts that
\begin{align}
\label{bootstrapset}
    \hat{{\Delta}}_{12}\,F_{\aa\bb}&=\Gamma\left(1-i\Omega_T\right)\,k_T^{-1}(-k_T\eta_0)^{i \Omega_T}\,c_{\aa\bb}+J_{\aa\bb}(k_{12},k_{34},s)\,,\\ 
      \hat{{\Delta}}_{34}\,F_{\aa\bb}&=\Gamma\left(1-i\Omega_T\right)\,k_T^{-1}(-k_T\eta_0)^{i \Omega_T}\,c_{\aa\bb}+J_{\aa\bb}(k_{12},k_{34},s)\,,
\end{align}
in which,   
\begin{align}
    c_{++}=\exp(-\pi\,\Omega_T/2)\,,\quad c_{--}=\exp(-\pi\,\Omega_T/2)\,,\quad c_{\pm\mp}=0\,,\quad \text{and}\,\quad \Omega_T=\Omega_L+\Omega_R\,,
\end{align}
and the operators $\hat{{\Delta}}_{12,34}$ are defined as
\begin{align}
   \hat{{\Delta}}_{12}&=(k_{12}^2-s^2)\partial^2_{k_{12}}+2(1-i\Omega_L)k_{12}\partial_{k_{12}}+\left[m_0^2+(1+i\,\Omega_L)(-2+i\Omega_L)\right]\,,\\ 
     \hat{{\Delta}}_{34}&=(k_{34}^2-s^2)\partial^2_{k_{34}}+2(1-i\Omega_R)k_{34}\partial_{k_{34}}+\left[m_0^2+(1+i\,\Omega_R)(-2+i\Omega_R)\right]\,,
\end{align}
and finally, 
\begin{align}
\label{Jppmm}
    J_{\aa\bb}(k_{12},k_{34})=\aa\bb\int \dfrac{{\rm{d}}\eta}{\eta^{2}}\, \dfrac{{\rm{d}}\eta'}{\eta'^{2}}\left(\dfrac{\eta}{\eta_0}\right)^{-i\Omega_L}\,\left(\dfrac{\eta'}{\eta_0}\right)^{-i\Omega_R}\Delta m^2(\eta)\,\,e^{i\aa k_{12}\eta} \,e^{i \bb k_{34}\eta'} G_{\aa\bb}(s, \eta, \eta')\,.
\end{align}
Without knowledge of $J_{\aa\bb}$, the equations \eqref{bootstrapset} would be of little use in finding $F_{\aa\bb}$. Fortunately, this gap can be closed by directly solving $J_{\aa\bb}$ in terms of $F_{\aa\bb}$. To achieve this, we Fourier transform the mass term $\Delta m^2$,
\begin{align}
    \Delta m^2=\dfrac{1}{2}m_0^2\int_{-\infty}^{+\infty} {\rm{d}}\omega\,\rho_\omega\,e^{i\omega t}=\dfrac{1}{2}m_0^2\int_{-\infty}^{+\infty} {\rm{d}}\omega\,\rho_\omega\,\left(\frac{\eta}{\eta_0}\right)^{-i\omega/H}\,,
\end{align}
where $\rho_\omega$ ($=\rho^*_{-\omega}$) is a dimensionless quantity.  Inserting the above transformation into \eqref{Jppmm} and using the identities
\begin{align}
    \left(\dfrac{\eta}{\eta_0}\right)^{-i\omega}&=\dfrac{e^{-\frac{\pi\omega}{2}}}{\Gamma(i\omega)}\int_0^\infty\dfrac{{\rm{d}}q}{q^{1-\epsilon}}\,(-q\eta_0)^{i\omega}e^{-\epsilon q}e^{iq\eta}\,\label{fracDerivativeFormula}\\ \nn
    &=\dfrac{e^{+\frac{\pi\omega}{2}}}{\Gamma(i\omega)}\int_0^\infty\dfrac{{\rm{d}}q}{q^{1-\epsilon}}\,(-q\eta_0)^{i\omega}e^{-\epsilon q}e^{-iq\eta}\,\qquad (\epsilon\to 0^+)\,,
\end{align}
we find an integral relation between $J_{\aa\bb}$ and $F_{\aa\bb}$. Note that the convergence of the time integrals requires the first and second lines to be substituted into $J_{+\pm}$ and $J_{-\pm}$, respectively.  An inspiring way to understand these identities is to take the analogy of integer power-shifting derivatives, which give e.g. $\partial_{k_{12}} e^{i k_{12}\eta}\sim \eta e^{i k_{12}\eta}$, and apply their generalisation to \textit{fractional derivatives}, yielding $D^{\alpha}_{k_{12}} e^{i k_{12}\eta}\sim \int_{k_{12}}^\infty \d p\, (p-k_{12})^{-1-\alpha}e^{ip\eta} \sim \eta^\alpha e^{i k_{12}\eta}$, essentially giving rise to \eqref{fracDerivativeFormula}. 

Plugging this relation back into \eqref{bootstrapset} finally yields a closed set of \textit{integro-differential equations} (IDEs) for $F_{\aa\bb}$:
\begin{keyeqn}
\begin{align}
\nn
    \hat{{\Delta}}_{12}\,F_{\pm\pm}&=\Gamma\left(1-i\Omega_T\right)\,k_T^{-1}(-k_T\eta_0)^{i \Omega_T}\,c_{\pm\pm}+\int_\omega \int_{0}^\infty {\rm{d}}q\,\Pi_{\pm\pm}(q,\omega)\,F_{\pm\pm}(k_{12}+q,k_{34},s)\\ 
    \label{IDEs}
     \hat{{\Delta}}_{12}\,F_{\mp\pm}&=\int_\omega\,\int_{0}^\infty {\rm{d}}q\,\Pi_{\mp\pm}(q,\omega)\,F_{\mp\pm}(k_{12}+q,k_{34},s)\,.
\end{align}
\end{keyeqn}
\noindent in which we have introduced the following kernels:  
\begin{align}
\nn
    \Pi_{+-}(q,\omega)=\Pi_{++}(q,\omega)&=-\dfrac{1}{2}\,m_0^2\left(\dfrac{e^{-\frac{\pi\omega}{2}}}{\Gamma( i\omega)}\,\rho_\omega\right)\dfrac{e^{-\epsilon q}}{q^{1-\epsilon}}(-q\eta_0)^{+i\omega}\,,\\ \label{kernels}
    \Pi_{-+}(q,\omega)=\Pi_{--}(q,\omega)&=-\dfrac{1}{2} m_0^2\left(\dfrac{e^{+\frac{\pi\omega}{2}}}{\Gamma(i\omega)}\,\rho_\omega\right) \dfrac{e^{-\epsilon q}}{q^{1-\epsilon}}(-q\eta_0)^{+i\omega}\,,
\end{align}
in addition to \eqref{IDEs}, a parallel set of equations follow from exchanging $\hat{{\Delta}}_{12}$ with $\hat{{\Delta}}_{34}$ on the left-hand side and simultaneously $F_{\aa\bb}(k_{12}+ q,k_{34},s)$ with  $F_{\aa\bb}(k_{12},k_{34}+ q,s)$ on the right-hand side. (An alternative bootstrap approach for exchange diagrams with monochromatic masses is presented in Appendix \ref{vertexbootstrap}, where an infinite set of recursive ordinary differential equations are derived for an infinite array of exchange diagrams whose vertices oscillate as $\exp(in\omega)$, with $n\in \mathbb{Z}$.)

A few remarks on our integro-differential equations are in order.
First, we note that unlike its Schwinger-Keldysh components $F_{\aa\bb}$, 
the full correlator $F=\sum_{\aa\bb}F_{\aa\bb}$ does not satisfy any IDE. This is due to the difference between the IDE kernels, i.e. $\Pi_{+\pm}\neq \Pi_{-\mp}$. In fact, it may seem discouraging that our IDEs are based on individual Schwinger-Keldysh components of the graph, which are not well-defined observables in isolation. As an alternative, one may choose to work with the \textit{wavefunction coefficient} $\psi_4$ associated with the same exchange graph, with the advantage of being an observable that satisfies an IDE identical to that of $F_{++}$. However, even after solving the IDE for $\psi_4$, one would still require the mode function at $\eta_{\text{emd}}$ in order to compute relevant boundary observables at the end of inflation, such as the power spectrum and the bispectrum. This justifies our preference for working with the Schwinger-Keldysh components $F_{\aa\bb}$, allowing us to directly extract the full correlator without any explicit reference to the mode function $\sigma(s,\eta_{\rm end})$, the closed-form of which is unknown.  See Section \ref{sec: bispectrum} for the derivations.

Second, we highlight again that the integro-differential equations \eqref{IDEs} are intrinsically non-local in kinematic space only in a restricted sense: At fixed $k_{34}$ and $s$, the behaviour of $F_{\aa\bb}(k_{12})$ on the LHS is only tied to four-point configurations that are more squeezed on the RHS, i.e. those with $k'_{12}=k_{12}+q>k_{12}$. This kinematic retardation is a direct avatar of the Bunch-Davies initial condition which is built into $F_{\aa\bb}$ via the $i\epsilon$ prescription. Indeed, in deriving the IDEs, the $i\epsilon$ prescription for $F_{+\pm}$($F_{-\pm}$) required us to decompose the mass perturbation $\Delta m^2$ into negative (positive) frequency plane-waves $e^{+iq\eta}$ ($e^{-iq\eta}$), before inserting them into $J_{+\pm}$($J_{-\pm}$) in \eqref{Jppmm}. Consequently, $F_{\aa\bb}(k_{12}+q,k_{34},s)$ always appears with $q>0$ in \eqref{IDEs}. 

Third, until now we have set the massive field's sound speed for simplicity to unity, but a time-dependent speed $c_s(t)$ can be easily incorporated into the integro-differential equation. Specifically, assuming the decomposition 
\begin{align}
    c_s^2(\eta)=c_{0}^2+\dfrac{1}{2}c_0^2\int_{-\infty}^{+\infty} {\rm{d}}\omega\,\tilde{\rho}_\omega\,\left(\frac{\eta}{\eta_0}\right)^{-i\omega/H}\,,
\end{align}
one can derive an analogous equation to \eqref{bootstrapset}, 
\begin{align}
\nn
    \tilde{{\Delta}}_{12}\,F_{\aa\bb}&=\Gamma\left(1-i\Omega_T\right)\,k_T^{-1}(-k_T\eta_0)^{i \Omega_T}\,c_{\aa\bb}+\int_\omega \int_{0}^\infty {\rm{d}}q\,\Pi_{\aa\bb}(q,\omega)\,F_{\aa\bb}(k_{12}+q,k_{34},s)\\ 
    &+\int_\omega \int_{0}^\infty {\rm{d}}q\,\tilde{\Pi}_{\aa\bb}(q,\omega)\times\,s^2\frac{\partial^2}{\partial k_{12}^2}\,F_{\aa\bb}(k_{12}+q,k_{34},s)\,,
\end{align}
where the new kernels $\tilde{\Pi}_{\aa\bb}$, after replacing $m_0^2\,\rho_\omega\to c_0^2\tilde{\rho}_\omega$, are given by the same expressions as \eqref{kernels}, and the new derivative operator on the LHS is defined as: 
\begin{align}
   \tilde{\Delta}_{12}=(k_{12}^2-c_0^2s^2)\partial^2_{k_{12}}+2(1-i\Omega_L)k_{12}\partial_{k_{12}}+\left[m_0^2+(1+i\,\Omega_L)(-2+i\Omega_L)\right]\,.
\end{align}
As a final technical comment, note that integrations over frequency $\omega$ and comoving momentum $q$ in the IDEs are not generically interchangeable. Doing so may cause an artificial UV divergence in the $\omega$ integral, which could be avoided by prior integration over $q$. Alternatively, one can rely on the assumption that physical time-dependences should only have a compact support in the frequency domain i.e. $\lim_{|\omega|\to\infty}|e^{\alpha \omega}\rho_\omega|=0$ for any $\alpha$ to avoid infinite energies.

\subsection{Monochromatic masses}

A particular case of interest is a monochromatic time-dependent mass, i.e. 
\begin{align}
\label{massmodulation}
    m^2&=m_0^2\left(1+g^2\cos[\omega(t-t_0)]\right)\\ \nn
    &=m_0^2+\dfrac{1}{2}g^2\,m_0^2
\,\left[\left(\frac{\eta}{\eta_0}\right)^{i\omega }+\left(\frac{\eta}{\eta_0}\right)^{-i\omega }\right]\,,
\end{align}
where $\eta_0=-1/a(t_0)$. 
To avoid tachyonic instabilities, we require $m^2$ to remain positive at all times, i.e. $g<1$. Cosinusoidal time-dependent masses like above could arise during inflation in  scenarios such as axion monodromy, where the continuous shift symmetry of the inflaton $\phi$ is broken due to non-perturbative effects down to the discrete subgroup  
\begin{align}
    \phi\to \phi+2\pi n\,f\,,
\end{align}
where $f$ is the axion field decay constant, and $n$ is an integer. See \cite{Chen:2008wn,Flauger:2014ana,Silverstein:2008sg, Flauger:2009ab,McAllister:2008hb} for an incomplete list of references. This opens the possibility of inflaton-dependent masses of the form $m^2=m_0^2[1+g^2\cos(\phi/f)]$, with the associated oscillation frequency  
\begin{align}
    \omega=\dot{\bar{\phi}}/f\,,
\end{align} 
around slow-roll backgrounds. 
Generically, similar oscillations would also appear in the cubic couplings, but to keep the problem's complexity under control, we take the vertices to be time-independent, setting $\Omega_L=\Omega_R=0$. Nevertheless, monochromatic vertices could be incorporated into our proposed formulation after minor modifications.   

Assuming time-evolving masses of the form \eqref{massmodulation}, the integro-differential equations for the exchange diagram \eqref{IDEs} simplify to
\begin{keyeqn}
\begin{align}
\label{inhomo}
    \hat{{\Delta}}_{12}\,F_{++}(k_{12},k_{34},s)&=\dfrac{1}{k_T}+\int_{0}^\infty {\rm{d}}q\,K(q)\,\,F_{++}(k_{12}+q,k_{34},s)\,,\\ 
    \label{homo}
     \hat{{\Delta}}_{12}\,F_{+-}(k_{12},k_{34},s)&=\int_{0}^\infty {\rm{d}}q\,K(q)\,F_{+-}(k_{12}+q,k_{34},s)\,,
\end{align}
\end{keyeqn}
\noindent where
\begin{align}
    K(q)=-\dfrac{1}{2}\,m_0^2 g^2\dfrac{e^{-\pi\omega/2}}{\Gamma( i\,\omega)}\dfrac{e^{-\epsilon q}}{q^{1-\epsilon}}(-q\eta_0)^{+i\omega}+(\omega\leftrightarrow-\omega)\,,
\end{align}
and 
\begin{align}
    \hat{{\Delta}}_{12}&=(k_{12}^2-s^2)\partial^2_{k_{12}}+2k_{12}\partial_{k_{12}}+\left(m_0^2-2\right)\,.
\end{align}
The remaining Schwinger-Keldysh components $F_{--}$ and $F_{-+}$ are governed by the same equations, complex conjugated. Let us stress once again that the full correlator $F=\sum_{\aa,\bb}F_{\aa\bb}$ does not satisfy any similar IDE since $K(q)$ is not a real function.\\

The $\pm\mp$ component could be further simplified by noting that the corresponding time integrals factorise, allowing them to be expressed as 
\begin{align}
    F_{+-}=f(k_{12},s)f^*(k_{34},s)\,,\qquad F_{-+}=f^*(k_{12},s)f(k_{34},s)\,,
\end{align}
in which $f$ is defined by 
\begin{align}
\label{timeintf}
    f(k_{12},s)=+i\int^0_{-\infty(1-i\epsilon)} \dfrac{{\rm{d}}\eta}{\eta^2}\,e^{ik_{12}\eta}\sigma_-(s,\eta)\,. 
\end{align} 
This object is proportional to the cubic wavefunction coefficient, $\psi_3=f(k_{12},s)/\sigma_-(s,\eta_{\rm end})$ and satisfies the same IDE as \eqref{homo}. While we will slightly misuse the terminology and refer to $f$ as the three-point function/building block, the precise relation to the actual three-point function is given by  $\langle\phi(\bm{k}_1)\phi(\bm{k}_2)\sigma(-\bm{s},\eta_0)\rangle'=\frac{\eta_{\rm end}^2}{2k_1 2k_2}\text{Re}\lbrace f(k_{12},s)\sigma_-(s,\eta_0)\rbrace $.
\\

\subsection{Symmetries, microcausality and analyticity}\label{symmetris-micro-anal}

Before solving the bootstrap equations, it is instructive to study the properties of their solutions based on general principles. In particular, we will focus in this section on the implications of \textit{symmetries}, \textit{microcausality}, and \textit{analyticity}.

\paragraph{Symmetries:} Mass oscillations explicitly break the continuous dilatation invariance of the background de Sitter, while preserving the discrete subset
\begin{align}
    \eta\to e^{-2\pi n/\omega} \eta\,,\qquad \bm{x}\to e^{-2\pi n/\omega}\bm{x}\,\quad (n\in \mathbb{Z})\,.
\end{align}
It is a simple exercise to show that $F_{\aa\bb}$ under this symmetry transforms as  
\begin{align}
    \bm{k}_i\to e^{2\pi n/\omega}\bm{k}_i\,,\qquad F_{\aa\bb}\to e^{-2\pi n/\omega}F_{\aa\bb}\,. 
\end{align}
The transformation rule for e.g. $F_{++}$ can be made manifest by writing
\begin{keyeqn}
\begin{align}
\label{Fppdecomposition}
    F_{++}(k_{12},k_{34},s)&=s^{-1}\sum_{l=-\infty}^{+\infty}(-s\eta_0)^{il\omega}\,F_l(u,v)\,,
\end{align}
\end{keyeqn}
where $u$ and $v$ are defined as
\begin{align}
   u=\dfrac{s}{k_{12}}\,,\qquad v=\dfrac{s}{k_{34}}\,. 
\end{align}
As a side note, the full four-point $F(k_{12},k_{34},s)$ of course admits the same decomposition as \eqref{Fppdecomposition}. However, to solve their integro-differential equations, it will be more useful to decompose each Schwinger-Keldysh component separately, as done above. 

\begin{figure}
    \centering
    \includegraphics[scale=2.0]{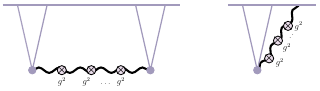}
    \caption{Four-point and three-point function with multiple mass insertions.}
    \label{fig:massinsertions}
\end{figure}

Similarly, the three-point function can be expressed as 
\begin{keyeqn}
    \begin{align}
    f(k_{12},s)=\dfrac{1}{\sqrt{s}}\,\sum_{l=-\infty}^{+\infty}(-s\eta_0)^{il\omega}\,f_l(u)\,.
\end{align}
\end{keyeqn}
Inserting the above mode expansions into the respective IDEs does not yield any major simplification, other than producing a recursive set of IDEs for $F_l$ and $f_l$. Despite this recursive nature, the decomposition becomes especially convenient in perturbation theory, where---up to a fixed order in $g^2$---it is sufficient to include only a finite number of harmonics in $l$. To make this simplification more manifest, one can associate the ${\cal O}(g^{2l})$ contribution with a perturbative graph, as in Figure \ref{fig:massinsertions}, consisting of $l$ mass insertion vertices of the form $\Delta m^2(t)\,\sigma^2$. This structure implies the following scaling behaviour: 
\begin{align}
    F_l(u)\sim {\cal O}(g^{2|l|})\,,\qquad f_l(u)\sim {\cal O}(g^{2|l|})\,,
\end{align}
which ensures that, to describe the exchange diagram $F_{++}$ and its three-point building block $f$ at order ${\cal O}(g^2)$, we only need to retain the three leading harmonics, $l=0,\pm 1$.

\paragraph{Microcausality:} Next we explore how the analytic structure of the exchange diagram is constrained by microcausality. A theory is micro-causal if the commutation relation of its local operators vanish outside the spacetime light cone, i.e. 
\begin{align}
    [{\cal O}(\eta,\bm{x}),{\cal O}(\eta',\bm{x}')]=0\,,\qquad |\bm{x}-\bm{x}'|>|\eta-\eta'|\,.
\end{align}
This condition must be preserved, even in the absence of the Poincaré symmetry, by every local operator in the theory. This includes the heavy field operator $\hat{\sigma}(\eta,\bm{x})$ in our cosmological Lorentz-violating setup. The vanishing of the commutator outside the light cone has profound consequences for the analytic structure of the retarded propagator in momentum space. It implies, in particular, that the Fourier transformation of the retarded propagator,   
\begin{align}
\nn
    G_R(\bm{k},\eta,\eta')&=\int {\rm{d}}^3\bm{x}\,e^{-i\bm{k}\cdot\bm{x}}\,\langle\Omega|\left[\hat{\sigma}(\eta',0),\hat{\sigma}(\eta,\bm{x})\right]|\Omega\rangle\,\theta(\eta'-\eta)\,,\\ 
    &=\Big[\sigma_+(k,\eta')\sigma_-(k,\eta)-\sigma_-(k,\eta')\sigma_+(k,\eta)\Big]\,\theta(\eta'-\eta)\,,
\end{align}
must be analytic everywhere in the complex plane of $\bm{k}$, including around the origin $\bm{k}=0$, see \cite{Hui:2025aja,Colas:2025ind} for a recent discussion. This analyticity directly translates to the analyticity of the four-point exchange diagram in the vicinity of $\bm{s}=0$, which can be shown, along the lines of \cite{Tong:2021wai,Qin:2023bjk,Ema:2024hkj}, by decomposing the $++$ Schwinger-Keldysh propagator as
\begin{align}
   G_{++}(s,\eta,\eta')=G_R(s,\eta,\eta')+\sigma_+(s,\eta)\sigma_-(s,\eta')\,. \label{eq: propagator_Cut}
\end{align}
Inserting this decomposition into the time integral \eqref{genericfourpoint}, we conclude that the non-analytic behaviour of $F_{++}$ as a function of $\bm{s}$ (or equivalently $s^2=\bm{s}\cdot\bm{s}$) could only be due to the second contribution above, which is devoid of any Heaviside step function. 

Replacing the Wightman propagator according to \eqref{factorisedFpp} within the Schwinger-Keldysh integral for $F_{++}$  yields
\begin{align}
\nn
    F_{++}(k_{12},k_{34},s)&=-\lambda^2\left(\int_{-\infty(1- i\epsilon)}^{0} \frac{{\rm{d}}\eta}{\eta^2}\,e^{+ik_{12}\eta}\sigma_+(s,\eta)\right)\times \left(\int_{-\infty(1- i\epsilon)}^{0}\frac{{\rm{d}}\eta'}{\eta'^2}\,  \,e^{+ik_{34}\eta'}\,\sigma_-(s,\eta')\right)\\ 
    \label{factorisedFpp}
    &+\text{analytic in}\,\, s^2\,.
\end{align}
The first line, generically featuring branch point singularities\footnote{Note that $s=\sqrt{\bm{s}^2}$ is itself non-analytic as a function of the spatial vector $\bm{s}$, therefore terms with odd powers of $s$ can only appear in the factorised fashion dictated by \eqref{factorisedFpp}.} in the soft limit $s\to 0$, can be expressed in terms of the three-point building block $f$,  
\begin{keyeqn}
\begin{align}
    F_{++}(k_{12},k_{34},s)=- [f(-k_{12}-i\epsilon,s)]^*\times f(k_{34}-i\epsilon,s)+\text{analytic in}\,\,s^2\,,\label{eq: soft_Fac2}
\end{align}
\end{keyeqn}
where we have assumed $k_{12},k_{34}$ and $s$ all to be positive. 
Quite remarkably, the above factorisation property is \textit{non-perturbative} in $g^2$ because it solely relies on the commutation of the heavy field outside the lightcone, independently of its time-dependent mass. In other words, the detailed knowledge of the mode function is not necessary to arrive at \eqref{eq: soft_Fac2}.

However, the derivation of \eqref{eq: soft_Fac2} relied on the critical assumption that the time integration itself does not introduce additional singularities in $\bm{s}$.  
One can verify this assumption within a sufficiently small radius around $\bm{s}=0$, where the bulk-bulk propagator $G_{++}$ can be expanded into a power series in $s^2$. This expansion typically includes both integer and fractional powers of $s^2$. After substituting the propagator with this Taylor expansion, the corresponding Schwinger-Keldysh integral, thanks to the $i\epsilon$ prescription, will still converge for $k_{12},k_{34}>0$, without introducing new singularities near $\bm{s}=0$. As a result, the final answer $F_{++}$ remains analytic around $\bm{s}=0$ except for branch-point singularities descending exclusively from the Wightman propagator at fixed conformal times $\eta,\eta'$. Notably, away from the origin $\bm{s}=0$, additional \textit{partial energy} singularities emerge from time integration, as will be discussed shortly. 

We note in passing that \eqref{eq: soft_Fac2} effectively generalises the cutting rule for non-local cosmological collider signals \cite{Tong:2021wai,Qin:2023bjk,Ema:2024hkj} to scale-breaking cases.   

For computing the bispectrum and the power spectrum in our setup, it is sufficient to solve the seed exchange diagram $F_{++}$ only for soft configurations with $\bm{k}_4=0$ (hence $k_{34}=k_3$ and $s=k_3$). So it is convenient to have a factorisation theorem around $\bm{k}_3=0$ directly for $F(k_{12},k_3,k_3)$. In fact, $F(k_{12},k_3,k_3)$ turns out to factorise in this limit in almost the same way as \eqref{eq: soft_Fac2}:  
\begin{keyeqn}
\begin{align}
    F_{++}(k_{12},k_{3},k_3)=- [f(-k_{12}-i\epsilon,k_3)]^*\times f(k_{3}-i\epsilon,k_3)+\text{analytic in}\,\,k_3\,, \label{eq: soft_Fac}
\end{align}
\end{keyeqn}
except that the second contribution is now analytic in $k_3$, not necessarily in $k_3^2$, as \eqref{eq: soft_Fac2} would have naively suggested. Indeed, since this non-factorisable contribution originates from replacing the Feynman propagator $G_{++}$ with the retarded one $G_R$ in $F_{++}$, it inherits odd powers of $k_3$ from the right-vertex bulk-boundary propagator($\propto e^{ik_{34}\eta'}=\,e^{ik_3\,\eta'}$). In more detail, the time ordering $\eta'>\eta$ enforced by the retarded propagator $G_R$ guarantees the convergence of the integral over $\eta'$, even after substituting $e^{ik_3\,\eta'}$ with its Taylor series within the Schwinger-Keldysh integrand. From this substitution, additional odd powers of $k_3=\sqrt{k_3^2
}$ appear, multiplying even powers of $k_3$ already present in $G_R$. This explains the difference between \eqref{eq: soft_Fac2} and \eqref{eq: soft_Fac}. Fortunately, odd powers of $k_3$ are the only type of non-analyticities that taking the soft limit $\bm{k}_4\to 0$ in $F_{++}$ can ever introduce; fractional powers of $k_3$ should still factorise in $F_{++}$ according to \eqref{eq: soft_Fac}. This factorisation will serve as a powerful bootstrap condition for solving the IDEs. 

\paragraph{Analyticity:} In addition to a branch point at $s=0$, Schwinger-Keldysh diagrams contain singularities in the complex plane arising from the UV regime of their defining time integrals. 
The analytic behaviour of diagrams drastically simplifies near these singularities, reducing in form to lower-point diagrams and/or scattering amplitudes defined by the same diagrams (or their sub-diagrams) in flat space. Let us start with the three-point building block $f(k_{12},s)$: analytically continuing $k_{12}$ to the lower complex half-plane,\footnote{Note that $f$ and $F_{++}$ are analytic functions in the lower complex half plane of the external energies, i.e. $\text{Im}(k_i)<0$. Therefore, singularities on the real axis, such as $k_T=0$ and $E_{L,R}=0$, should be always approached from below.} a singularity emerges by taking
\begin{align}
    k_T=k_{12}+s\to 0\,. 
\end{align}
Near this \textit{total-energy singularity}, $f$ behaves as 
\begin{align}
    \lim_{k_T\to 0}f(k_{12},s)=-\dfrac{i}{\sqrt{2s}}\log(k_{12}+s)\,.\label{eq: f_full_kTLimit}
\end{align}
Note that this singular behaviour is not sensitive to mass modulations since it originates from the early time limit of the bulk integral \eqref{timeintf}. Apart from the branch cut along $k_{12}+s<0$, the three-point function must be regular in the rest of the complex plane, in particular around the \textit{folded} limit $k_{12}\to s$. Analyticity in this region follows from the Bunch-Davies initial condition and provides a crucial boundary condition for solving the IDEs in the next Section.  

As for $F_{++}$, the total energy singularity is approached by sending $k_T=k_{12}+k_{34}\to 0$, leading to the following asymptotic behaviour: 
\begin{align}
    \lim_{k_T\to 0}F_{++}=-\dfrac{2k_T\log k_T}{s^2-k_{34}^2}\,.  
\end{align}
As is known, the coefficient of the total energy singularity is proportional to the scattering amplitude defined by the same diagram in flat space---in this case, the $s$-channel, two-to-two amplitude ${\cal A}_4=(s^2-k_{34}^2)^{-1}$. In addition, $F_{++}$ diverges as either of its partial energies, $E_L=k_{12}+s$ or $E_R=k_{34}+s$, are taken to zero:    
\begin{align}
    \lim_{E_L\to 0}F_{++}=-\dfrac{i}{\sqrt{2s}}\log(E_L)\,f^*(-k_{34}-i\epsilon,s)\,,\quad \lim_{E_R\to 0}F_{++}=-\dfrac{i}{\sqrt{2s}}\log(E_R)\,f^*(-k_{12}-i\epsilon,s)\,.
\end{align}
Once again, assuming that $m^2(t)$ grows no faster than the kinetic term in the UV, the mass time-dependence does not affect the above behaviour. Moreover, as in the three-point case, the Bunch-Davies initial condition requires $F_{++}$ to be regular at all physical configurations, particularly in the \textit{collinear limits} defined by $k_{12}=s$ and $k_{34}=s$.

\section{Perturbative solutions for small mass modulations} \label{Sec: PertSolution}

Solving the presented integro-differential equations is a formidable task, even for cosinusoidal masses. In this section, we resort to perturbation theory at leading order in mass modulations, ${\cal O}(g^2)$, to find closed-form solutions to the exchange diagram. Although our bootstrap computation will be highly technical, its complexity should be contrasted with that of the in-in/Schwinger-Keldysh calculation.
Indeed, the equivalent bulk computation involves evaluating a double-exchange diagram, i.e. Figure \ref{fig:massinsertions} with a single mass insertion. Consisting of two fixed-mass propagators joining at an oscillating vertex, this diagram represents a nested, triple time integral with four factors of Hankel functions, the evaluation of which is analytically intractable. With appropriate boundary conditions imposed by microcausality and analyticity, we will illustrate the advantage of using the integro-differential equations for computing this diagram over explicitly performing the in-in bulk integral.    
\subsection{Three-point contact diagram}\label{Three-point contact diagram}

We begin with the homogeneous integro-differential equation for the three-point function $f$: 
\begin{align}
\label{IDEforf}
     \hat{{\Delta}}_{12}\,f(k_{12},s) &=\int_{0}^\infty {\rm{d}}q\,K(q)\,f(k_{12}+q,k_{34},s)\,, 
\end{align}
which we aim to solve at leading order in $g^2$. As noted earlier, it is sufficient at this order to incorporate only three harmonics ($l=0,\pm 1$) in the mode expansion of $f$, therefore    
\begin{align}
\label{finalfk12s}
    f(k_{12},s)=\dfrac{1}{\sqrt{s}}\left(f_0(u)+\sum_{\pm}f_{\pm1}(u)\,x_0^{\pm i\omega}+{\cal O}(g^4)\right)\,,
\end{align}
where $x_0=-s\eta_0$. Plugging this expansion into the IDE, we find
\begin{align}
    \hat{\Delta}_u\,f_0(u)&=0+{\cal O}(g^4)\,,\label{eq: D_f0}\\
    \hat{\Delta}_u\,f_{\pm 1}(u)&=\int_0^\infty {\rm{d}}x\,K_\pm(x)f_0\left(\dfrac{u}{1+u\,x}\right)+{\cal O}(g^4)\,,\label{eq: D_fpm1}
\end{align}
in which we have defined 
\begin{align}
    x=q/s\,,\qquad K_{\pm}(x)=-\dfrac{1}{2}\,m_0^2 g^2\,\dfrac{e^{\mp\frac{\pi\omega}{2}}}{\Gamma( \pm i\,\omega)}\dfrac{e^{-\epsilon x}}{x^{1-\epsilon}}x^{\pm i\omega}\,,
\end{align}
and 
\begin{align}
\hat{\Delta}_u\equiv \left(u^2-u^4\right)\partial^2_u-2u^3\partial_u+\left(\mu^2+\frac{1}{4}\right)\,,\qquad \mu^2=m_0^2-9/4\,.
\end{align}
Notice that $\hat{\Delta}_u$ here differs from its conventional definition in \eqref{bootstrapODE} by the inclusion of the mass term.

The scale-invariant sector $f_0$, which is governed by \eqref{eq: D_f0}, receives no contribution\footnote{Note that, despite breaking the dilation symmetry, mass oscillations starting at order $g^4$ do contribute to the scale-invariant part of the three-point function because with an even number of mass insertions in the perturbative expansion of $f$, the comoving scale $s$ can partially cancel out between the oscillatory vertices. This will lead to a finite contribution to the scale-invariant component $f_0(u)$ at each even order in $g^2$ which can be partially understood as mass corrections (see Appendix \ref{LightFieldSection}).}
at this order from mass oscillations. Therefore, it is equal to the corresponding three-point function in de Sitter (see e.g. \cite{Arkani-Hamed:2015bza, Arkani-Hamed:2018kmz}), 
\begin{align}
    f_0(u)= \frac{i\,\Gamma(\tfrac{1}{2}-i\mu)\Gamma(i\mu)}{\sqrt{2\pi}}\times\left(\frac{u}{2}\right)^{\frac 12-i\mu}\pFq{2}{1}{\tfrac{1}{4}-\tfrac{i\mu}{2}, \tfrac{3}{4}-\tfrac{i\mu}{2}}{1- i \mu}{u^2}+(\mu\leftrightarrow-\mu)+{\cal O}(g^4)\,. \label{eq: f0}
\end{align}
\\ 
Inputting $f_0$ into the right-hand side of \eqref{eq: D_fpm1}, we arrive at an ordinary, sourced differential equation for $f_{\pm 1}(u)$. This equation can be solved in the following steps: 
\begin{itemize}
    \item We exploit the Taylor series of the Hypergeometric function around the origin and expand $f_0$ in the squeezed limit, namely around $u=0$. The net result schematically looks like: 
    \begin{align}
        f_0(u)=\sum_{k=0}^\infty c_k(\mu)\,u^{\frac{1}{2}+2k-i\mu}+(\mu\leftrightarrow -\mu)\,.
    \end{align}
    Inserting this series expansion into the integrand of \eqref{eq: D_fpm1}, we then integrate over $x$ using the identity: 
    \begin{align}
        \int_0^\infty \dfrac{dx}{x^{1-\epsilon}}\,x^{i\omega}\left(\dfrac{u}{1+u\,x}\right)^{\frac 12+2k-i\mu}=u^{\frac 12+2k-i\mu-i\omega}\,\Gamma\left[\bgm \frac 1 2+2k-i\mu-i\omega,i\omega\\
        \frac 12+2k-i\mu\edm\right]\,, \label{eq: momentum_int_identity}
    \end{align}
    in which the $\Gamma[\dots]$ symbol is defined by \eqref{gammasymbol} in terms of the Gamma functions. After integration over $x$, the result will be readily organised as a power series in $u$ with the exponents $\frac 12+2k\pm i\mu\pm i\omega$. 
    \item  We then adopt a suitable power series ansatz for $f_{\pm1}(u)$ which should take the form 
    \begin{align}
        f^{\text{ansatz}}_{\pm 1}(u)=\sum_{k=0}^\infty\sum_{n=0}^\infty d^\pm_{k,n}(\mu,\omega)\,u^{\frac{1}{2}+2k+2n}u^{-i\mu\mp i\omega}+(\mu\leftrightarrow -\mu)\,.
    \end{align}
    Plugging this ansatz into the IDE for $f_{\pm 1}$, \eqref{eq: D_fpm1}, yields a  recursive relation between $d^{\pm}_{k,n+1}$ and $d^{\pm}_{k,n}$ which can be easily solved. Inserting these coefficients back into $f^{\text{ansatz}}_{\pm 1}$ and summing over $n$, we arrive at 
\begin{keyeqn}
    \begin{align}
    \nn
        f^{\text{ansatz}}_{\pm 1}(u)&=g^2m_0^2\sum_{k=0}^{\infty}b_k^\pm (\mu,\omega)\,u^{\frac{1}{2}+2k-i\mu\mp i\omega}\pregFq{3}{2}{1,\tfrac{1}{4}+k-\tfrac{i\mu}{2}\mp \tfrac{i\omega}{2},\tfrac{3}{4}+k-\tfrac{i\mu}{2}\mp\tfrac{ i\omega}{2}}{1+k\mp \tfrac{ i\omega}{2},1+k- i\mu\mp \tfrac{ i\omega}{2}}{u^2}\\ 
        \label{fansatz}
        &+(\mu\to -\mu)\,,
\end{align}
\end{keyeqn}
where
\begin{keyeqn}
\begin{align}
    b_k^{\pm}(\mu,\omega)=&-\frac{\sqrt{\pi}}{16}\,e^{\mp\frac{\pi\omega}{2}}\text{csch}(\pi \mu)\,2^{-2k+i\mu}\nonumber\\
    &\times \Gamma\left[\bgm k\mp \frac{i\omega}{2},k- i\mu\mp \frac{i\omega}{2},\frac{1}{2}+2k- i\mu\mp i\omega\\
    1+k,1+k- i\mu\edm\right]\,.
\end{align}
\end{keyeqn}
\item Up to order $g^2$, $f^{\text{ansatz}}_{\pm 1}$ satisfies the IDE \eqref{eq: D_fpm1}, however, it exhibits a divergence in the folded limit $u=1$ which is not compatible with the Bunch-Davies initial condition. This spurious singularity must be cancelled by adding an appropriate solution of the homogeneous equation $\hat{{\Delta}}_u\,f^{\text{hom}}_{\pm 1}=0$, which is of the form
\begin{align}
    f^{\text{hom}}_{\pm 1}(u)= \chi^\pm_1\,\left(\frac{u}{2}\right)^{\frac{1}{2}+  i\mu}\pFq{2}{1}{\tfrac{1}{4}+\tfrac{ i\mu}{2}, \tfrac{3}{4}+\tfrac{ i\mu}{2}}{1+ i \mu}{u^2}+\chi^\pm_2\,\left(\frac{u}{2}\right)^{\frac{1}{2}-  i\mu}\pFq{2}{1}{\tfrac{1}{4}-\tfrac{ i\mu}{2}, \tfrac{3}{4}-\tfrac{ i\mu}{2}}{1- i \mu}{u^2}\,.\label{eq: fpm1_hom}
\end{align}
\item The cancellation of the folded singularity leaves a specific linear combination of $\chi^{\pm}_1$ and $\chi^{\pm}_2$ undetermined. This ambiguity can be fixed by imposing the flat space limit of the total three-point $f(k_{12},s)$, according to \eqref{eq: f_full_kTLimit}. Since the zeroth order solution $f_0(u)$ already saturates this limit, it follows that  
\begin{align}
    \lim_{k_{12}+s\to 0}f_{\pm 1}=\text{finite}\,. \label{eq: TotalEnergyLimit}
\end{align}
Imposing this constraint, we finally arrive at 
\begin{align}
    \chi_1^{\pm}(\mu,\omega)=\chi_2^{\pm}(-\mu,\omega)=\frac{g^2m_0^2\pi}{4\sqrt{2}}\,\frac{1-\tanh(\pi\mu)}{e^{\pm\pi\omega}-e^{-2\pi\mu}}\,\Gamma\left[\bgm  -i\mu,\,\frac 1 2\pm\frac{i\omega}{2},\,\mp \frac{i\omega}{2}\\
    \frac{1}{2}-i\mu,\,1- i\mu\pm \frac{i\omega}{2},\,1+i\mu\pm \frac{i\omega}{2}\edm\right].\label{eq: f0_final_coeffs}
\end{align}
\end{itemize}
We refer the reader to Appendix \ref{sec: derivations} for the details of these calculations.

\paragraph{Bulk understanding of the three-point contact diagram} Before turning to the exchange diagram, we take a bulk perspective to gain intuition about the behaviour of the contact three-point function, focusing on the soft limit of the external massive field (i.e. $s\to 0$). 
In fact, $f$ is entirely dictated at leading order in this soft limit by the asymptotic form of the heavy field at late times. To elucidate the corresponding mode function evolution at late times, we decompose it as
\begin{align}
    \sigma_-=\sigma_-^{(0)}+\Delta\sigma^{\text{part}}_-+\Delta\sigma^{\text{hom}}_-\,,
\end{align}
where $\sigma_-^{(0)}$ denotes its zeroth-order part while $\Delta\sigma^{\text{part}}_-$ and $\Delta\sigma^{\text{hom}}_-$ add up to its ${\cal O}(g^2)$ correction.  Specifically, $\Delta\sigma^{\text{part}}_-$ is governed by the perturbed equation of motion at order $g^2$, 
\begin{align}
\label{eqsigmapart}
    (\eta^2\partial_\eta^2-2\eta\partial_\eta+s^2\eta^2+m_0^2)\,\Delta\sigma^{\text{part}}_-=-g^2m_0^2\cos\left(\omega\log\frac{\eta}{\eta_0}\right)\sigma_-^{(0)}\,,
\end{align}
which is an inhomogeneous ODE sourced by the time-dependent component of the mass term, $\Delta m^2\,\sigma$. By contrast, $\Delta\sigma^{\text{hom}}_-$ satisfies the same homogeneous equation as the unperturbed mode function $\sigma_-^{(0)}$. 

Let us study the super-horizon limit of each component separately, starting with the zeroth-order piece which exhibits the well-known scaling behaviour
\begin{align}
\label{zerth}
    \lim_{\eta \to 0}\sigma^{(0)}_-= a^{-\frac{3}{2}}\left[\dfrac{\alpha_0}{\sqrt{2\mu}}(-s\eta)^{-i\mu}+\dfrac{\beta_0}{\sqrt{2\mu}}(-s\eta)^{+i\mu}\right]\,,
\end{align}
with $\alpha_0$ and $\beta_0$ denoting the Bogoliubov coefficients 
\begin{align}
   \alpha_0(\mu) &=\frac{(1+i) \, e^{\frac{\pi  \mu }{2}} \sqrt{\mu }\, \Gamma (i
   \mu )}{2^{1-i\mu}\sqrt{\pi }}\,,\qquad 
      \beta_0(\mu) =\frac{(1+i) \sqrt{\pi }  e^{\frac{\pi  \mu }{2}} (\coth (\pi 
   \mu )-1)}{2^{1+i\mu}\,\sqrt{\mu }\, \Gamma (i \mu )}\,,
\end{align}
and the exponents $\frac{3}{2}\pm i\mu$ reflect the late-time oscillation and dilution of the massive field. 
By contrast, the particular piece presents the following asymptotic form: 
\begin{align}
\label{sigmapart}
      \Delta\sigma_-^{\text{part}}=\dfrac{g^2m_0^2}{2\sqrt{2\mu}}\,\sum_{\pm} (-\eta)^{\frac{3}{2}}\left(\frac{\eta}{\eta_0}\right)^{\pm i\omega}\left[\dfrac{\beta_0}{\omega(\omega\pm 2\mu)}(-s\eta)^{i\mu}+\dfrac{\alpha_0}{\omega(\omega\mp 2\mu)}(-s\eta)^{-i\mu}\right]\,,
\end{align}
featuring four distinct power laws with prefactors that are fixed by the equation of motion; see Section \ref{UVresonances} for the derivation. 

Finally, the homogeneous part is similar in the super-horizon era with $\sigma_-^{(0)}$,
\begin{align}
\label{homocorrect}
    \lim_{\eta\to 0}a^{\frac{3}{2}}\Delta\sigma^{\text{hom.}}_-=\dfrac{\Delta \alpha}{\sqrt{2\mu}}\left(-s\eta\right)^{-i\mu}+\dfrac{\Delta \beta}{\sqrt{2\mu}}\left(-s\eta\right)^{+i\mu}\,,
\end{align}
where we have defined $\Delta\alpha$ and $\Delta\beta$ as the first order corrections to the Bogoliubov coefficients $\alpha_0$ and $\beta_0$.\footnote{Strictly speaking, one can refer to $\alpha=\alpha_0+\Delta\alpha$ and $\beta=\beta+\Delta\beta_0$ as Bogoliubov coefficients only if mass oscillations switch off at future infinity, so that the particular solution $\Delta\sigma_{\text{part}}$ also vanishes asymptotically. For us, using this otherwise intuitive terminology is harmless as we will not impose the Wronskian condition $|\alpha|^2-|\beta|^2=1$, which is not valid with ever-present mass oscillations. } At linear order in $g^2$, these coefficients are associated with \textit{particle production} induced by mass oscillations on the right-hand side of \eqref{eqsigmapart}; see the discussion below and Section \ref{UVresonances}.  
For future convenience, the scale-dependence of these coefficients can be made manifest by the following decompositions: 
\begin{align}
\label{bogoldeco}
  \Delta\alpha &=\Delta\alpha_+(\mu,\omega) x_0^{i\omega}+\Delta\alpha_-(\mu,\omega) x_0^{-i\omega}\,,\\ \nn
  \Delta\beta &=\Delta\beta_+(\mu,\omega) x_0^{i\omega}+\Delta\beta_-(\mu,\omega) x_0^{-i\omega}\,,
\end{align}
where higher order modes, i.e. factors of $x_0^{il\omega}$ with $|l|>1$, have been neglected at this order in $g^2$.  

Having understood the mode function in the infrared, we are ready to investigate the three-point function in the $s\to 0$ limit. We begin with the scale-invariant component  $f_0(u)$, which exhibits the soft behaviour
\begin{align}
\label{f0soft}
    \lim_{s\to 0}f_0=c_0(-\mu)u^{\frac{1}{2}+i\mu}+c_0(\mu)u^{\frac{1}{2}-i\mu}\,.
\end{align}
This power-law profile is inherited from the late-time expansion of $\sigma_-^{(0)}$ in \eqref{zerth} by integration over time. 

Now consider the scale-dependent ${\cal O}(g^2)$ correction to $f$, which is naturally partitioned into
\begin{align}
\label{Deltafdecomp}
    \Delta f=\underbrace{\frac{1}{\sqrt{s}}\sum_{\pm} x_0^{\pm i\omega}f^{\text{ansatz}}_{\pm 1}(u)}_{\Delta f^{\text{part}}}+\underbrace{\frac{1}{\sqrt{s}}\sum_{\pm} x_0^{\pm i\omega}f^{\text{hom}}_{\pm 1}(u)}_{\Delta f^{\text{hom}}}\,,
\end{align}
where $\Delta f^{\text{part}}$ and $\Delta f^{\text{hom}}$ are identified with the particular and homogeneous solutions to the IDE \eqref{IDEforf}, respectively, which are given in terms of $f^{\text{hom}}_{\pm 1}(u)$ in \eqref{fansatz} and $f^{\text{part}}_{\pm 1}(u)$ in \eqref{eq: fpm1_hom}. From a bulk perspective, $\Delta f^{\text{part}}$ can be written as,      
\begin{align}
\label{bulkparticular}
    \Delta f^{\text{part}}=\frac{1}{\sqrt{s}}\sum_{\pm} x_0^{\pm i\omega}f^{\text{ansatz}}_{\pm 1}(u)=+i\int_{-\infty(1-i\epsilon)}^{0} \dfrac{{\rm{d}}\eta}{\eta^2}\,e^{ik_{12}\eta}\Delta \sigma^{\text{part}}_-(s,\eta)\,.
\end{align}
Using the asymptotic expansion of $\Delta \sigma_-^{\text{part}}$ in \eqref{sigmapart}, we get that  
\begin{align}
\label{particularsoft}
   \lim_{u\to 0}\,f^{\text{ansatz}}_{\pm 1}(u)= d_{0,0}^\pm (\mu,\omega)\,u^{\frac{1}{2}-i\mu\mp i\omega}+d_{0,0}^\pm (-\mu,\omega)u^{\frac{1}{2}+i\mu\mp i\omega}\,,
\end{align}
consistent with the soft behaviour of the explicit result in \eqref{fansatz}. As a corollary, note that non-analyticities proportional to $s^{\pm i\omega}$ cannot appear in $\Delta f^{\text{part}}(k_{12},s)$ because those in the particular component $\Delta \sigma^{\text{part}}_-(s,\eta)$ are only of the $s^{\pm i\mu}$ type. Therefore, all factors of $s^{\pm i\omega}$, despite the appearance of \eqref{bulkparticular}, have precisely cancelled in $\Delta f^{\text{part}}$ between the power law $u^{\pm i\omega}$ in $f^{\text{ansatz}}_{\pm 1}(u)$ \eqref{fansatz} and the prefactor $x_0^{\pm i\omega}$.

The second contribution to the three-point $f$ involves the homogeneous components $f^{\text{hom}}_{\pm 1}$ and corresponds to the following bulk integral:    
\begin{align}
\label{homintegral}
    \Delta f^{\text{hom}}=\frac{1}{\sqrt{s}}\sum_{\pm} x_0^{\pm i\omega}f^{\text{hom}}_{\pm 1}(u)=+i\int_{-\infty(1-i\epsilon)}^0 \dfrac{{\rm{d}}\eta}{\eta^2}\,e^{ik_{12}\eta}\Delta \sigma^{\text{hom}}_-(s,\eta)\,. 
\end{align}
Substituting \eqref{homocorrect} into the equation above shows that $f_{\pm 1}^{\text{hom}}$ exhibits the same soft behaviour as $f_0(u)$ in \eqref{f0soft}. In contrast with the particular solution $\Delta f^{\text{part}}$, branch-point singularities proportional to $s^{\pm i\omega}$ do appear in the homogeneous part $\Delta f^{\text{hom}}$, through the prefactors $x_0^{\pm i \omega}$. From the bulk perspective, these branch points are identical to those appearing in the homogeneous component $\Delta \sigma^{\text{hom}}$, through the scale-dependent phases of the Bogoliubov coefficients in \eqref{bogoldeco}. 
\begin{figure}
    \centering
    \includegraphics[scale=0.82]{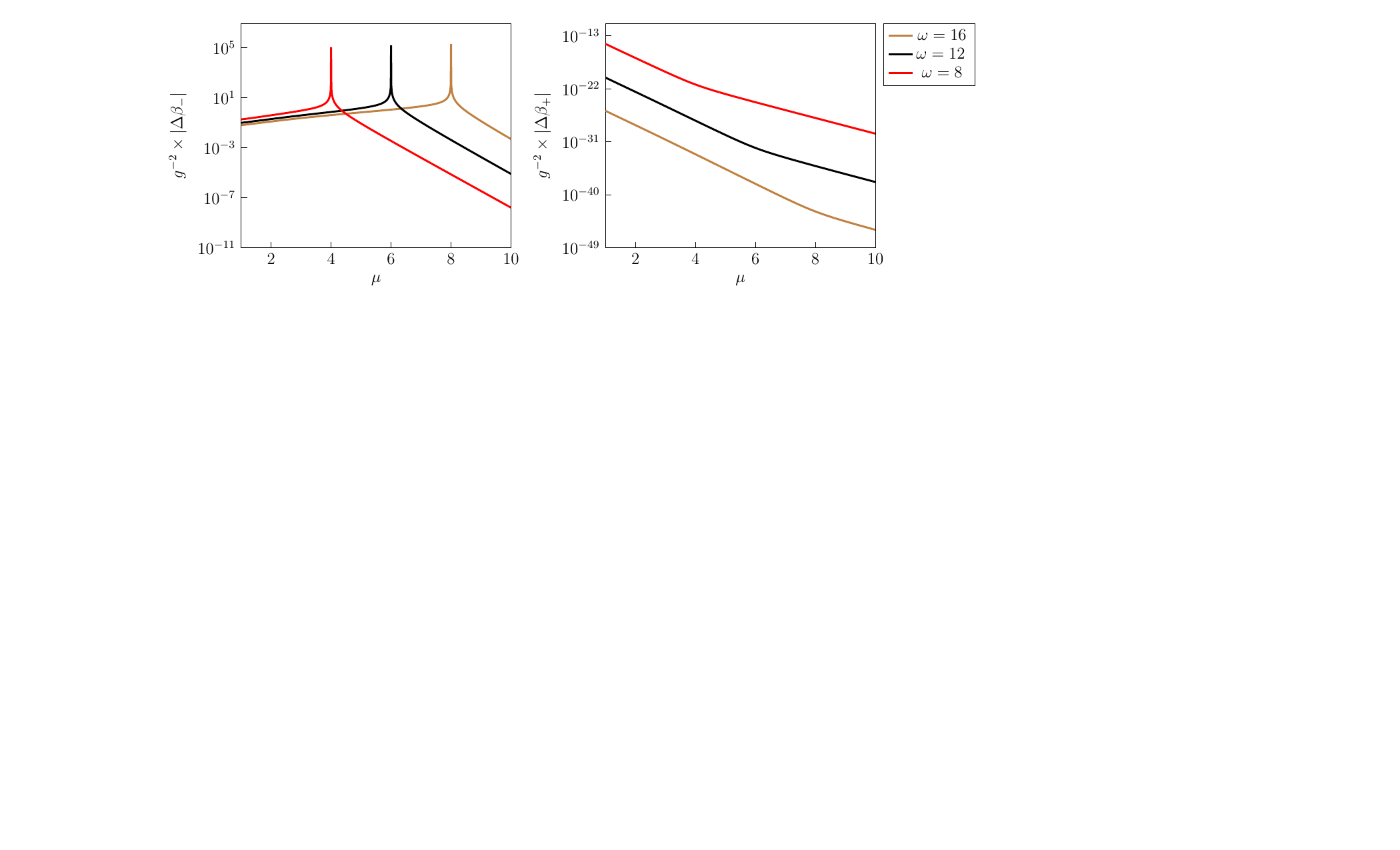}
    \caption{The plots depict two components of the Bogolyubov coefficient $\Delta\beta$, defined by \eqref{homocorrect}--\eqref{bogoldeco}, as functions of $\mu$. See the discussions around \eqref{alphabetalink}. We observe that the negative frequency component $\Delta\beta_-$ (\textit{left}) is enhanced within the mass window $H\ll \mu\lesssim \omega$, due to mass oscillations, while the positive frequency component $\Delta\beta_+$ (\textit{right}) exponentially decays for large masses $\mu\gtrsim 1$, regardless of the frequency $\omega$. Moreover, $\Delta\beta_-$ exhibits a singularity at $\mu=\omega/2$, which is a precursor to a non-perturbative resonance effect in the infrared. See Section \ref{IRresonanceSubSect}. }
    \label{fig:Bogol}
\end{figure}

Our soft limit analysis has so far remained qualitative, going from the asymptotic expansion of the mode function near $\eta\approx0$ to the general form of the three-point $f$ around $s\approx0$. More quantitatively, there is a relation between the coefficients of the soft factors $u^{1/2\pm i\mu}$ (in the three-point components $f_0$ and $f^{\text{hom}}_{\pm 1}$) and the rate of \textit{particle production}, which is driven not only by the expanding background but also by the vibrating mass. 
The most familiar example is the Bogoliubov coefficient $\beta_0$ which quantifies particle creation in de Sitter (i.e. without mass oscillations). For large masses, this rate is suppressed by the familiar Boltzmann factor,   
\begin{align}
    \lim_{\mu\to \infty}|\beta_0(\mu)|=e^{-\pi\mu}\,,
\end{align}
which is carried over to the prefactor of the soft term $u^{1/2+i\mu}$ in \eqref{f0soft} by integration over time. As a result,  
\begin{align}
    \big|c_0(-\mu)\big|=\dfrac{1}{\sqrt{2\mu}}\big|\beta_0(\mu)\big|\times  e^{\pi\mu/2}\big|\Gamma(1/2+i\mu)\big|\underbracket[0 ex]{\approx}_{\mu\gtrsim 1} \sqrt{\dfrac{\pi}{\mu}}\big|\beta_0(\mu)\big|\,
\end{align}
is also exponentially small. In fact, both coefficients in \eqref{f0soft} are exponentially suppressed but for different reasons: the $u^{1/2+i\mu}$ term is small because it is proportional to $\beta_0$, while the integration over the late time oscillations of the heavy field suppresses the coefficient of the $u^{1/2-i\mu}$ term. 

In our setup, particles can be efficiently produced by mass oscillations alone if the frequency exceeds the (average) mass, i.e. $\omega\gtrsim m_0$.    
Similarly with the $g=0$ case, the corresponding particle production rate is encoded in the soft limit of $f^{\text{hom}}_{\pm 1}$, which goes as  
\begin{align}
\label{softexpansion}
    f^{\text{hom}}_{\pm 1}(u)\to \chi_1^{\pm}(\mu,\omega)\left(\dfrac{u}{2}\right)^{1/2+i\mu}+\chi_1^{\pm}(-\mu,\omega)\left(\dfrac{u}{2}\right)^{1/2-i\mu}\,.
\end{align}
Inputting the asymptotic form of $\Delta\sigma^{\text{hom}}_-$ into the bulk integral \eqref{homintegral}, we find that
\begin{align}
\label{alphabetalink}
    |\chi^{\pm}_1(\mu,\omega)|=\dfrac{1}{\sqrt{\mu}}|\Delta \beta_\pm(\mu,\omega)|\times  e^{\pi\mu/2}|\Gamma(1/2+i\mu)|\,\underbracket[0 ex]{\approx}_{\mu\gtrsim 1}\, \sqrt{\dfrac{2\pi}{\mu}}\,|\Delta \beta_\pm(\mu,\omega)|\,.
\end{align}
Using the above relation, we can read off the particle production rate from \eqref{eq: f0_final_coeffs}.

Let us focus on the mass range $H\ll m_0\lesssim \omega$. In this parametric regime, the Bogoliubov coefficient $\Delta\beta$ need not be exponentially small since particles even heavier than $H$ can still be produced by drawing energy from mass oscillations (or equivalently, from the inflaton's kinetic energy). Indeed, the explicit formula for $\chi^-_1$ shows a power-law enhancement in $\Delta \beta_-$:    
\begin{align}
\label{bogoinferred}
    |\Delta\beta_-|\approx \sqrt{\dfrac{\pi}{2}} \dfrac{g^2\,m_0^2}{H^{1/2}\omega^{3/2}}\,,\qquad\text{when}\qquad  H\ll m_0\ll \omega\,.
\end{align}
This enhancement is illustrated in Figure \ref{fig:Bogol} (left panel). 
In contrast, $\Delta\beta_+$ ($\propto \chi^+_1$) does not exhibit such growth and remains exponentially suppressed in this mass range, as seen in Figure \ref{fig:Bogol} (right panel). The contrasting sizes of $\Delta\beta_\pm$ can be understood by explicitly solving the mode function to linear order in $g^2$. As will be shown in Section \ref{UVresonances}, $\Delta\beta_-$ is dominated by an ultraviolet saddle-point at $|s\eta|\sim \omega/2$, corresponding to the resonance between the heavy field's kinetic energy and mass oscillations. By contrast, the alternative component $\Delta\beta_+$ does not experience such a resonance and thus remains exponentially small. Based on energy conservation considerations, particle production eventually dies off as the mass $m_0$ increases above the frequency $\omega$. The corresponding damping tails can be deduced from the large mass limit of \eqref{eq: f0_final_coeffs}, 
\begin{align}
   |\Delta\beta_\pm|\approx \dfrac{g^2}{8\sqrt{\pi}}\,\mu\,e^{-\pi\mu}\times e^{\mp\pi\omega}\,\bigg|\Gamma\left(\mp\frac{i\omega}{2}\right)\Gamma\left(\frac{1}{2}\pm \frac{i\omega}{2}\right)\bigg|\quad (m_0\gg \omega,H)\,.
\end{align}
For completeness, let us also analyse the prefactors $d_{0,0}^{\pm}$ entering the soft expansion of the particular solution $f_{\pm 1}^{\text{ansatz}}$ in \eqref{particularsoft}, for large values of $\mu$ and $\omega$. Besides an artificial divergence at $\mu=\omega/2$, which will be discussed shortly, the coefficients $d_{0,0}^{\pm}(-\mu,\omega)$ are exponentially suppressed for generic $\mu\gg 1$, regardless of the oscillation frequency $\omega$. This suppression is due to the Boltzmann factor $\beta_0$ in front of the first term in \eqref{sigmapart}. By contrast, $d_{0,0}^{+}(\mu,\omega)$ is exponentially diminished for $\mu\gg 1$ by the rapid oscillations of the second term in this bracket, integrated over time.  Conversely, $d_{0,0}^{-}(\mu,\omega)$ grows as
\begin{align}
    |d_{0,0}^{-}(\mu,\omega)|\sim g^2\dfrac{\sqrt{\pi}\mu^{3/2}}{2\omega^2}\,\quad \text{for}\qquad H\ll m_0\ll \omega\,.
\end{align}
This amplification is due to the saddle point of the time integral \eqref{bulkparticular} at $|\eta|\sim (\omega-\mu)/k_{12}$. Around this characteristic moment, the sub-horizon oscillations of the external conformally coupled fields, behaving as $\exp(ik_{12}\eta)$, are in partial resonance with the super-horizon oscillations of the massive field---those going as $(-\eta)^{i\omega}(-s\eta)^{-i\mu}$ in \eqref{sigmapart}---thereby enhancing the time integral \eqref{bulkparticular}.

Finally, we highlight an apparent divergence in the three-point function $f(k_{12},s)$ when $\omega$ approaches $2\mu$ affecting both the homogeneous and the particular parts in \eqref{fansatz}--\eqref{eq: fpm1_hom}. Indeed, at this characteristic frequency, 
the mode function at order $g^2$ also exhibits a divergence, as can be seen from the numerical plots of the Bogolyubov coefficient $\Delta \beta_-$ in Figure \ref{fig:Bogol}, and also from the analytical expression for the particular component $\Delta\sigma_-^{\text{part}}$ in \eqref{sigmapart}. The underlying reason for this divergence is an infrared resonance between the late-time oscillations of the heavy field and its time-dependent mass. As will be shown in the dedicated Section \ref{IRresonanceSubSect}, this divergence is an artefact of linear perturbation theory; the final result converges once the expansion in $g^2$ is properly resummed. 
\\

\subsection{Three-point exchange diagram}\label{Three-point exchange diagram}

Next, we consider $F_{++}$, which satisfies the inhomogeneous IDE \eqref{inhomo}. To simplify the problem, we shall solve this equation only for soft configurations where $k_4\to 0$ (implying $k_{34}\to s$). This allows us to ignore the other permutation of the IDE in \eqref{inhomo}, involving the operator $\hat{\Delta}_{34}$. As we will see in Section \ref{sec: bispectrum}, this external soft limit is suited for extracting the bispectrum $B(k_1,k_2,k_3)$.  
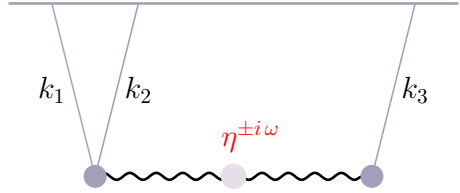
\begin{figure}[ht]
    \centering
\begin{tikzpicture}[scale=2.3,transform shape]
\tikzset{snake it/.style={decorate, decoration=snake}}
\draw[snake it,decoration={amplitude=1.3},line width=1.0pt] (0,0) -- (1.6,0);
\draw[CadetBlue!55,line width=0.6pt] (-0.25,1.0) -- (0,0);
\draw[CadetBlue!60,line width=0.6pt] (0,0) -- (0.25,1.0);
\draw[CadetBlue!60,line width=0.6pt] (1.6,0) -- (1.85,1.0);
\filldraw [CadetBlue!60] (0,0) circle (1.8pt);
\filldraw [CadetBlue!60] (1.6,0) circle (1.8pt);
\filldraw [lightgray!45] (0.8,0) circle (2.0pt);
\node[scale=0.35] at (0.9,0.23){\textcolor{red}{\Large$\eta^{\pm i\,\omega}$}};
\node[scale=0.35] at (-0.25,0.5){\Large$k_1$};
\node[scale=0.35] at (0.25,0.5){\Large$k_2$};
\node[scale=0.35] at (1.85,0.5){\Large$k_3$};
\draw[CadetBlue!60,line width=1.0pt] (-0.5,1.0) -- (2.1,1.0);
\end{tikzpicture}
\caption{The exchange diagram corresponding to $F_{++}(k_{12},k_3,k_3)$ at linear order in $g^2$.}
\label{massinsertion}
\end{figure}

The derivation of the exchange diagram parallels that of the three-point contact diagram $f(k_{12},s)$. We start by decomposing $F_{++}$ according to \eqref{inhomo}. Up to linear order in $g^2$
\begin{align}
    F_{++}(k_{12},k_{34},s)|_{k_4=0}=\dfrac{1}{s}\left(F_0(u,1)+\sum_{\pm}x_0^{\pm i\omega}F_{\pm1}(u,1)+{\cal O}(g^4)\right)\,,\label{Fexchange}
\end{align}
where we have set $v=1$, therefore $u=k_3/k_{12}$. To ease our notation, we henceforth drop the second argument in $F_{0,\pm 1}(u,1)$. The IDE for $F_{++}$ implies that 
\begin{align}
    \hat{{\Delta}}_u F_0(u)&=\dfrac{u}{1+u}+{\cal O}(g^4)\,,\label{eq: F0g0}\\
    \label{DFpm1}
    \hat{{\Delta}}_u F_{\pm 1}(u)&=\int_0^\infty dx\,K_\pm(\omega,x)F_0\left(\dfrac{u}{1+u\,x}\right)+{\cal O}(g^4)\,.
\end{align}
The first equation corresponds to the standard single-exchange diagram with a fixed-mass intermediate line. As was pointed out in \cite{Qin:2023ejc}, the computation of this piece greatly simplifies if one switches to the variable 
\begin{align}
    U=\dfrac{2u}{1+u}\,.
\end{align}
In terms of this new variable, 
\begin{align}
    &\hat{{\Delta}}_u\to \hat{{\Delta}}_U=U^2(1-U)\partial_{U}^2-U^2\partial_{U}+\left(\mu^2+1/4\right)\,,
\end{align}
while the right-hand side of the IDE for $F_{++}$ transforms into
\begin{align}
     &\int_0^\infty dx\,K_\pm(\omega,x)F_0\left(\dfrac{u}{1+u\,x}\right)\to \int_0^\infty dx\,K_\pm(\omega,x)F_0\left(\dfrac{U}{1+U\,x/2}\right)\,. \label{eq: Source_int}
\end{align}
For algebraic simplicity, hereafter we use $U$ to express our results. Using this variable, the zeroth order three-point exchange diagram can be expressed as \cite{Qin:2023ejc}
\begin{align}
\nonumber
    F_0(U)&=\frac{U}{2(\mu^2+1/4)}\,\, 
    \pFq{3}{2}{1,1,1}{\frac32- i\mu,\frac12+ i\mu}{U}\\ \label{F0U}
    &+\frac{ i}{2\sqrt{2\pi}}\,\Gamma\left[\bgm \frac12- i\mu,\frac12+ i\mu\edm\right]\Big(e^{\pi\mu}{\mathcal Y}_1(U)+e^{-\pi\mu}{\mathcal Y}_2(U)\Big)\,,
\end{align}
where 
\begin{align}\label{hom}
&\mathcal Y_1(U) =\mathcal{Y}^*_2(U)= 2^{- i\mu}\left(\frac{U}2\right)^{1/2+i\mu}
\Gamma\Big[\frac12+ i\mu,- i\mu\Big] \times
\pFq{2}{1}{\frac12+ i\mu,\frac12+ i\mu}{1+ 2 i\mu}{U}\,,
\end{align}
furnish a basis of solutions to the homogeneous bootstrap equation \eqref{eq: D_f0}. 

Let us recall some basic properties of the scale-invariant components $F_0(U)$ starting with the first line of \eqref{F0U}, which is analytic around $U=0$ (equivalently around $k_3=0$). The large-mass asymptotic behaviour of this part is captured by an EFT expansion in which the exchange diagram is substituted with an infinite tower of contact diagrams. These diagrams are characterised by higher derivative vertices of the form $\dfrac{1}{(\mu^2+1/4)^{n+1}}\vpi^2\Box^n \vpi^2$. By contrast, the second line of \eqref{F0U} captures the particle production contribution, which the EFT cannot reproduce and is suppressed by the Boltzmann factor $\exp(-\pi\mu)$. Around $U=0$, this part display non-analytic oscillations as $U^{\pm i\mu}$, corresponding to the \textit{cosmological collider signal} \cite{Arkani-Hamed:2015bza}, see \eqref{F0partcollider}. 

According to \eqref{DFpm1}, the scale-breaking part of the three-point exchange diagram, characterised by $F_{\pm 1}(U)$, is sourced by the scale-invariant part $F_0(U)$.  
To solve this equation, we adopt a strategy similar to the three-point contact case. Here we only highlight the most salient features of the derivation, while deferring the details to Appendix \ref{sec: derivations}:  
\begin{itemize}
    \item First, we Taylor expand the $\mathcal{O}(g^0)$ solution $F_0(U)$ around $U=0$, obtaining the following form:
    \begin{align}
    \label{F0partcollider}
        F_0(U)= \sum_{k=0}^{\infty}\,c^{\text{part}}_{k}(\mu)\times U^{1+k}+\left[c^{\text{hom}}_{k}(\mu)\times U^{\frac{1}{2}+i\mu+k}+(\mu\to-\mu)\right]\,,
    \end{align}
    in which the coefficients satisfy $c^{\text{part}}_k(\mu)=c^{\text{part}}_k(-\mu)$, ensuring the whole expression is symmetric under $\mu\to-\mu$.
    The first and second terms above originate from the first and second lines of \eqref{F0U}, respectively.  
    \item Substituting this series into \eqref{eq: Source_int} and performing the momentum integral using an identity analogous to \eqref{eq: momentum_int_identity}, we reduce the IDE~\eqref{DFpm1} to an ordinary differential equation whose right-hand side contains a series of sources. Although the first part of $F_0(U)$ is fully analytic around $U=0$, its convolution through the right-hand side of IDE generates oscillatory terms of the form $U^{\pm i\omega}$.
    Motivated by the structure of the source terms, we propose the following ansatz:
    \begin{align}
        F^{\text{ansatz}}_{\pm1}(U)= \sum_{k,n=0}^{\infty} d^{\,\text{part}}_{k,n}(\mu,\pm \omega)\, U^{1+k+n\mp i\omega}+\left[d^{\,\text{hom}}_{k,n}(\mu,\pm\omega)\, U^{\frac{1}{2}+k+n+i\mu\mp i\omega}+(\mu\to-\mu)\right]\,.
    \end{align}
    Plugging this ansatz into the boundary differential equation \eqref{DFpm1} yields separate recursive relations for $d^{\,\text{part}}_{k,n}$ and $d^{\,\text{hom}}_{k,n}$. The general solution then follows straightforwardly by solving these recursions. Similar to the contact‑diagram case, although the ansatz involves two layers of infinite series, one of them can be resummed, leading to  
    \begin{keyeqn}
    \begin{align}
        F^{\text{ansatz}}_{\pm1}(U)=&\sum_{n=0}^{\infty}\,\mathcal{A}^{\pm}_n(\mu,\omega)\times U^{1+n\mp i\omega}\,\pFq{3}{2}{1,1+n\mp i\omega,1+n\mp i\omega}{\frac{3}{2}+n-i\mu\mp i\omega,\frac{3}{2}+n+i\mu\mp i\omega}{U}\nonumber\\
        +&\sum_{n=0}^{\infty}\Big(\mathcal{B}_n^{\pm}(\mu,\omega)\times U^{\frac{1}{2}+n+i\mu\mp i\omega}\,\pFq{3}{2}{1,\frac{1}{2}+n+i\mu\mp i\omega,\frac{1}{2}+n+i\mu\mp i\omega}{1+n\mp i\omega,1+n+2i\mu\mp i\omega}{U}\nonumber\\
        &\qquad+(\mu\to-\mu)\Big)\,, \label{eq: Fpm_part}
    \end{align}
    \end{keyeqn}
    where 
    \begin{align}
    &\mathcal{A}^{\pm}_n(\mu,\omega)\equiv-\frac{m_0^2g^2\pi}{4}\frac{2^{\pm i\omega}e^{\mp\frac{\pi\omega}{2}}\,{\rm{sech}}(\pi\mu)}{(\frac{1}{2}+n+i\mu\mp i\omega)(\frac{1}{2}+n-i\mu\mp i\omega)}\Gamma\left[\bgm 1+n,1+n\mp i
    \omega\\\frac{3}{2}+n-i\mu,\frac{3}{2}+n+i\mu\edm\right],\\
    &\mathcal{B}^{\pm}_n(\mu,\omega)\equiv\frac{m_0^2g^2\pi}{4}\frac{2^{\pm i\omega}e^{\mp\frac{\pi\omega}{2}+\pi\mu}\,{\rm{csch}}(2\pi\mu)}{(n\mp i\omega)(n+2i\mu\mp i\omega)}\Gamma\,\left[\bgm \frac{1}{2}+n+i\mu,\frac{1}{2}+n+i\mu\mp i
    \omega\\1+n,1+n+2i\mu\edm\right]\,.\label{BnCoeff}
    \end{align}
    The particular solution \eqref{eq: Fpm_part} represent two qualitatively different contributions to the exchange diagram, $F_{++}=x_0^{\pm i\omega}F_{\pm}^{\text{ansatz}}$, as a function of the intermediate momentum $k_3$. The first line's contribution is analytic around $k_3=0$ since factors of $k_3^{\pm i\omega}$ are precisely cancelled between the prefactor $x_0^{\pm i\omega}$ and $F^{\text{ansatz}}_{\pm1}$. However, the second line in \eqref{eq: Fpm_part} induces a branch point at $k_3=0$ in the final four-point function. We shall soon come to the physical interpretation of each analytic behaviour below.  These coefficients have a pole at $\omega=2\mu$, indicating a breakdown of perturbative calculation. As will be shown in Section \ref{IRresonanceSubSect}, this singluar behaviour is in fact a manifestation of resonance that is a non-perturbative effect beyond the reach of perturbative solutions. Further discussion of the radius of convergence of the perturbative expansion is provided in Appendix  \ref{LightFieldSection}.
    \item The particular solution \eqref{eq: Fpm_part} exhibits a spurious singularity around $U=1$, or equivalently at $s=k_{12}$. Such a folded pole is incompatible with the BD vacuum choice and must be removed by adding an appropriate solution to the homogeneous equation, namely 
    \begin{align}
        F_{\pm1}(U)=F^{\text{ansatz}}_{\pm1}(U)+F_{\pm1}^{\text{hom}}(U)\,,
    \end{align}
where 
    \begin{keyeqn}
    \begin{align}
    \label{Fhomfinal}
        F_{\pm1}^{\text{hom}}(U)= \xi_1^{\pm} (\mu,\omega)\cdot\mathcal{Y}_{1}(U)+\xi_2^{\pm} (\mu,\omega)\cdot\mathcal{Y}_{2}(U)\,,
    \end{align}
\end{keyeqn}
and $\mathcal{Y}_{1,2}(U)$ are given by \eqref{hom}.
We need at least two independent constraints to fix the coefficients $\xi_1$ and $\xi_2$. One constraint is already provided by the cancellation of the folded pole from $F_{\pm 1}$. Since the three-point contact diagram is already known, we take the \textit{microcausality} factorisation, given by \eqref{eq: soft_Fac}, as our second bootstrap constraint. 
In particular, at $\mathcal{O}(g^2)$, this property requires that 
\begin{align}
    \lim_{\boldsymbol{s}\to 0} F^{\text{hom}}_{\pm 1}(U)=-\left[f^{\text{hom}}_{\mp 1}(-u+i\epsilon)\right]^*\times f_0(1)-\left[f_0(-u+i\epsilon)\right]^*\times f_{\pm1}(1)\,.\label{eq: fact_u^imu}
\end{align}
Matching the $u^{1/2\pm i\mu}$ terms on both sides yield
    \begin{align}
    &\xi_2^{\pm}(\mu,\omega)=\xi_1^{\pm}(-\mu,\omega)\,,\\
    &\xi_1^{\pm}(\mu,\omega)=\frac{g^2m_0^2\pi}{16\sqrt{2}}\frac{1+\coth(\pi(\mu\pm \frac{\omega}{2}))}{\cosh \pi\mu}\Gamma\left[\bgm \frac 1 2\pm\frac{i\omega}{2},\,\mp \frac{i\omega}{2}\\
    1- i\mu\pm \frac{i\omega}{2},\,1+i\mu\pm \frac{i\omega}{2}\edm\right]+\frac{e^{\pi\mu}}{2\sqrt{\pi}}\,f_{\pm1}(1)\,. 
\end{align}
In this expression, the numerical factor $f_{\pm1}(u=1)$ can be evaluated in closed form by resumming the series in \eqref{fansatz} for $u=1$ and adding $f_{\pm 1}^\text{hom}(u=1)$, leading to
\begin{align}
     &f_{\pm1}(1)\nonumber\\
     &=- \frac{i\,\pi g^2m_0^2\,e^{\mp\frac{\pi\omega}{2}} }{2^{\frac{3}{2}\mp i\omega}\cosh(\pi\mu)} \Gamma\left[\frac{1}{2}-i\mu\mp i\omega,\frac{1}{2}+i\mu\mp i\omega\right]\,\pregFq{4}{3}{1,1,\frac{1}{2}-i\mu\mp i\omega,\frac{1}{2}+i\mu\mp i\omega}{\frac{3}{2}-i\mu,\frac{3}{2}+i\mu,1\mp i\omega}{1}\,,\label{fpm1_ClosedForm}
\end{align}
in which ${}_p\tilde{\rm{F}}_q$ is the regularised hypergeometric function defined in \eqref{eq: reg_pFq}. Let us stress that while it was sufficient to take the leading soft limit $k_3 \to 0$ to fix the free coefficients $\xi_{1,2}$, the factorisation in \eqref{eq: soft_Fac} holds to all orders in $k_3$. We shall elaborate on this point below. Additional intermediate steps leading to the above results are spelled out in Appendix \ref{sec: derivations}.
\end{itemize}

\paragraph{EFT limit and factorisation} Let us run a consistency check on our result by taking the limit where the average mass $m_0$ is much greater than the frequency $\omega$ and the Hubble rate $H$. Since particle production (whether due to expansion or oscillations) is negligible in this limit, the heavy field can be integrated out in favour of a single-field EFT for the conformally coupled scalar $\vpi$. At tree-level, the interacting part of this EFT is given by  
\begin{align}
    {\cal L}_{\text{EFT}}&=\dfrac{1}{2}\,\sqrt{-g}\,\vpi^2\,\left(1+\dfrac{1}{m^2(t)}\Box\right)^{-1}\,\dfrac{1}{m^2(t)}\vpi^2\\ \nn
    &=\dfrac{1}{2}\,\sqrt{-g}\,\vpi^2\,\left(\dfrac{1}{m^2(t)}-\dfrac{1}{m^2(t)}\Box\,\dfrac{1}{m^2(t)}+\dots \right)\,\vpi^2\,.
\end{align}
The leading order term at ${\cal O}(g^2)$, namely 
\begin{align}
    {\cal L}_{\text{EFT}}&=
    \dfrac{1}{2}\sqrt{-g}\,\dfrac{1}{m_0^2}\left[1-g^2\cos(\omega(t-t_0))\right]\vpi^4+{\cal O}(g^4)\,,
\end{align}
induces the following four-point function: 
\begin{align}
    F_{++}=\dfrac{1}{m_0^2\,k_T}-\dfrac{g^2}{2 m_0^2 k_T}\left[(-ik_T\eta_0)^{-i\omega}\Gamma(1+i\omega)+\text{c.c.}\right]+{\cal O}(g^4)\,.
\end{align}
Using the decomposition \eqref{Fexchange} and taking the soft limit $k_4\to 0$, this four-point function can be expressed in terms of 
\begin{align}
    F^{\text{EFT}}_{0}(U)=\dfrac{U}{2(\mu^2+9/4)}\,,\qquad F^{\text{EFT}}_{\pm 1}(U)=-\dfrac{g^2}{2(\mu^2+9/4)}\left(\dfrac{U}{2}\right)^{1\mp i\omega}\,e^{\mp \pi\omega/2}\,\Gamma(1\mp i\omega)\,.
\end{align}
As a non-trivial check, let us verify that the behaviour of our exchange diagram in the large mass limit $m_0\to \infty$ matches the EFT prediction above. Beginning with $F_0(U)$: while the contribution in the second line of \eqref{F0U} exponentially vanishes as $\mu\to \infty$, that in the first line asymptotically converges towards $F_0^{\text{EFT}}(U)$, up to order $\mu^{-4}$ corrections.  As for $F_{\pm 1}$, the terms characterised by the coefficients $\mathcal{B}_n^{\pm}$ in the particular ansatz \eqref{eq: Fpm_part} are exponentially small, whereas those with the coefficient $\mathcal{A}^{\pm}_{n>0}$ are only power-law suppressed. Specifically,  
\begin{align}
  \lim_{\mu\to \infty}\mathcal{A}^{\pm}_{0}(\mu,\omega)=-\dfrac{g^2}{\mu^2}2^{-2\pm i\omega}e^{\mp \pi\omega/2}\Gamma(1\mp i\omega)\,,
\end{align}
while $\mathcal{A}^{\pm}_{n>0}$ are of order $1/\mu^4$ or higher. By plugging the asymptotic form of $\mathcal{A}^{\pm}_{0}$ into \eqref{eq: Fpm_part} and neglecting the remaining terms, we recover the leading order EFT term $F^{\text{EFT}}_{\pm 1}$. Similarly, higher order EFT contributions could be isolated in \eqref{eq: Fpm_part} by keeping more powers of $1/\mu^2$ in the prefactors $\mathcal{A}^{\pm}_{n}$, and by expanding $_3F_2$ near $U=0$. In this series, only a finite number of terms need to be included at a given order in the EFT expansion, because the coefficient of the $U^k$ term is already of order ${\cal O}(\mu^{-2k})$.

Indeed, it is not an accident that the first line in \eqref{eq: Fpm_part} resembles the four-point function in the EFT, but rather a natural expectation from its analytic structure. Specifically, the corresponding contribution to the four-point function, contained in $\sum_{\pm} x_0^{\pm i\omega}F_{\pm 1}^{\text{ansatz}}$, is analytic in the magnitude of the exchanged momentum $k_3$. As a result, these parts correspond to the \textit{local} imprints of the heavy field exchange. {As we elaborated below \eqref{eq: soft_Fac}, this contribution originates from substituting the time-ordered propagator within $F_{++}$ with the retarded propagator $G_R$, thereby capturing all \textit{local} effects associated with the heavy field exchange. In terms of the four-point function kinematics, this contribution would be analytic in the exchanged momentum $\bm{s}$. However, after sending $\bm{k}_4\to 0$, analyticity remains only in $\sqrt{\bm{s}.\bm{s}}=k_3$.} By contrast, the remaining terms in $F^{\text{ansatz}}$ and the homogeneous part $F^{\text{hom}}_{\pm}(U)$ display non-analyticities proportional to $s^{\pm i\mu}$ and $s^{\pm i\mu\pm i\omega}$, respectively. These are, correspondingly, the \textit{non-local} imprints of the exchange of the heavy field, which remain invisible at any finite order in the EFT expansion. However, as explained in Section \ref{symmetris-micro-anal}, these non-analytic terms are uniquely fixed in terms of the three-point building block $f$, thanks to the factorisation property \eqref{factorisedFpp}.

Already at leading order in the soft limit $s\to 0$, factorisation played an essential role in determining the free coefficients in the homogeneous piece $F^{\text{hom}}_{\pm 1}$. However, its implications go above and beyond the leading order soft behaviour, fixing \textit{all} terms in $F_{++}$ that are irrational in $k_3$. This includes the second part of $F_0(U)$ in \eqref{F0U}, the second part of $F^{\text{ansatz}}_{\pm 1}$ in \eqref{eq: Fpm_part}, as well as the entire homogeneous piece $F^{\text{hom}}$ in  \eqref{Fhomfinal}, all of which by virtue of this property are fixed in terms of the three-point function $f$.  

Starting at zeroth-order in $g$, the factorisation of the $F_0(U)$'s non-analytic part becomes evident by using the quadratic transformations of the hypergeometric function $_2{\rm{F}}_1(a,b,2b,z)$ to recast ${\cal Y}_{1,2}$ in \eqref{hom} as:      
\begin{align}
\label{Yrewrite}
     {\cal Y}_1 (U)={\cal Y}^*_2 (U)=2^{- i\mu}\,u^{\frac 12+i\mu}\,\Gamma\left[\bgm \frac12+i\mu,-i\mu\edm\right]\,\pFq{2}{1}{\tfrac{1}{4}+ \tfrac{i\mu}{2}, \tfrac{3}{4}+ \tfrac{i\mu}{2}}{1+ i \mu}{u^2}\,.
\end{align}
Inserting these into $F_0(U)$ and noting that   
\begin{align}
    f_0(1)=\dfrac{i\pi}{\sqrt{2}\cosh(\pi\mu)}\,,\quad\text{with} \qquad (-u+i\epsilon)^{\frac{1}{2}\pm i\mu}=i\,u^{\frac{1}{2}\pm i\mu}\exp(\mp \pi \mu)\,,
\end{align}
we arrive at the desired identity 
\begin{align}
        F_0(U)|_{\text{non-analytic}}&=\frac{ i}{2\sqrt{2\pi}}\,\Gamma\left[\bgm \frac12- i\mu,\frac12+ i\mu\edm\right]\Big(e^{\pi\mu}{\mathcal Y}_1(U)+e^{-\pi\mu}{\mathcal Y}_2(U)\Big)\\ \nn
        &=-[f_0(-u+i\epsilon)]^*\times\,f_0(1)
        \,.
\end{align}
Similarly plugging the transformed version of ${\cal Y}_{1,2}(U)$ above into the homogeneous part \eqref{Fhomfinal}, yields    
\begin{align}
    F^{\text{hom}}_{\pm 1}(U)=-[f^{\text{hom}}_{\mp 1}(-u+i\epsilon)]^*\times f_0(1)-[f_0(-u+i\epsilon)]^*\times f_{\pm1}(1)\,. \label{eq: fact_homo}
\end{align}
Finally, the particular ansatz must factorise as 
\begin{align}
\label{ansatzfactor}
    F^{\text{ansatz}}_{\pm 1}(U)|_{\text{non-analytic}}=-[f^{\text{ansatz}}_{\mp 1}(-u+i\epsilon)]^*\times f_0(1)\,,
\end{align}
which, unlike the previous two cases, is entirely obscured within the structure of $F^{\text{ansatz}}$ in \eqref{eq: Fpm_part}. In particular, factorisation requires all power laws of the form $u^{2k+1}u^{\frac{1}{2}\pm i\mu\pm i\omega}$ to cancel from $F^{\text{ansatz}}$, leaving only terms with even powers. This cancellation becomes evident at the level of the final answer after Taylor-expanding the generalised hypergeometric functions around $U=0$, and even then, only by combining terms at different orders in $n$. It can be shown that the remaining terms match order by order in $u$ on both sides of \eqref{ansatzfactor}.\\
\begin{figure}
    \centering
    \includegraphics[scale=0.9]{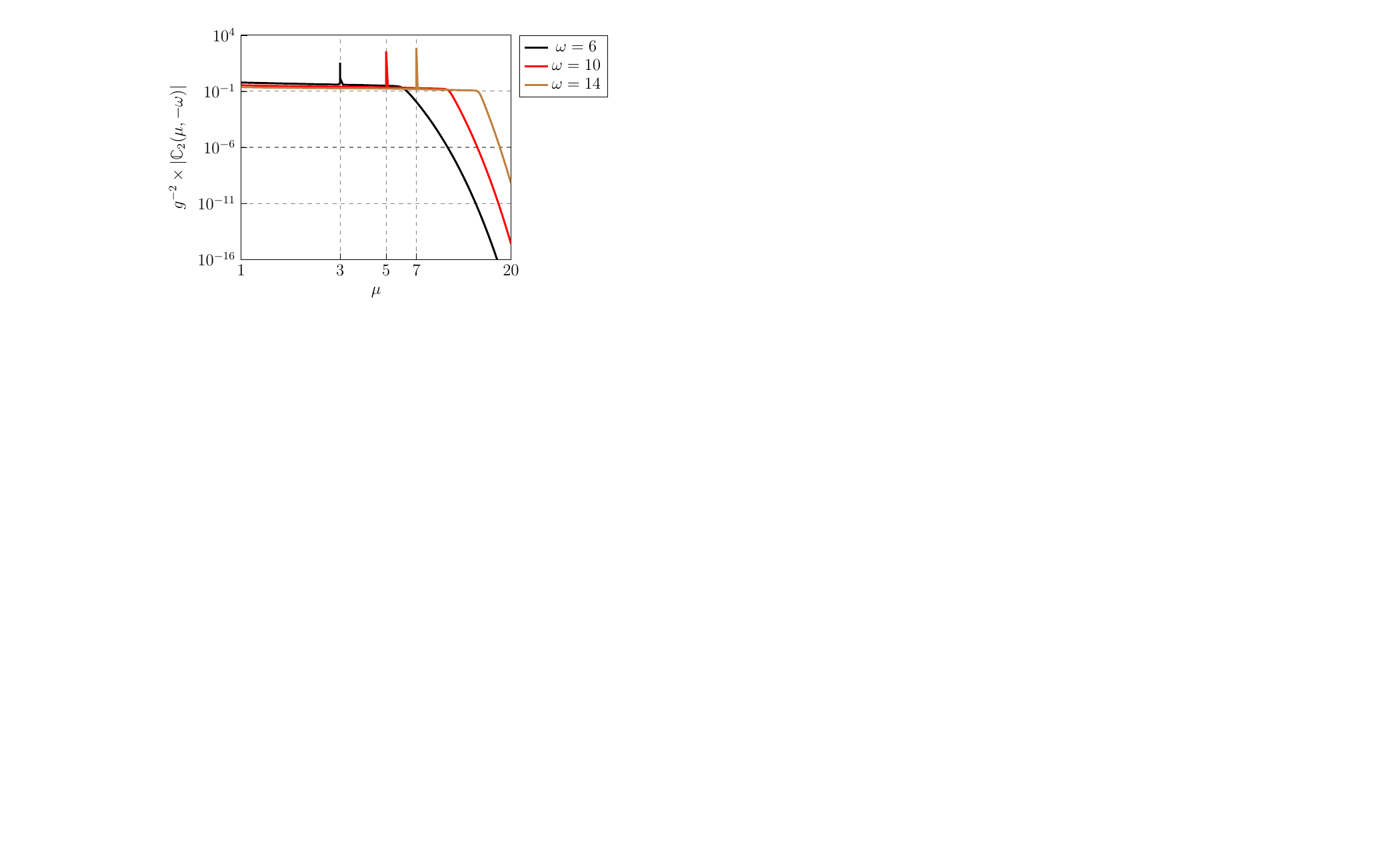}
    \caption{The modulus of the prefactor of the $u^{1/2+i\mu}\,x_0^{-i\omega}$ term in the resonant cosmological collider signal in \eqref{fullF4}, $|\mathbb{C}_2(\mu,-\omega)|$, for three sample values of the oscillation frequency $\omega$, as a function of $\mu$. Thanks to the mass oscillation induced particle production, the prefactor is amplified in the mass range $m_0\lesssim\omega$, exhibiting almost a flat behaviour apart from a sharp peak at $\mu=\omega/2$, which corresponds to the IR resonance discussed in Section \ref{IRresonanceSubSect}. Conversely, the alternative components, $\mathbb{C}_2(\mu,\omega)=\mathbb{C}^*_2(-\mu,-\omega)$,  are exponentially suppressed for $\mu\gtrsim 1$, irrespectively of the choice of the frequency $\omega$. }
    \label{fig:C2}
\end{figure}

\paragraph{Squeezed-limit oscillations} We have all the necessary ingredients to compute the $s$-channel exchange diagram $F$ by summing over its Schwinger-Keldysh components $F_{\pm\pm}$. It is particularly instructive to simplify the result in the $u=k_3/k_{12}\to 0$ limit from which, as will be discussed in the next section, the bispectrum in the squeezed regime ($k_3\ll k_{1,2}$) follows. In terms of the previously computed components $F_{0,\pm 1}$, $f_{0,\pm 1}$, the seed function is given by 
\begin{align}\label{ThreeptsSeed}
    &F(k_{12}, k_3)=\dfrac{1}{k_3}\left[F_0(u)+F_0^*(u)+f_0(u)f_0^*(1)+f_0^*(u)f_0(1)\right]\\ \nn
    &+\dfrac{1}{k_3}\sum_{\pm}\left[F_{\pm 1}(u)+F^*_{\mp 1}(u)+f_{\pm 1}(u)f_0^*(1)+f_0(u)f_{\mp 1}^*(1)+f_{\mp 1}^*(u)f_0(1)+f_0^*(u)f_{\pm 1}(1)\right]\,x_0^{\pm i\omega}\,,
\end{align}
where the first line is identical to the ordinary single-exchange seed function (with a constant intermediate mass), and the second line captures the ${\cal O}(g^2)$ correction due to mass oscillations. In the soft limit $u\to 0$, the result simplifies to 
\begin{align}
\label{fullF4}
    &\lim_{k_3\to 0} F(k_{12}, k_3)=\dfrac{1}{k_3}\left[\mathbb{C}_0(\mu)u^{1/2+i\mu}+\mu\to -\mu\right]\\ \nn
    &+\sum_{\pm}\dfrac{1}{k_3}\left[\mathbb{C}_1(\mu,\mp \omega)\,u^{1/2+i\mu\pm i\omega}+\mathbb{C}_2(\mu,\mp \omega)\,u^{1/2+i\mu}+\mu\to -\mu\right]\,x_0^{\mp i\omega}\,,
\end{align}
where 
\begin{align}
    \mathbb{C}_0(\mu) &=\dfrac{\sqrt{\pi}(1+ie^{\pi\mu})}{2^{\frac{1}{2}+i\mu}(i+e^{\pi\mu})}\Gamma\left[\frac{1}{2}+i\mu,-i\mu\right]\underbracket[0 ex]{\approx}_{\mu\gtrsim 1}-\dfrac{(1-i)\pi^{3/2}}{2^{i\mu}\sqrt{\mu}}e^{-\pi\mu}\,\\ 
    \mathbb{C}_1(\mu,\omega)&=\dfrac{\sqrt{\pi}\,g^2 m_0^2\,\text{sech}(\pi\mu)}{2^{3/2+i\mu}\omega (\omega-2\mu)}\Gamma\left[\frac{1}{2}+i\mu-i\omega,-i\mu\right]\,\Big(\cosh(\frac{\pi\omega}{2})+i\sinh(\pi \mu-\frac{\pi \omega}{2})\Big)\,,\label{coeffC1}\\ \nn
    \mathbb{C}_2(\mu,\omega)&=\dfrac{e^{\pi\mu}-ie^{\pi\omega}}{2^{i\mu}\sqrt{\pi}(i+e^{\pi\mu})}\cosh(\pi\mu)\Gamma\left[\frac{1}{2}+i\mu,-i\mu\right]\,f_{+1}(1)\\ 
    &+\dfrac{\pi g^2 m_0^2\,e^{\pi\mu}\left(-i+e^{\pi(\mu+\omega)}\right)}{2^{5/2+i\mu}(i+e^{\pi\mu})\left(-1+e^{\pi(2\mu+\omega)}\right)}\Gamma\left[\bgm \frac{1}{2}+i\mu,-i\mu,\,\frac{1}{2}+\frac{i\omega}{2}\,,\,-\frac{i\omega}{2}\\ 1-i\mu+\frac{i\omega}{2},1+i\mu+\frac{i\omega}{2}\edm\right]\,.\label{coeffC2}
\end{align}
Note that the reality of the seed correlator $F$ implies $\mathbb{C}_0(-\mu)=\mathbb{C}_0^*(\mu)$ and $\mathbb{C}_{1,2}(\mp\mu,\mp\omega)=\mathbb{C}_{1,2}^*(\pm\mu,\pm\omega) $. In \eqref{fullF4}, the first line corresponds to the ordinary, scale-invariant cosmological collider signal in the squeezed limit, whereas the second line introduces a distinct, \textit{resonant cosmological collider signal} which breaks scale invariance; see \cite{Chen:2022vzh,Pajer:2024ckd, Wang:2025qww} and the discussion below. 

An interesting parametric region is where the oscillation frequency $\omega$ is much greater than the mass $m_0$ and the expansion rate $H$. As discussed in Section \ref{Three-point contact diagram}, particle production in this mass range is exponentially enhanced by mass oscillations relative to the ever-present pair creation rate in an expanding background. This enhancement is also reflected in the size of the cosmological collider oscillations in the squeezed limit of $F$. Indeed we observe in Figure \ref{fig:C2} that the prefactors of the oscillatory terms $u^{1/2\pm i\mu}x_0^{\mp i\omega}$, i.e. $\mathbb{C}_2(\mu,-\omega)=\mathbb{C}_2^*(-\mu,+\omega)$, are of order $g^2\times {\cal O}(0.1)$, within the mass range $H\ll m_0\lesssim \omega$.  By contrast, the coefficients of the remaining terms in the resonant collider signal, i.e. $u^{1/2\pm i\mu\mp i\omega }x_0^{\pm i\omega}$, are exponentially suppressed,  
\begin{align}
    |\mathbb{C}_1(\mu,\omega)|\approx |\mathbb{C}_1(\mu,-\omega)|\approx\dfrac{g^2\pi^{3/2}}{4\sqrt{2}}\dfrac{m_0^2\,e^{-\pi\mu}}{\sqrt{\mu}\,\omega^2}\,\quad(1\ll \mu\ll \omega)\,.
\end{align}
This implies that the squeezed limit of the seed function $F$ is dominated by the middle term in the second line of \eqref{fullF4}. Consequently, this term induces the principal cosmological collider signal in the bispectrum of curvature perturbation for $H\ll m_0\lesssim \omega$, as will be discussed in the next section.

Let us compare our cosmological collider signals with those predicted in related previous works\cite{Bodas:2020yho, Chen:2022vzh,Pajer:2024ckd, Wang:2025qww, Pinol:2023oux, Werth:2023pfl}. These studies focus on diagrams with oscillatory cubic vertices and fixed-mass intermediate lines. In particular, it has been shown that vertex oscillations---unlike our mass modulations at order $g^2$---can boost the scale-invariant part of the cosmological collider signal \cite{Bodas:2020yho, Pajer:2024ckd, Wang:2025qww}. Meanwhile, the dominant scale-dependent resonant collider signal in these studies takes similar forms to the first and/or the second resonant contribution in \eqref{fullF4}, depending on the specifics of the vertex couplings\footnote{We thank Xingang Chen for related discussions on the \textit{classical} cosmological collider signal (see \cite{Chen:2022vzh} and references therein).}. Away from the ultra-squeezed limit, the waveform of our cosmological collider signal departs from those presented in previous works (with vertex oscillations), as new distinctive patterns of oscillations and scaling behaviours emerge towards the equilateral limit, see e.g. Figure \ref{figSchannelB_Figure_Total}.

\subsection{Observational signatures} \label{sec: bispectrum}
As was advocated in Section \ref{Setup}, the single-exchange diagrams of the two- and three-point functions of $\pi$ can all be obtained by acting with appropriate weight-shifting operators on the corresponding seed exchange diagram. In this section, for concreteness, we derive these weight-shifting operators assuming $\pi'\sigma$ and $\pi'^2\sigma$ are the dominant interacting terms in the Lagrangian. Using these vertices, we will go ahead and evaluate the bispectrum using our analytical results for the seed exchange diagram, and for sample values of the parameters $\lbrace g, m_0,\omega\rbrace$. Finally, we provide explicit formulas for the resonant cosmological collider signal in the squeezed limit of the bispectrum in our setup.  

\paragraph{Weight-shifting operators} Consider the single-exchange diagram depicted in Figure \ref{fig:seedweight} involving three external massless legs, assuming $\lambda_R(\eta)\dot{\pi}_c\sigma$ and $\lambda_L(\eta)\dot{\pi}_c^2\sigma$ as its right/left vertices, where $\pi_c$ is the canonically normalised Goldstone. For now, we leave their time-dependence generic, but later restrict to constant vertices. 
The corresponding Schwinger-Keldysh integrand
can be mapped onto that of a seed four-point function with the simpler vertices $\lambda_{L,R}(\eta)\vpi^2\sigma$. To make this point concrete, let us look at the integrand of the bispectrum diagram, 
\begin{align}
\nn
    \mathbb{B}(\eta,\eta',k_1,k_2,k_3)&=\sum_{\aa,\bb=\pm}\left(\prod_{i=1}^3\pi^{\aa}_c(k_i,\eta_0)\right)\\ \nn
    &\times\aa\bb\dfrac{\lambda_L(\eta)}{\eta^4}\dfrac{\lambda_R(\eta')}{\eta^4}(\eta\,\partial_{\eta}\pi^{\aa*}_c(k_1))\,(\eta\,\partial_{\eta}\pi^{\aa*}_c(k_2))G_{\aa\bb}(k_3,\eta,\eta')(\eta'\,\partial_{\eta'}\pi^{\bb*}_c(k_3))\,\\ 
    &=\left(\prod_{i=1}^3 \dfrac{1}{2k_i}\right)\sum_{\aa,\bb=\pm}\dfrac{1}{\eta'^2}\lambda_L(\eta)\lambda_R(\eta')e^{i \aa k_{12}\eta}G_{\aa\bb}(k_3,\eta,\eta')e^{i \bb k_{3}\eta'}\,,
\end{align}
in which several kinematic-independent prefactors are dropped for simplicity; they will be restored in the final formula \eqref{PowerBispectrum}. Comparing this expression with the integrand of the seed four-point function $F(\{k\},s)$ defined in \eqref{fourpointall}, henceforth denoted by $\mathbb{F}(k_{12},k_{34},s;\eta,\eta')$, we find that
\begin{align}
    \mathbb{B}(\eta,\eta'; k_1,k_2,k_3)=\left(\prod_{i=1}^3 \dfrac{1}{2k_i}\right)\dfrac{\partial^2}{\partial (k_{12})^2}\mathbb{F}(k_{12},k_{34},s;\eta,\eta')\bigg|_{k_4=0,s=k_3}\,.
\end{align}
Notably, this relation involves the derivative operation $\frac{\partial^2}{\partial (k_{12})^2}$ with respect to the external kinematics and the soft limit $\bm{k}_4\to 0$, which also sends $s\to k_3$. Integrating over the conformal times $\eta$ and $\eta'$, the same relation holds between the final diagrams, namely the bispectrum $B(k_1,k_2,k_3)$ and the seed $F(k_{12},k_3)$. 

Another example is the single-exchange diagram for the power spectrum in Figure \ref{fig:seedweight}, with the vertices $\lambda_{L,R}(t)\dot{\pi}\sigma$. The associated integrand is given by 
\begin{align}
    \mathbb{P}(k;\eta,\eta')=-\frac{1}{2k^3}\sum_{\aa,\bb=\pm}\aa\bb\dfrac{\lambda_L(\eta)}{\eta^4}\dfrac{\lambda_R(\eta')}{\eta'^4}(\eta\,\partial_{\eta}\pi^{\aa*}_c(k))G_{\aa\bb}(k,\eta,\eta')(\eta'\,\partial_{\eta'}\pi^{\bb*}_c(k))\,\,,
\end{align}
which, by taking the double soft-limit $\bm{k}_3,\bm{k}_4\to 0$ (implying $s,k_2\to k_1=k$), can be algebraically related to $\mathbb{F}$,  
\begin{align}
    \mathbb{P}(k;\eta,\eta')=\dfrac{1}{4k^2}\mathbb{F}(k_{12},k_{34},s;\eta,\eta')\bigg|_{k_4=k_3=0,s=k_1=k_2=k}\,.
\end{align}
Similarly, the integrands associated with other exchange diagrams (with different vertex structures such as $\pi\dot{\pi}\sigma, (\partial_i\pi)^2\sigma,\text{etc}$.) can be mapped onto $\mathbb{F}$, leading to analogous relations between the corresponding power spectrum or the bispectrum and $F$.  

We further simplify our setup by assuming that the vertices are time-independent. So the leading vertices in the EFT are:
\begin{align}
\label{actioncubic}
    S=\int\dfrac{\mathrm{d}\eta\,\mathrm{d}^3\bm{x}}{\eta^4}\,\left(\rho\, \eta\,\pi'_c\sigma-\dfrac{1}{\Lambda}\eta^2\pi_c'^2\sigma-\dfrac{1}{\Lambda'}\eta^2(\partial_i\pi_c)^2\sigma\right)\,,
\end{align}
where $\rho,\Lambda$ are two independent energy scales, while $\Lambda'$ is related to $\rho$ via the non-linearly realised time diffeomorphism, 
\begin{align}
    \dfrac{1}{\Lambda'}=\dfrac{\rho}{2(2|\dot{H}|)^{1/2}\Mpl}\,.
\end{align}
We require the quadratic mixing $\dot{\pi}\sigma$ to be perturbative\footnote{Though pushing us outside the perturbative realm, a strong quadratic mixing could make for interesting phenomenology, see e.g. \cite{Jazayeri:2023xcj,Werth:2023pfl,Pinol:2023oux}.} at Hubble crossing, which provides the upper bound $\rho\lesssim H$, up to order one prefactors\cite{Jazayeri:2022kjy}. This bound translates into 
\begin{align}
\label{moststringent}
    \Lambda'\gtrsim \dfrac{H}{\sqrt{\Delta_\zeta}}\,.
\end{align}
Moreover, the cubic action must be weakly coupled across the relevant energy scales of the problem. This includes the oscillation frequency $\omega$, corresponding to the UV resonance; the mass of the heavy field, $m_0$;  and finally the Hubble rate, $H$. To put it short, perturbative unitarity concerning the sizes of the $\pi'^2\sigma$ and $(\partial_i\pi)^2\sigma$ operators demands (at an order of magnitude level):   
\begin{align}
\label{lessstringent}
    \text{min}\lbrace\Lambda',\Lambda\rbrace\gtrsim \text{max}\lbrace H,m_0,\omega\rbrace\,.
\end{align} 
Furthermore, although having been ignored in our tree-level computation, there are additional mixings between $\pi$ and $\sigma$ induced by the ever-present operator, 
\begin{align}
    -\dfrac{1}{2}g^2\,m_0^2\cos(\phi/f)\,\sigma^2\,,
\end{align}
which should be under perturbative control for our setup to be consistent. Such mixings contribute for instance to the double-massive exchange diagram with the $\pi\sigma^2$ type vertex, or one-loop contributions to the power spectrum and higher point functions, as studied in \cite{Flauger:2016idt}.

The conservative value for the
cutoff associated with this operator is roughly $f=\dot{\phi}/\omega$, which is a naive upper bound on the scale of new physics merely on dimensional grounds. A more rigorous derivation in the single-field setup has shown a higher cutoff \cite{Hook:2023pba}. While it is beyond the scope of this work, it would be interesting to revisit this strong coupling scale using $n\to m$ amplitudes in our setup, along the lines of \cite{Creminelli:2025tae}. In the following, we adopt this naive cutoff, which requires us to impose
\begin{align}
\label{fbound}
    \text{max}\lbrace H,m_0,\omega\rbrace\lesssim f\sim \dfrac{H^2}{2\pi\omega}\Delta_\zeta^{-1}\,, 
\end{align}
at an order of magnitude level.

From \eqref{lessstringent} and the inequality above, we get that the condition \eqref{moststringent}---which is essential for keeping the linear mixing perturbative---is the most stringent lower bound on $\Lambda'$. By contrast, the coefficient of the $\dot{\pi}^2\sigma$ operator, being independent of the linear mixing, is only constrained by perturbative unitarity, leading to the less stringent bound \eqref{lessstringent} on $\Lambda$. 
In other words, without an approximate boost symmetry tying their coefficients, the cubic term $\dot{\pi}^2\sigma$ consistent with these bounds can be much greater than the $(\partial_i\pi)^2\sigma$ term, making exchange diagrams involving the latter comparatively negligible. Motivated by this observation, we henceforth only keep the $\dot{\pi}\sigma$ and $\dot{\pi}^2\sigma$ vertices in \eqref{actioncubic}, while dropping the gradient term $(\partial_i\dot{\pi})^2\sigma$ for concreteness. We also note that the EFT cubic operator $\pi\pi'\sigma$ in \eqref{Smixing} need not be included, as the vertices \eqref{Smixing} under consideration are time-independent. However, around generic rapidly oscillating backgrounds, this term could dominate the cubic action, as recently noted in \cite{Pajer:2024ckd}. Both terms can nevertheless be included in our computation by appropriately adjusting the weight-shifting operator discussed below. 

To summarise, the first two vertices in \eqref{actioncubic}
constitute the dominant contributions to the power spectrum and the bispectrum in our setup. 
As elaborated above, these diagrams can be written in terms of the seed function \eqref{ThreeptsSeed} as 
\begin{align}
    \Delta P(k)&=(2\pi^2\Delta_\zeta^2\,)\dfrac{r_1}{k^2}F(k,k)\,,\\ \label{PowerBispectrum}
    B(k_1,k_2,k_3)&=
    (4\pi^4\Delta_\zeta^4\,)\dfrac{r_2}{k_1k_2k_3}\dfrac{\partial^2}{\partial k_{12}^2}F(k_{12},k_3)+(t\text{-}\,\, \text{and}\,\,u\text{-channels})\,,
\end{align}
where $\Delta P$ is the contribution of the exchange diagram to the power spectrum, $\Delta^2_\zeta=\frac{H^4}{8\pi^2|\dot{H}|\Mpl^2}$ denotes the amplitude of the scalar power spectrum in free theory, and the parameters
\begin{align}
    r_1 &=\dfrac{\rho^2}{2H^2}\,,\\
    r_2 &=\Delta_\zeta^{-1}\dfrac{\rho}{2\pi\Lambda}\,,
\end{align}
characterising the overall sizes of the signals. 
Note that the relationships \eqref{PowerBispectrum}, which link the seed  function to the desired massless exchange diagrams, are identical to those obtained in the constant-mass case, see \cite{Jazayeri:2022kjy}(Eqs. 3.31 and 3.33a). These relations are invariant under changes in the mass because the corresponding weight-shifting operators only act on the external lines without touching the internal propagator.

\paragraph{Power spectrum and the bispectrum} For the convenience of our numerical computations, we expand the power spectrum and the bispectrum directly in terms of the seed function components, $f_{0,\pm 1}(u)$ and $F_{0,\pm 1}(u)$, which are already computed in Sections \ref{Three-point contact diagram} and \ref{Three-point exchange diagram}. Using these components,
\begin{align}
    \Delta P(k) &=(4\pi^2\Delta_\zeta^2\,r_1)\times \dfrac{|\mathscr{P}(\mu,\omega)|}{k^3}\cos\left[\omega \log\left(\frac{k}{k_0}\right)+\text{arg}\left(\mathscr{P}(\mu,\omega)\right)\right]\,,\\ \nn
    B(k_1,k_2,k_3) &=
     \frac{8\pi^4\Delta_\zeta^4\,r_2}{k_1k_2k_3 k_{12}^3}\left[\text{Re}\,\mathscr{B}(u)\cos\left(\omega\log\frac{k_3}{k_0}\right)-\text{Im}\,\mathscr{B}(u)\sin\left(\omega\log\frac{k_3}{k_0}\right)\right]\\ 
     &+(t\text{-}\,\, \text{and}\,\,u\text{-channels})\,,\label{BispectrumFinal}
\end{align}
with $k_0$ denoting the fiducial co-moving scale $1/|\eta_0|$ and 
\begin{align}
    \mathscr{P}(\mu,\omega)&=F_{+1}(1)+F_{-1}^*(1)+2f_{+1}(1)f_0^*(1)+2f_0(1)f_{-1}^*(1)\,,\\ \nn 
    \mathscr{B}(u)&=(2\partial_u+u\,\partial_u^2)\Big[F_{+1}(u)+F^*_{-1}(u)+\\ 
    &+f_{+1}(u)f_0^*(1)+f_0(u)f_{-1}^*(1)+f_{-1}^*(u)f_0(1)+f_0^*(u)f_{+1}(1)\Big]\,,
\end{align}
where the functional dependence on $\mu$ and $\omega$ is implicit through factors of $F_{\pm1}$ and $f_{0,\pm 1}$. In writing the formulae above, we have only retained the scale-dependent contribution to the power spectrum and the bispectrum. This is because the scale-invariant part effectively is just another exchange diagram with a fixed intermediate mass, which is extensively studied in the literature and carries no information about oscillations. Nevertheless, it is the sum of the two contributions that should be compared with observations to constrain the parameter space of our model.    

Under the squeezed limit, bispectrum is dominated by the $s$-channel contribution and simplifies to
\begin{align}
\label{SqueezedBispectrum}
\lim_{k_3\ll k_{1}\sim k_2}B(k_1,k_2,k_3) &\approx r_2\,P(k_1)\,P(k_3)\left(\dfrac{k_3}{k_1}\right)^{3/2}\\ \nn
&\,\times \left\lbrace|\mathcal{A}_1(\mu,\omega)|\cos\left[(\mu-\omega)\log(\frac{k_3}{2k_1})+\omega \log(-k_3\eta_0)+\vartheta_1(\mu,\omega)\right]\right.\\ \nn
&~~~\left.+|\mathcal{A}_2(\mu,\omega)|\cos\left[\mu\log(\frac{k_3}{2k_1})+\omega \log(-k_3\eta_0)+\vartheta_2(\mu,\omega)\right]+(\mu\to -\mu)\right\rbrace\,,
\end{align}
where $P(k)$ is the scale invariant power spectrum, and $\vartheta_i$ is the phase of the complex coefficients $\mathcal{A}_i$, which are given by
\begin{align}
\mathcal{A}_1(\mu,\omega) &=\dfrac{\mathbb{C}_1(\mu,\omega)}{2^{3/2}}\left(\omega-\mu+\frac{3}{2}i\right)\left(\omega-\mu+\frac{1}{2}i\right),\\ 
\mathcal{A}_2(\mu,\omega) &=\dfrac{\mathbb{C}_2(\mu,\omega)}{2^{3/2}}\left(\frac{3}{4}-\mu^2+2i\mu\right)\,.
\end{align}
\begin{figure}
    \centering
    \includegraphics[width=\textwidth]{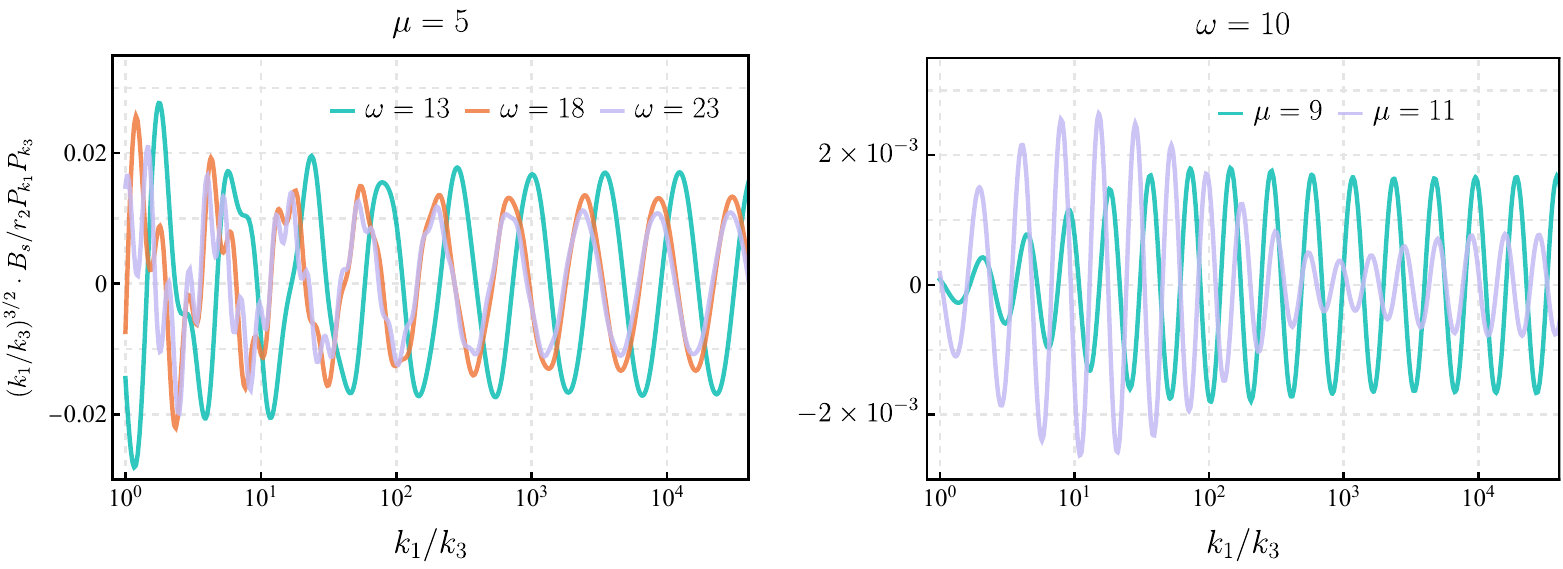}
    \caption{The $s$-channel bispectrum \eqref{BispectrumFinal}, normalised by two power spectrum $P(k_1)\,P(k_3)$ and the coupling parameter $r_2$, with an additional factor of $(k_1/k_3)^{3/2}$ for the better visualisation. The soft momentum $k_3$ is fixed as we vary $k_1$ and the scale-breaking phase is also at $x_0 =-k_3\eta_0 = 10$. \textit{Left panel:} the mass is fixed as $\mu=5$ while the oscillation frequency varies, $\omega = 13, 18, 23$, with coupling $g = 0.1$. \textit{Right panel:} the frequency is fixed at $\omega = 10$ while the mass varies, $m = 9$ and $11$, with the coupling $g = 0.05$. When the frequency $\omega$ exceeds the mass $\mu$, the oscillation becomes sufficiently energetic to excite more particles and overcome the Boltzmann suppression. This explains why the \textcolor{figure-green}{green} curve ($\omega>\mu$) is large than the \textcolor{figure-purple}{purple} ($\omega<\mu$) one in the squeezed limit.}\label{figSchannelB_Figure_Total}
\end{figure}
the expressions of $\mathbb{C}_1$ and $\mathbb{C}_2$ can be found in \eqref{coeffC1} and \eqref{coeffC2} respectively. In light of the discussion presented in the final part of Section \ref{Three-point exchange diagram}, let us concentrate on the regime of $H\ll m_0\ll \omega$ in which mass oscillations are expected to compensate for the otherwise Boltzmann-suppressed particle production in de Sitter space. This enhancement should manifest itself at the level of the bispectrum, where the squeezed-limit cosmological collider oscillations are typically proportional to the rate of particle production in the bulk. 

In Figure \ref{figSchannelB_Figure_Total}, we plot the $s$-channel bispectrum for various choices of the  parameters $\mu$ and $\omega$. Remarkably, even for very large masses $\mu \geq 5$, the cosmological collider signal remains sizeable, owing to the exponential enhancement of the particle production rate by mass oscillations. In contrast, in the scale-invariant case, this cosmological collider signal is strongly suppressed by the Boltzmann factor $e^{-\pi\mu}$, making it difficult to observe even for moderately large masses $\mu\gtrsim3$. In the \textit{left panel}, near the equilateral configuration, the intricate wiggly pattern arises from the superposition of multiple oscillatory modes\footnote{See \cite{Pinol:2023oux} for similar equilateral beating patterns present in setups with oscillatory vertices. We thank Sébastien Renaux-Petel for related discussions.}. As one moves toward the squeezed limit, the bispectrum becomes  dominated by the contribution \eqref{SqueezedBispectrum}. Clearly, in the regime of interest where $\omega>\mu$, the oscillation frequency is determined by $\mu$, regardless of how $\omega$ is varied. This is because only the second oscillatory component with frequency $\mu$ and amplitude $\mathcal{A}_2(\mu,-\omega)$ is enhanced, as we have illustrated several times above.

We now turn to estimating how detectable this enhanced cosmological collider could be in forthcoming CMB and LSS observations. To this end, we first introduce the dimensionless bispectrum shape function $S$, defined as
\begin{align}
    S(k_1,k_2,k_3)\equiv\frac{(k_1 k_2 k_3)^2}{(2\pi\Delta_\zeta)^{4}}\times B(k_1,k_2,k_3)\,.
\end{align}
A commonly used measure of the bispectrum is the amplitude of its shape function evaluated in the equilateral configuration, i.e. $f_{\rm{NL}}={10}/{9}\times|S(k,k,k)|$. With all analytical expressions at hand to evaluate the full bispectrum, we readily find that its magnitude is
\begin{align}
    &\mathcal{O}(g^0):\quad f_{\rm{NL}} \sim\mathcal{O}(0.1)\times\frac{r_2}{\mu^2}\,,\label{fNLg0}\\
    &\mathcal{O}(g^2):\quad f_{\rm{NL}} \sim \mathcal{O}(1)\times r_2\, g^2\,,\label{fNLg2}
\end{align}
where we consider the parameter regime $\omega > \mu \gg H$. The first term \eqref{fNLg0} comes from the scale-invariant contribution associated with the exchange of a fixed intermediate mass, since the cosmological collider signal is highly suppressed by the Boltzmann factor, the amplitude $f_{\rm NL}$ can be estimated from the EFT contribution, scaling as $f_{\rm NL} \sim \mu^{-2}$. The second term \eqref{fNLg2} comes from our new signals. The overall amplitude should be the summation of these two contributions, and when the  parameter $g\gtrsim \mu^{-1}$, the second term  takes over as the dominant one. 

These resonant features cannot be arbitrarily large, as the correction to the power spectrum is strongly constrained by observations. To obtain the two-point function, we simply take the limit of the three-point seed function $F({k_{12},s})$ as $k_3\to1$, following the same procedure used in deriving the two-point contact one $f_{\pm1}(1)$, whose details are provided in the Appendix \ref{sec: derivations}. Unlike the case of $f_{\pm1}(1)$, which can be resummed and expressed compactly in terms of one generalised hypergeometric function of higher weight \eqref{fpm1_ClosedForm}, $F(k,k)$ still retains one layer of summation. We find that this series converges rather slowly when the oscillation frequency $\omega$ far exceeds the mass scale $\mu$, as the expression involves extensive products of Gamma and hypergeometric functions. It becomes increasingly difficult to evaluate for larger parameters. We leave the systematic refinement of the series and a search for the closed-form expression through resummation for future work.

After applying the relevant relation \eqref{PowerBispectrum}, we can finally estimate the power spectrum as
\begin{align}
    &\mathcal{O}(g^2):\quad \frac{\Delta P(k)}{P(k)} \sim \mathcal{O}(0.1)\times r_1\,.g^2\,,\label{dPg2}
\end{align}
These features are tightly constrained by CMB experiments, with their amplitude required to be less than a few percent\cite{Akrami:2018odb}. This constraint translates into the bound  $r_1 g^2<10^{-1}$. The bispectrum amplitude, nevertheless, is determined by the parameter $r_2$ rather than $r_1$, with the two related through $r_2=(\Delta_{\zeta}^{-1}H/\Lambda\sqrt{2}\pi)\,\sqrt{r_1}$. Thanks to the enhancement factor $\Delta_{\zeta}^{-1}\sim10^{4}$, the resulting bispectrum can still be appreciably large despite the tight constraint on the features of the power spectrum. We can then roughly estimate the amplitude of bispectrum as
\begin{align}
    f_{\rm{NL}}\sim10^4\times \left(r_1g^2\right)^{\frac{1}{2}}\times\left(\frac{ gH}{\sqrt{2}\pi\Lambda}\right) \lesssim \mathcal{O}(10^4)\times\left(\frac{gH}{\Lambda}\right)\,,
\end{align}
where $0\leq g<1$ to avoid tachyonic instabilities. In this work, we consider $g\sim\mathcal{O}(0.1)$, while $\Lambda$ is constrained by \eqref{lessstringent},  typically  corresponding to $\mathcal{O}(10)H$. Consequently, over a broad range of parameter space, the bispectrum can reach a sizeable amplitude (e.g. $f_{\rm{NL}}\sim\mathcal{O}(10-10^2)$), that is within the range of sensitivity for future observations.

\section{Parametric resonance}\label{ParaResSection}

\begin{figure}[h!]
    \centering
    	\begin{tikzpicture}[scale=1.0, >=to]
		\draw[->, thick] (0,0) -- (7,0) node[right] {$t$};
		\draw[->, thick] (0,0) -- (0,5) node[above] {$w_{\rm{eff}}(t)$};
		\node at (3.8,2.5)
		{\small{$w_{\mathrm{eff}} (t) = \sqrt{\dfrac{k^2}{a^2(t)}+\mu^2}$}};
        \draw (-0.1,1.2) -- (0.0,1.2);
		\fill[red3,opacity=0.15] (0.01,0.8) rectangle (6.8,1.6);
		\node[red3] at (7.5,1.2) {${w_\mathrm{IR}}/{2}$};	
		\fill[cyan!80!black,opacity=0.2] (0.01,3.5) rectangle (6.8,4.3);
		\node[blue3] at (7.5,3.9) {$w_{\mathrm{UV}}/2$};
        \node at (-0.4,1.2) {$\mu$};
		\draw[thick, black, domain=0.5:6.8, samples=150]
		plot(\x,{sqrt(1.2 + 12*exp(-1.0*(\x-1)))});
	\end{tikzpicture}
    \caption{Schematic illustration of the two types of parametric resonances in our model. The black curve denotes the evolution of the effective frequency $w_{\rm eff}(t)$ of a massive particle with cosmic time, where it is dominated by the kinetic energy $k/a(t)=-k\eta$ at early times (UV) and by the rest mass $\mu$ at late times (IR). Parametric resonances can be triggered when the effective frequency sweeps across the resonance bands in the UV (\textcolor{blue3}{blue}) or in the IR (\textcolor{red3}{red}). Note that the UV resonances are always transient whereas the IR resonances are persistent.}
    \label{figResonancesBandsCartoon}
\end{figure}
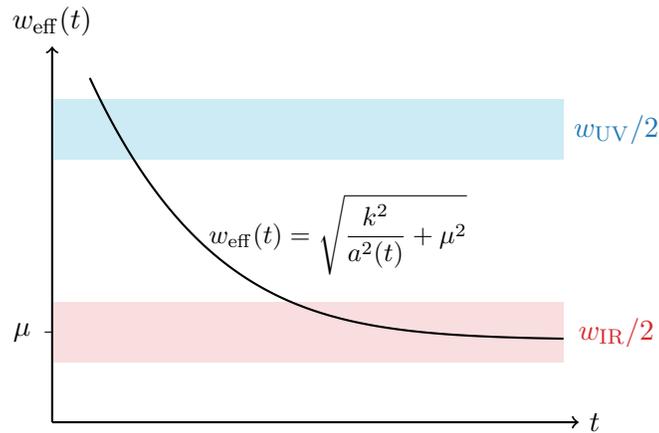

A dynamical system of oscillators typically exhibits novel resonance phenomena if the parameters of the system are also oscillating in time. These parametric resonances are widely studied throughout science and engineering. In cosmology, it has been applied to models of preheating \cite{Kofman:1994rk,Kofman:1997yn,Greene:1997fu}, particle production \cite{Bezrukov:2008ut,Garcia-Bellido:2008ycs,Lozanov:2017hjm}, gravitational waves \cite{Cai:2020ovp,Cai:2021yvq,Brandenberger:2022xbu}, fuzzy dark matter \cite{Hertzberg:2018zte,Dror:2018pdh,Brandenberger:2023idg}, and primordial black hole generation \cite{Cai:2018tuh,Chen:2020uhe,Zhou:2020kkf}. In our model, the oscillatory mass of the heavy field leads to different types of parametric resonances with dramatic consequences and colourful phenomenology as well. This is because the effective frequency $w_{\rm eff}(t)=\sqrt{k^2/a^2(t)+\mu^2}$ of the massive field varies with time due to the redshift of physical momentum, and depending on the frequency of mass oscillations $\omega$, parametric resonances can happen either in the UV or IR, where the effective frequency of the heavy particle is dominated by the kinetic energy and rest mass, respectively (see Figure \ref{figResonancesBandsCartoon} for illustration). In the following two subsections, we shall discuss them separately.

\subsection{IR resonances}\label{IRresonanceSubSect}

The most noticeable type of parametric resonance is when the mass oscillation frequency is close to the average mass, i.e. $\omega\sim \mu$. This effect can be viewed from both the bulk and boundary perspectives.

\paragraph{Bulk perspective} The equation of motion for the heavy field in the late-time (IR) limit is
\begin{align}
    \left[\frac{\partial^2}{\partial t^2}+m_0^2\left(1+g^2 \cos(\omega t)\right)-\frac{9}{4}\right](a^{3/2}\sigma(s,t))\approx 0~,
\end{align}
where we have dropped the redshifted momentum term $s^2/a^2\ll 1$. Equivalently, we can put the equation of motion into the standard form of a Mathieu equation,
\begin{align}
    \left[\frac{\partial^2}{\partial t^2}+\mu^2\left(1+\tilde{g}^2 \cos(\omega t)\right)\right](a^{3/2}\sigma(s,t))\approx 0~,\quad \tilde{g}^2\equiv g^2 \left(1+\frac{9}{4\mu^2}\right)~.
\end{align}
The standard analysis of the Mathieu equation shows the existence of narrow resonance bands in the weak coupling limit $\tilde g\ll 1$, corresponding to the resonant frequencies near $\omega=\omega_n = 2 \mu/n+\mathcal{O}(\tilde{g}^4)$, $n=1,2,\cdots$.\footnote{Note that in the weak-coupling limit $\tilde g\to 0$, the width of the resonance bands shrinks to zero faster than the central resonance frequency approaching $2 \mu/n$. Therefore to accurately characterise the narrow resonances in the weak-coupling limit, one needs to include the higher-order terms in $\omega_n=2 \mu/n+\mathcal{O}(\tilde{g}^4)$.} The width of the instability bands is given by
\begin{align}
    \frac{\Delta \omega_n}{\omega_n}\equiv\frac{1}{n^2}\frac{(n^2\tilde g^2/2)^{n}}{\left[2^{n-1}(n-1)!\right]^2}\,.\label{bandwidthFromEoM}
\end{align}
Near these resonance frequencies, the combination $a^{3/2}\sigma(t)\propto e^{\lambda_n t}$ grows exponentially in cosmic time at a rate\footnote{See Section 4.91 of \cite{mclachlan1951theory}.}
\begin{align}
    \lambda_n\equiv \frac{n}{4} \Delta\omega_n~.\label{RateWidthRelation}
\end{align}
For the primary resonance, $n=1$ and the growth rate reads
\begin{align}
    \lambda_1=\frac{\tilde g^2}{4} \mu =\frac{g^2}{4\mu}\left(\mu^2+\frac{9}{4}\right)~. \label{growthRateFromEoM}
\end{align}
Since the mode function of the heavy field experiences a continuous exponential amplification, its total particle number also increased exponentially in cosmic time. Consequently, the scaling exponents of the cosmological collider signal are expected to be altered by such resonances, giving rise to relative exponential growth in the squeezed limit,
\begin{align}
    F(k_{12},s)\sim\left(\frac{k_{12}}{s}\right)^{-1/2+\lambda_1\pm i \mu}~.
\end{align}

\paragraph{Boundary perspective} One might be concerned that the late-time analysis in the bulk captures only the qualitative features of the resonance but may not be quantitatively precise since it only includes the dynamics of the massive field but not the external inflaton field. In the following, we shall analyse the squeezed-limit behaviour of $F(k_1,k_2,s)$ starting directly from the boundary bootstrap equation and show that the boundary perspective predicts a consistent result.

We begin with an ansatz for $F\equiv F_{++}+F_{+-}$ that takes the form of a Fourier series with an unknown exponent $\alpha\in \mathbb{C}$,
\begin{align}
    F(k_{12},s)=\sum_{n=-\infty}^{\infty} \xi_n u^{-\alpha-i n \omega}~,\quad u\equiv \frac{s}{k_{12}}~,\label{squeezedAnsatz}
\end{align}
and plug it into the squeezed-limit ($u\ll 1$) bootstrap equation (see \eqref{inhomo} and \eqref{homo})
\begin{align}
    \left[u^2\partial^2_u+\left(\mu^2+\frac{1}{4}\right) \right]F(u)=\int_0^\infty \d x\, K(x) F\left(\frac{u}{1+ux}\right)~.
\end{align}
We then arrive at a three-term recurrence relation
\begin{align}
    {\sf A}_n \xi_n= {\sf B}_n \xi_{n+1} +{\sf C}_n \xi_{n-1}~, \quad n\in\mathbb{Z}~,
\end{align}
with
\begin{align}
    {\sf A}_n&\equiv \left(\alpha+\frac{1}{2} + i n \omega +i \mu \right) \left(\alpha+\frac{1}{2} +i n \omega -i \mu \right)~,\\
    {\sf B}_n&\equiv -\frac{1}{2}g^2\left(\mu^2+\frac{9}{4}\right)\frac{e^{\frac{\pi  \omega }{2}} \Gamma (-\alpha -i n \omega ) (-s\eta_0)^{-i \omega } }{\Gamma (-\alpha -i (n+1) \omega )}~,\\
    {\sf C}_n&\equiv -\frac{1}{2}g^2\left(\mu^2+\frac{9}{4}\right)\frac{e^{-\frac{\pi  \omega }{2}} \Gamma (-\alpha -i n \omega ) (-s\eta_0)^{i \omega
   }}{\Gamma (-\alpha -i (n-1) \omega )}~.
\end{align}
Near the primary resonance $\omega\sim\omega_1=2\mu$, we expect only the three leading modes with $n=0,\pm 1$ are relevant and all the overtones are subdominant. Thus we set $\xi_{\pm 2}=\xi_{\pm 3}=\cdots=0$ as an approximation that is valid only near the primary resonance. We are then left with three homogeneous linear equations of three variables $c_0, c_{\pm 1}$. In order that the equations admit a non-trivial solution, the determinant must vanish,
\begin{align}
    \,\left|
\begin{array}{ccc}
 {\sf A}_{-1} & {\sf B}_{-1} & 0 \\
 {\sf C}_0 & {\sf A}_0 & {\sf B}_0 \\
 0 & {\sf C}_1 &  {\sf A}_{1}\\
\end{array}
\right|=0~,
\end{align}
giving rise to a non-trivial constraint equation for $\alpha$,
\begin{align}
    \nonumber&\alpha ^6+3 \alpha ^5+\alpha ^4 \left(3 \mu ^2+2 \omega ^2+\frac{15}{4}\right)+\alpha ^3 \left(6 \mu ^2+4
   \omega ^2+\frac{5}{2}\right)\\
   \nonumber& +\alpha ^2 \left(-\frac{1}{32} g^4 \left(4 \mu ^2+9\right)^2+3 \mu
   ^4+\frac{9 \mu ^2}{2}+\omega ^4+3 \omega ^2+\frac{15}{16}\right)\\
   \nonumber& +\alpha  \left(-\frac{1}{32} g^4
   \left(4 \mu ^2+9\right)^2+\frac{3}{16} \left(4 \mu ^2+1\right)^2+\omega ^4+\omega
   ^2\right)\\
   & +\frac{1}{128} \left(2 \left(4 \mu ^2+1\right) \left(16 \mu ^4+\mu ^2 \left(8-32 \omega
   ^2\right)+\left(4 \omega ^2+1\right)^2\right)-g^4 \left(4 \mu ^2+9\right)^2 \left(4 \mu ^2-4 \omega
   ^2+1\right)\right)=0~.\label{boundaryResExponentConstraintEq}
\end{align}
Notice that the dependence on $\eta_0$ miraculously cancel out in the characteristic equation, showing that the resonance is not sensitive to the phase of mass oscillations.

This algebraic equation turns out to be exactly solvable, yielding six roots for the characteristic exponent $\alpha$. We focus on the weak-coupling regime and expand them in powers of $g\ll 1$ and around $|\omega-2\mu|\lesssim g^2 \mu$. The real parts of the first four roots are corrected at order $\mathcal{O}(g^2)$,
\begin{align}
    \alpha_{\rm I}^{(\sf ab)}\equiv -\frac{1}{2}+{\sf a}\frac{i \omega }{2}+\frac{{\sf b}}{2} \sqrt{\frac{g^4 }{4 \mu ^2}\left(\mu ^2+\frac{9}{4}\right)^2-(\omega -2 \mu )^2}+\mathcal{O}(g^4)~,\quad {\sf a,b}=\pm~,\label{alphaIsolutions}
\end{align}
whereas those of the other two roots are not,
\begin{align}
    \alpha_{\rm II}^{\pm}\equiv -\frac{1}{2}\pm  i (\omega+\mu)+\mathcal{O}(g^4)~.
\end{align}
Near $\omega_1=2\mu$, the $\alpha_{\rm I}$ modes correspond to positive- and negative-frequency modes that either grow or decay in the squeezed limit $u\ll 1$. The maximal growth rate is achieved at $\omega_1=2\mu$, where
\begin{align}
    \lambda_1=\frac{1}{2}+\Re \alpha_{\rm I}^{\pm +}=\frac{g^2}{4\mu}\left(\mu ^2+\frac{9}{4}\right)~.\label{growthRateFromBoundaryEq}
\end{align}
This IR parametric resonance disappears when the square-root term in \eqref{alphaIsolutions} vanishes, setting the width of the resonance to be
\begin{align}
    \frac{\Delta \omega_1}{\omega_1}=\frac{g^2}{2}\left(1+\frac{9}{4\mu^2}\right)~.\label{bandwidthFromBoundaryEq}
\end{align}
Comparing \eqref{growthRateFromBoundaryEq}, \eqref{bandwidthFromBoundaryEq} to \eqref{growthRateFromEoM} and \eqref{bandwidthFromEoM}, we see that the boundary bootstrap equation predict the same scaling exponent with our naive late-time analysis in the bulk, now taking into account the external inflaton dynamics. Unsurprisingly, the IR parametric resonance is a non-perturbative phenomenon that cannot be fully understood at the level of our $\mathcal{O}(g^2)$ perturbation theory in Section \ref{BdyIDEg2BootstrapSection}. This is because resonant growth requires the coherent resummation of an \textit{infinite} number of oscillating mass insertions. At any given order in the perturbation theory, one only observes a pole at $\omega=\omega_n$ (see e.g. \eqref{BnCoeff} at $\omega=\omega_1=2\mu$), indicating the need of resummation. However, we shall indeed confirm our resonance analysis using numerical bootstrap later in Section \ref{numericalBootstrapSection}.

The $\alpha_{\rm II}$ modes, on the other hand, are not growing nor decaying in the squeezed limit. Rather, they oscillate at an overtone frequency $\Im \alpha_{\rm II}^{\pm}=\pm 3\mu$, and can be neglected whenever there is a resonant growing mode.

To move on to the secondary resonance at $n=2$, we simply need to push forward the cutoff in $\xi_n$ and demand $\xi_{\pm 3}=\xi_{\pm 4}=\cdots=0$. The characteristic exponent is then constrained by
\begin{align}
    \,\left|
\begin{array}{ccccc}
 {\sf A}_{-2}& {\sf B}_{-2} & 0 & 0 & 0 \\
 {\sf C}_{-1}& {\sf A}_{-1} & {\sf B}_{-1} & 0 & 0 \\
 0 & {\sf C}_0 & {\sf A}_0 & {\sf B}_0 & 0\\
 0 & 0 & {\sf C}_1 &  {\sf A}_{1} & {\sf B}_1\\
 0 & 0 & 0 & {\sf C}_2 &  {\sf A}_{2} 
\end{array}
\right|=0~,
\end{align}
similar analysis gives the growth rate
\begin{align}
    \lambda_2=\frac{g^4\mu}{8}\left(1+\frac{9}{4\mu^2}\right)^2\label{growthRateSecondary}
\end{align}
at the secondary resonant frequency
\begin{align}
    \omega_2=\mu\left[1-\frac{g^4}{12}\left(1+\frac{9}{4\mu^2}\right)^2\right]~.
\end{align}
The bandwidth also agrees with the general formula \eqref{RateWidthRelation}. Higher overtones can be analysed in an analogous fashion.

\paragraph{Instabilities in the IR and the sensitivity to the UV} The narrow resonances in the IR source exponential production of the massive particles and could lead to instabilities if the production rate overcomes cosmic dilution. As the massive field amplitude grows as $\sigma\propto a^{-3/2+\lambda_n}$ near the $n$-th resonance, the energy density stored in the $\sigma$-sector evolves as $\rho_\sigma\sim \mu^2 \sigma^2\propto a^{-3+2\lambda_n}$. Requiring the inflationary background stability therefore constrains
\begin{align}
    \text{Background stability: }\lambda_n<\frac{3}{2}~.
\end{align}
Examining the regularity of the boundary bootstrap equation shows a stronger constraint, since the integral over kinematics near the resonances goes as
\begin{align}
    \int_0^\infty \d x\, K(x)\, F\left(\frac{u}{1+ux}\right)\sim \int_0^\infty \d x\, \frac{1}{x}\, x^{-3/2+\lambda_n}~,
\end{align}
whose manifest convergence requires
\begin{align}
    \text{Perturbation stability: }\lambda_n<\frac{1}{2}~.
\end{align}
This bound can also be interpreted on bulk side as a requirement on the late-time convergence of $\varphi$ correlators.

Notice that interestingly, one might expect that stability should always be maintained as long as the coupling is weak i.e. $g\ll 1$. This is indeed the case away from the IR resonances where the oscillation effect conducted to the massive field is negligible. However, at the resonances, the scaling exponents $\lambda_n$ depend on both the coupling $g$ and the mass $\mu$, as seen from \eqref{growthRateFromBoundaryEq} and \eqref{growthRateSecondary}. Thus somewhat counter-intuitively, increasing the mass (and the oscillation frequency) does not decouple high-energy processes from the EFT at Hubble scale, but rather destabilises it further through copious particle production. 
Of course, this bizarre behaviour is a non-perturbative effect exclusively within the resonance bands, whose relative widths $\Delta\omega_n/\omega_n$ do shrink to zero at weak coupling (see e.g. \eqref{bandwidthFromBoundaryEq}). The general lesson to be learnt here is that low-energy EFTs decouple from high-energy physics for typical parameter choices in theory space, but could become UV-sensitive in special cases.

\subsection{UV resonances}\label{UVresonances}

\begin{figure}[h!]
    \centering
    \includegraphics[scale=0.9]{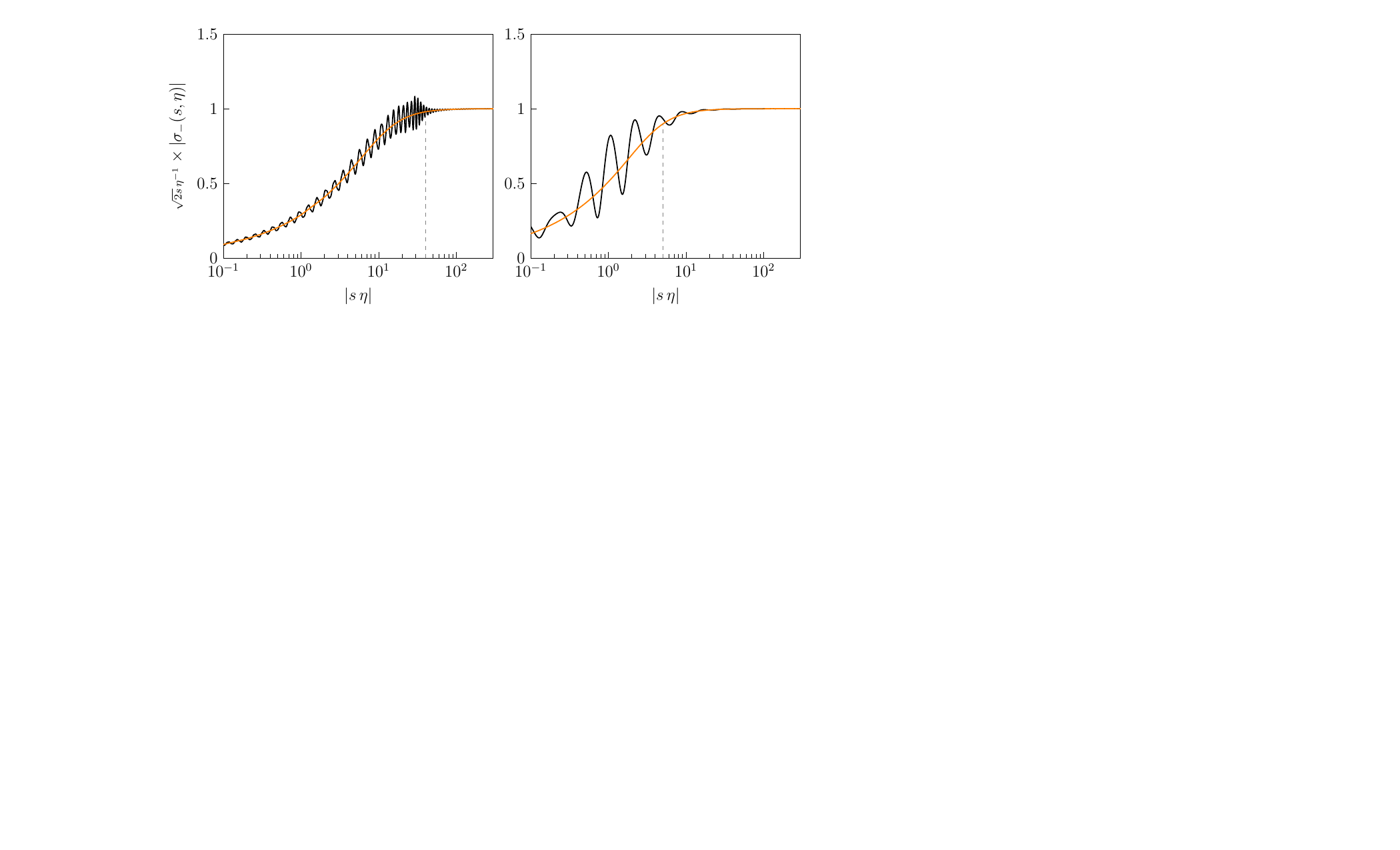}
    \caption{The time evolution of the modulus of the mode function (with comoving momentum $s$), from deep inside the horizon ($|s\eta|\gg 1$) to the end of inflation ($|s\eta|\to 0$), evaluated for the parameter values:   
    $(\omega=80.0,\,m_0=12.0,\,g^2=0.36)$ [\textit{left} \textbf{black} curve] and $(\omega=10.0,\,m_0=4.0,\,g^2=0.36)$ [\textit{right} \textbf{black} curve]. The \textcolor{orange}{orange} line in each case corresponds to $g=0$ (with the same mass $m_0$). The dashed lines mark the UV resonance at $|s\eta|=\omega/2$, roughly coincident with the onset of oscillations in the modulus $|\sigma_-|$, induced by particle production. These oscillations are (relative to the benchmark orange curves) of order $g^2$ [e.g. \eqref{bogoinferred}], consistent with the transient nature of the UV resonance
    . See Section \ref{UVresonances} for discussions.  
    By contrast, the oscillations in the \textcolor{orange}{orange} curves are invisibly small due to the Boltzmann suppression. 
    }
    \label{fig: modefunctionresonance}
\end{figure}

The parametric resonance considered so far only occurs for finely tuned masses near $\mu\sim \omega/2$.  Beyond this infrared effect and more generally, the heavy field equation of motion for $\omega\gg m_0$ points to another resonance at early times, when the heavy field's physical momentum crosses the characteristic energy scale of oscillations, $\omega$. Similar resonances are frequently encountered in reheating scenarios after inflation, with inflaton-dependent interactions (e.g. $\phi^2\,\sigma^2$) driving high-frequency mass oscillations in the heavy sector \cite{Kofman:1997yn}. Through the corresponding parametric resonance, the particle occupation number can grow exponentially within certain instability bands. 
On the other hand, our setup concerns the evolution of the massive field during inflation; every mode exits the would-be instability band shortly after entry. Consequently (for $g<1$), the expanding background prevents the parametric resonance from fully developing during inflation, with no exponential amplification expected to arise. Still, the heavy field undergoes a mild enhancement of order $g^2$, within the reach of perturbation theory, as its momentum crosses the frequency scale (i.e. $s/a(\eta)\sim \omega/2$). See Figure \ref{fig: modefunctionresonance} for an illustration. This ensures that our perturbative computation for the exchange diagram is reliable and free from contamination by any sizeable non-perturbative effects. Therefore, our results should be reproducible by an explicit bulk computation in which mass oscillations are treated perturbatively within the heavy field's equation of motion (as long as $\mu$ is not close to $\omega/2$).   

We have already employed the bulk picture in Section \ref{Three-point contact diagram} to shed light on the soft behaviour of the three-point function $f$.  Here, we provide further details on the perturbed mode function and identify specific components of its late-time oscillations that are enhanced by the UV resonance noted above.

We begin with the equation of motion for $\Delta\sigma_-$, 
\begin{align}
        (\eta^2\partial_\eta^2-2\eta\partial_\eta+s^2\eta^2+m_0^2)\,\Delta\sigma_-=-\Delta m^2(\eta)\,\sigma_0^{(0)}\,,
\end{align}
where $\Delta m^2=g^2m_0^2\cos[\omega(t-t_0)]$. The initial condition for $\Delta\sigma_-$ can be set by noting that the unperturbed component $\sigma^{(0)}_-(s,\eta)$ already saturates the Bunch-Davies behaviour as $\eta\to -\infty$. This implies that $\Delta\sigma_-$ must diverge more slowly than $\eta$ at early times,    
\begin{align}
    \lim_{\eta\to -\infty} \dfrac{\Delta\sigma_-(s,\eta)}{\eta\,\exp(is\eta)}=0\,.
\end{align}
This boundary condition is automatically satisfied by writing
\begin{align}
\label{retarded}
    \Delta\sigma_-(s,\eta)=-i\int_{-\infty}^{0} \dfrac{{\rm{d}}\eta'}{\eta'^4}\,G_R(\eta,\eta')\,\Delta m^2(\eta')\sigma_-(s,\eta')\,,
\end{align}
where
\begin{align}
  G_R(\eta,\eta')=\theta(\eta-\eta')\left[\sigma^{(0)}_-(s,\eta')\sigma^{(0)}_+(s,\eta)-\sigma^{(0)}_-(s,\eta)\sigma^{(0)}_+(s,\eta')\right]\,, 
\end{align}
is the retarded Green function. 
Hereafter we drop the superscript on $\sigma_{-}^{(0)}$ for notational simplicity. 
As discussed in Section \ref{Three-point contact diagram}, it is instructive to split the perturbed mode function into a particular and a homogeneous part. This can be achieved at the level of the time integral \eqref{retarded} by decomposing the retarded propagator as
\begin{align}
    G_R(\eta,\eta')=[\sigma_-(s,\eta')\sigma_+(s,\eta)-\sigma_-(s,\eta)\sigma_+(s,\eta')]-G_A(\eta,\eta')\,,
\end{align}
where $G_A$ is the \textit{advanced propagator}. By way of this decomposition, $\Delta\sigma_-$ separates into: 
\begin{align}
\label{sigmaI}
    \Delta\sigma^{\text{part}}_-(s,\eta)&=i\int_{\eta}^{0} \dfrac{{\rm{d}}\eta'}{\eta'^{4-\epsilon}}\,G_A(\eta,\eta')\,\Delta m^2(\eta')\sigma_-(s,\eta')\,,\\
    \label{sigmaII}
    \Delta\sigma^{\text{hom}}_-(s,\eta)&=I_1(s)\,\sigma_+(s,\eta)+I_2(s)\sigma_-(s,\eta)\,,
\end{align}
where
\begin{align}
\nn
      I_1(s)&=-i\int^{0}_{-\infty(1-i\epsilon)}\dfrac{{\rm{d}}\eta'}{\eta'^{4-\epsilon}}\sigma^2_-(s,\eta')\Delta m^2(\eta')\,,\\ \label{Iplusminus}
    I_2(s)&=i\int^{0}_{-\infty(1-i\epsilon)}\dfrac{{\rm{d}}\eta'}{\eta'^{4-\epsilon}}\sigma_-(s,\eta')\sigma_+(s,\eta')\Delta m^2(\eta')\,.
\end{align}
Note that, in the $\eta\to 0$ limit, the convergence of the time integrals above is ensured by changing the volume factor from $\eta^{-4}$ to $\eta^{4-\epsilon}$. Meanwhile, the total mode function $\Delta\sigma _-=\Delta\sigma^{\text{part}} _-+\Delta\sigma^{\text{hom}} _-$ 
is unaffected by this infrared regulation because the retarded Green function vanishes in any case for $\eta'>\eta$, ensuring the convergence of \eqref{retarded} at finite $\eta$. 

At late times $\eta\to 0$, the time integral associated with the particular piece $\Delta\sigma^{\text{part}}_-$ in \eqref{sigmaI} can be easily evaluated, yielding the power-law behaviour in \eqref{sigmapart}. 
Plugging this into the three-point function bulk integral \eqref{bulkparticular}, as we explained in Section \ref{Three-point contact diagram}, reproduces the soft limit $u\to 0$ of the particular ansatz in \eqref{fansatz}.

Now we proceed to the computation of the homogeneous component $\Delta\sigma^{\text{hom}}_-$ at late times. It is useful to rewrite \eqref{sigmaII} in terms of the Bogolyubov coefficients defined in \eqref{homocorrect}, which are, as functions of $I_{1,2}(s)$,  
\begin{align}
    \Delta\alpha &=\alpha_0\,I_2(s)+\beta_0^*\,I_1(s)=\Delta\alpha_-\,x_0^{-i\omega}+\Delta\alpha_+\,x_0^{+i\omega}\,,\\ \nn
    \Delta\beta&=\alpha_0^*\,I_1(s)+\beta_0\,I_2(s)=\Delta\beta_-\,x_0^{-i\omega}+\Delta\beta_+\,x_0^{+i\omega}\,.
\end{align}
Their harmonic components $\Delta\alpha_\pm$ and $\Delta\beta_\pm$ can thus be expressed as 
\begin{align}
\nn
    \Delta \beta_{\pm}&=-\dfrac{i\alpha_0^*\,g^2 m_0^2}{2}\int^{0}_{-\infty(1-i\epsilon)}\dfrac{{\rm{d}}\eta'}{\eta'^{4-\epsilon}}\sigma^2_-(s,\eta')(-s\eta)^{\mp i\omega}\\ \label{betapmbulk}
    &+\dfrac{i\beta_0 g^2 m_0^2}{2}\int^{0}_{-\infty(1-i\epsilon)}\dfrac{{\rm{d}}\eta'}{\eta'^{4-\epsilon}}\sigma_-(s,\eta')\sigma_+(s,\eta')(-s\eta)^{\mp i\omega}\,,\\ \nn
    \Delta \alpha_{\pm}&=-\dfrac{i\beta_0^*\,g^2 m_0^2}{2}\int^{0}_{-\infty(1-i\epsilon)}\dfrac{{\rm{d}}\eta'}{\eta'^{4-\epsilon}}\sigma^2_-(s,\eta')(-s\eta)^{\mp i\omega}\\  \label{alphapmbulk}
    &+\dfrac{i\alpha_0 g^2 m_0^2}{2}\int^{0}_{-\infty(1-i\epsilon)}\dfrac{{\rm{d}}\eta'}{\eta'^{4-\epsilon}}\sigma_-(s,\eta')\sigma_+(s,\eta')(-s\eta)^{\mp i\omega}\,.
\end{align}
Even though for generic values of the parameters $\mu$ and $\omega$ these integrals can be evaluated in closed form, it is more helpful to concentrate on their large frequency limit ($\omega\gg H,m$), where the impact of the ultraviolet resonance becomes more transparent. In particular, for $\Delta\alpha_-$ and $\Delta\beta_-$, the integrals in the first lines of \eqref{betapmbulk} and \eqref{alphapmbulk} are dominated in the large frequency regime by a saddle point at $|s\eta|=\omega/2$, where the sub-horizon and mass oscillations of the heavy field are in resonance. 
In contrast, no such UV saddle point exists for the second integrals in these expressions nor for the components $\Delta\alpha_+$ and $\Delta\beta_+$. Accordingly, these contributions and components exponentially decay as $\omega\to \infty$. 

Near the aforementioned saddle point, the mode function is well approximated by its early time limit,  
\begin{align}
    \sigma_-(s,\eta)\sim -\dfrac{H\eta}{\sqrt{2s}}\exp(ik\eta)\,,\qquad (|s\eta|\sim \omega/2\gg H)\,.
\end{align}
Substituting this into \eqref{betapmbulk} and using the asymptotic formula\footnote{Note that the right-hand side of this identity is independent of the upper bound $x_{\rm end}<0$ as long as the resonance (occurring at $x'\sim -\omega/2$) is enclosed within the integration range.} 
\begin{align}
    \int_{-\infty}^{x_{\rm end}} \dfrac{dx'}{x'^2}\exp(2ix')(-x')^{+ i\omega}\approx 2\omega^{-3/2}\sqrt{2\pi}\exp(-i\pi/4)\exp(-i\,\omega)\exp(i\omega\log(\omega/2))\qquad (\omega\to \infty)\,,
\end{align}
we arrive at
\begin{align}
    \Delta \beta_{-}&\approx -\dfrac{i\alpha_0^*\,g^2 m_0^2}{4}\int^{s\eta_{\rm end}}_{-\infty}\dfrac{dx}{x^2}e^{2ix}\,(-x)^{+i\omega}
    \approx -i\sqrt{\frac{\pi}{2}}g^2 m_0^2 \alpha_0^*\,e^{-i(\omega+\pi/4)}{\omega}^{-\frac{3}{2}+i\omega}\,2^{-i\omega}\,,\\ \nn
     \Delta\alpha_- &\approx \dfrac{\beta_0^*}{\alpha_0^*}\Delta\beta_-\,.
\end{align}
Note that these coefficients are only power-law suppressed in the large frequency regime. In contrast, their counterparts $\Delta\alpha_+$ and $\Delta\beta_+$ exponentially fall off as $\omega\to \infty$ since they are not enhanced through the resonance mechanism. Indeed, after setting $|\alpha_0|\sim 1$ for $m_0\gg H$, our approximate formula for $\Delta\beta_-$ matches with the particle production rate we had inferred from the soft limit of the three-point function in  \eqref{bogoinferred}.


\section{Numerical bootstrap}\label{numericalBootstrapSection}

To complement the analytical picture from perturbative methods, in this section, we directly bootstrap the scalar bispectrum in kinematic space using numerical tools. Our strategy is to first discretise the integro-differential equation and then apply the Finite Difference Method (FDM) to obtain non-perturbative numerical solutions. 

\subsection{Finite difference approach}

Consider the bootstrap equation for $F\equiv F_{++}+F_{+-}$ as an integro-differential equation of the Volterra type (see \eqref{inhomo} and \eqref{homo}),
\begin{align}
     \left[(k_{12}^2-s^2)\partial^2_{k_{12}}+2k_{12}\partial_{k_{12}}+\left(\mu^2+\frac{1}{4}\right)\right]F(k_{12},s) =\frac{1}{k_{12}+s}+\int_{0}^\infty {\rm{d}}q\,K(q)\,F(k_{12}+q,s)\,.\label{kspaceBootstrapEq}
\end{align}
Motivated by our analytical analysis, we anticipate non-trivial behaviours of correlators to lie in the squeezed limit $k_{12}/s \gg 1$, where the bispectrum exhibits oscillations uniform in the logarithm of the momentum ratio. Therefore, we perform a change of variables,
\begin{align}
    \frac{k_{12}}{s}\equiv e^r~,\quad r\in [0,\,\infty)~.
\end{align}
The bootstrap equation translates to
\begin{align}
    \left[\left(1-e^{-2r}\right)\partial_r^2+\left(1+e^{-2r}\right)\partial_r+\left(\mu^2+\frac{1}{4}\right)\right]F(r;x_0)=\frac{1}{1+e^r}+ \int_r^\infty \d r'\, \mathcal{K}(r,r';x_0) \,F(r';x_0)~,\label{rspaceBootstrapEq}
\end{align}
where the kernel reads
\begin{align}
     \mathcal{K}(r,r';x_0)\equiv -\frac{1}{2}g^2 \left(\mu^2+\frac{9}{4}\right)\left[\frac{e^{-\pi\omega/2}}{\Gamma( i\omega)}  \frac{x_0^{ i\omega}(e^{r'}-e^r)^{i\omega}}{1-e^{-(r'-r)}}+(\omega\to-\omega)\right]~.\label{KernelFormInrSpace}
\end{align}
To avoid the cluttering of symbols, we will omit the functional dependence on $x_0$ and focus on that on the momentum ratio $r$. 

\paragraph{Discretisation} To numerically solve this integro-differential equation, we first truncate the kinematic space by setting a cutoff on the momentum ratio i.e. $k_{12}/s<e^L$ and restrict $r\in [0,L]$. We then discretise the interval $[0,L]$ into $N$ sections,
\begin{align}
    r\in [0,L]\to r_i=\frac{i}{N}L~,~i=0,1,\cdots, N~.
\end{align}
The integral on the right-hand side is replaced by a finite sum under the trapezoidal rule of Newton-Cotes quadrature,
\begin{align}
    \int_r^L dr' \mathcal{K}(r,r') F(r')\to \frac{L}{2N} \mathcal{K}(r_i,r_i) F(r_i) +\sum_{j=i+1}^{N-1} \frac{L}{N} \mathcal{K}(r_{i},r_j) F(r_{j})+\frac{L}{2N} \mathcal{K}(r_i,r_N)  F(r_N)~.\label{NewtonCotesQuadratureFormOriginal}
\end{align}
The derivatives on the left-hand side translate to finite differences under the midpoint rule, 
\begin{subequations}
    \begin{align}
        \partial_r F(r)&\to \frac{F(r_{i+1})-F(r_{i-1})}{2L/N}~,\\
        \partial_r F(0)&\to \frac{F(r_{1})-F(r_{0})}{L/N}~,\quad \partial_r F(L)\to \frac{F(r_{N})-F(r_{N-1})}{L/N}~,\\
        \partial_r^2 F(r)&\to \frac{F(r_{i+1})-2F(r_i)+F(r_{i-1})}{L^2/N^2}~,\\
        \partial_r^2 F(0)&\to \frac{F(r_{2})-2F(r_1)+F(r_{0})}{2L^2/N^2}~,\quad\partial_r^2 F(L)\to \frac{F(r_{N})-2F(r_{N-1})+F(r_{N-2})}{2L^2/N^2}~.
    \end{align}
\end{subequations}
After discretisation, \eqref{rspaceBootstrapEq} translates to a matrix equation
\begin{align}
    \mathcal{D} \mathbf{F}=\bm{\mathcal{S}}+ \mathcal{Q} \mathbf{F}~,\label{matrixFormBootstrapEqRaw}
\end{align}
where $\mathbf{F}=\Big(F(r_0),\cdots, F(r_L)\Big)^T$. The difference matrix on the left-hand side is given by
\begin{align}
    \mathcal{D}\equiv \left(1-e^{-2R}\right)D_2+\left(1+e^{-2R}\right)D_1+\left(\mu^2+\frac{1}{4}\right)D_0~,
\end{align}
where
\begin{align}
    R=\left(\begin{array}{ccccc}
		r_0 & ~ & ~ & ~ & ~\\
		~ & r_1 & ~ & ~ & ~\\
        ~ & ~ & \ddots & ~ & ~\\
        ~ & ~ & ~ & r_{N-1} & ~\\
        ~ & ~ & ~ & ~ & r_N\\
	\end{array}\right)~,\quad D_0=\left(\begin{array}{ccccc}
		1 & ~ & ~ & ~ & ~\\
		~ & 1 & ~ & ~ & ~\\
        ~ & ~ & \ddots & ~ & ~\\
        ~ & ~ & ~ & 1 & ~\\
        ~ & ~ & ~ & ~ & 1\\
	\end{array}\right)
\end{align}
and
\begin{align}
    D_1=\frac{N}{L}\left(\begin{array}{ccccc}
		-1 & 1 & ~ & ~ & ~\\
		-\frac{1}{2} & 0 & \frac{1}{2} & ~ & ~\\
        ~ & ~ & \ddots & ~ & ~\\
        ~ & ~ & -\frac{1}{2} & 0 & \frac{1}{2}\\
        ~ & ~ & ~ & -1 & 1\\
	\end{array}\right)~,\quad D_2=\frac{N^2}{L^2}\left(\begin{array}{ccccc}
		\frac{1}{2} & -1 & \frac{1}{2} & ~ & ~\\
		1 & -2 & 1 & ~ & ~\\
        ~ & ~ & \ddots & ~ & ~\\
        ~ & ~ & 1 & -2 & 1\\
        ~ & ~ & \frac{1}{2} & -1 & \frac{1}{2}\\
	\end{array}\right)~.
\end{align}
The quadrature matrix on the right-hand side reads
\begin{align}
    \mathcal{Q}=\frac{L}{N}\left(\begin{array}{cccccc}
		\frac{1}{2} \mathcal{K}_{0,0} & \mathcal{K}_{0,1} & \mathcal{K}_{0,2} & \cdots & \mathcal{K}_{0,N-1} & \frac{1}{2} \mathcal{K}_{0,N}\\
		~ & \frac{1}{2} \mathcal{K}_{1,1} & \mathcal{K}_{1,2} & \cdots & \mathcal{K}_{1,N-1} & \frac{1}{2} \mathcal{K}_{1,N}\\
        ~ & ~ & \ddots & \cdots & \vdots & \vdots\\
        ~ & ~ & ~ & ~ & \frac{1}{2}\mathcal{K}_{N-1,N-1} & \frac{1}{2}\mathcal{K}_{N-1,N}\\
        ~ & ~ & ~ & ~ & ~ & 0\\
	\end{array}\right)~,\label{QuadratureMatrixOriginal}
\end{align}
where we have short-handed $\mathcal{K}_{i,j}\equiv \mathcal{K}(r_i,r_j)$. The source vector is
\begin{align}
    \bm{\mathcal{S}}=\frac{1}{1+e^R}\Big(1~1\cdots 1~1\Big)^T~.
\end{align}

\paragraph{Boundary conditions} Note that naively the matrix equation \eqref{matrixFormBootstrapEqRaw} can be formally solved by
\begin{align}
    \mathbf{F}=\left(\mathcal{D}-\mathcal{Q}\right)^{-1}\bm{\mathcal{S}}~,
\end{align}
as the matrix $\mathcal{D}-\mathcal{Q}$ appears invertible. However, this does not seem to leave freedom for implementing boundary conditions. This apparent dilemma is resolved by noticing that the discretised equations near the boundary $r=r_0, r_N$ are flawed and do not reflect the correct boundary conditions. Therefore the appropriate implementation is to remove the first and last rows of \eqref{matrixFormBootstrapEqRaw} and extend two extra equations representing the correct boundary conditions. In terms of matrix equations, we define a projection matrix
\begin{align}
    \mathcal{P}=\left(\begin{array}{ccccc}
		0 & ~ & ~ & ~ & ~\\
		~ & 1 & ~ & ~ & ~\\
        ~ & ~ & \ddots & ~ & ~\\
        ~ & ~ & ~ & 1 & ~\\
        ~ & ~ & ~ & ~ & 0\\
	\end{array}\right)~,
\end{align}
and the truncated equation takes the form
\begin{align}
    \mathcal{P}(\mathcal{D}-\mathcal{Q})\mathbf{F}=\mathcal{P} \bm{\mathcal{S}}~,
\end{align}
where $\mathcal{P}(\mathcal{D}-\mathcal{Q})$ is henceforth no longer invertible. To solve the system, we extend the system by adding the boundary conditions,
\begin{align}
    \left[\mathcal{P}(\mathcal{D}-\mathcal{Q})+\mathcal{C}_{\rm b.c.}\right]\mathbf{F}=\mathcal{P} \bm{\mathcal{S}}+\bm{\mathcal{S}}_{\rm b.c.}~,
\end{align}
where
\begin{align}
    \mathcal{C}_{\rm b.c.}=\left(\begin{array}{ccccc}
		c_0^{\rm I} & c_1^{\rm I} & \cdots & c_{N-1}^{\rm I} & c_N^{\rm I}\\
		0 & 0 & ~ & ~ & ~\\
        ~ & ~ & \ddots & ~ & ~\\
        ~ & ~ & ~ & 0 & 0\\
        c_0^{\rm II} & c_1^{\rm II} & \cdots & c_{N-1}^{\rm II} & c_N^{\rm II}\\
	\end{array}\right)~,\quad \text{and}\quad 
    \bm{\mathcal{S}}_{\rm b.c.}=\left(\begin{array}{c}
		s_0^{\rm I} \\
		0 \\
        \vdots \\
        0 \\
        s_0^{\rm II} \\
	\end{array}\right)~.
\end{align}
Now the discretised bootstrap equation can be readily solved as
\begin{align}
    \mathbf{F}_{\rm sol}=\left[\mathcal{P}(\mathcal{D}-\mathcal{Q})+\mathcal{C}_{\rm b.c.}\right]^{-1}\left(\mathcal{P} \bm{\mathcal{S}}+\bm{\mathcal{S}}_{\rm b.c.}\right)~,\label{formalFDMSolutionwithBCbeforeReg}
\end{align}
subjected to the boundary conditions I and II.

In numerics, however, we do not have analytical control over the behaviour of the solution near the folded ($r\to 0$), squeezed ($r\to \infty$) or factorisation limit ($r\to -i\pi$), therefore the boundary conditions corresponding to the correct Bunch-Davies vacuum are difficult to implement. The lack of boundary conditions is unlikely to be settled by discretising the bootstrap equation over physical kinematics alone, but might require a full analysis on the complex plane.\footnote{One can try to eliminate the folded-limit pole by minimising $F(0)$, but there is still a one-(complex) parameter family of solutions left undetermined.} As a result, instead of determining the boundary conditions from first principles, we directly implement two matching conditions at $r=r_{\rm I},r_{\rm II}$ using the analytical solution at order $\mathcal{O}(g^2)$,\footnote{This fixes $c_i^{\rm I,II}=\delta_{i,i_{\rm I,II}}$ and $s_i^{\rm I,II}=F^{(g^2)}(r_{i_{\rm I,II}})\delta_{i,i_{\rm I,II}}$ with $r_{i_{\rm I,II}}=r_{\rm I,II}$.}
\begin{subequations}
    \begin{align}
    F(r_{\rm I})&=F^{(g^2)}(r_{\rm I})~,\\
    F(r_{\rm II})&=F^{(g^2)}(r_{\rm II})~.
\end{align}\label{BCatrArB}
\end{subequations}
Consequently, this introduces an $\mathcal{O}(g^4)$ \textit{systematic error} in $f_{\rm sol}$, suggesting that our numerical solution is a good tracer of the true Bunch-Davies solution only in cases with $g\ll 1$. Despite the systematic error for Bunch-Davies solutions, we stress that $f_{\rm sol}$ is always some valid solution of the bootstrap equation, albeit with a non-Bunch-Davies initial condition for $g\gtrsim 1$. One can then scan the solution space by varying the boundary conditions \eqref{BCatrArB}, in hope of finding universal behaviours that are properties of the equation rather than the boundary/initial conditions. Interestingly, as we shall discuss below, we do find such universal behaviours near the IR resonance.

\paragraph{Regularisation} The validity of the numerical solution can be tested via convergence at large $N$ and $L$. Unfortunately the convergence of the above algorithm \eqref{formalFDMSolutionwithBCbeforeReg} appears bad for large $N$. This is due to the logarithmic divergence of the integral \eqref{rspaceBootstrapEq} at the threshold $r'=r$, i.e. around $q=0$ in \eqref{kspaceBootstrapEq}. A closer inspection shows that the oscillating factor $(r'-r)^{\pm i\omega}$ should automatically regulate the integral near the threshold as long as the function $F(r)$ is smooth. Alternatively, one can explicitly turn on a decaying factor $q^{\epsilon}$ as in \eqref{fracDerivativeFormula}. To regularise this spurious threshold divergence, we split the integral into two parts,
\begin{align}
    \int_r^\infty \d r'\, \mathcal{K}(r,r') F(r')=\int_r^{r+\delta} \d r'\, \mathcal{K}(r,r') F(r')+\int_{r+\delta}^\infty \d r'\, \mathcal{K}(r,r') F(r')~,
\end{align}
where $\delta\ll 1$ is a small gap isolating the threshold contribution. Assuming the smoothness of $F(r)$, we approximate the first term by
\begin{align}
    \nonumber\int_r^{r+\delta} \d r'\, \mathcal{K}(r,r') F(r')&\approx \int_r^{r+\delta} \d r'\, \mathcal{K}(r,r')\times F(r)\\
    &=-\frac{1}{2}g^2 \left(\mu^2+\frac{9}{4}\right)\left[-i \frac{e^{-\pi \omega/2}}{\Gamma(i\omega)} \frac{x_0^{i\omega}}{\omega}\left(e^{\delta }-1 \right)^{i \omega }e^{i \omega r}+(\omega\to -\omega)\right]\times F(r)~,
\end{align}
where the systematic error of the first step is suppressed by $\mathcal{O}(\delta)$ and in the second step, we have finished the integral of the kernel, regularising the oscillatory divergence at the threshold. Thus \eqref{NewtonCotesQuadratureFormOriginal} is replaced by
\begin{align}
    \nonumber\int_r^L dr' \mathcal{K}(r,r') F(r')\to\,& -\frac{1}{2}g^2 \left(\mu^2+\frac{9}{4}\right)\left[-i \frac{e^{-\pi \omega/2}}{\Gamma(i\omega)} \frac{x_0^{i\omega}}{\omega}\left(e^{\delta }-1 \right)^{i \omega }e^{i \omega r_i}+(\omega\to -\omega)\right]\times F(r_i)\\
    &+\frac{L}{2N} \mathcal{K}(r_i,r_{i_\delta}) F(r_{i_\delta}) +\sum_{j=i_{\delta}+1}^{N-1} \frac{L}{N} \mathcal{K}(r_i,r_{j}) F(r_{j})+\frac{L}{2N} \mathcal{K}(r_i,r_N)  F(r_N)~,
\end{align}
where $i_\delta$ is determined via $r_{i_\delta}=r_i+\delta$.\footnote{Without loss of generality, one can set $\delta$ to be an integer multiple of $L/N$.} The quadrature matrix $\mathcal{Q}$ becomes
\begin{align}
    \nonumber&\widehat{\mathcal{Q}}={\rm diag} \{\lambda_0,\lambda_1,\cdots,\lambda_N\}\\
    &+\frac{L}{N}\left(\begin{array}{ccccccccc}
		0 &\cdots & 0 &\frac{1}{2} \mathcal{K}_{0,0_\delta} & \mathcal{K}_{0,0_\delta+1} & \mathcal{K}_{0,0_\delta+2} & \cdots & \mathcal{K}_{0,N-1} & \frac{1}{2} \mathcal{K}_{0,N}\\
		0 & 0 &\cdots & 0 & \frac{1}{2} \mathcal{K}_{1,1_\delta} & \mathcal{K}_{1,1_\delta+1} & \cdots & \mathcal{K}_{1,N-1} & \frac{1}{2} \mathcal{K}_{1,N}\\
        0 & 0 & \cdots & 0 & 0 & \ddots & \cdots & \vdots & \vdots\\
        0 & 0 & \cdots & 0 & 0 & \cdots &  0 & \frac{1}{2}\mathcal{K}_{N(1-\delta/L)-1,N-1} & \frac{1}{2}\mathcal{K}_{N(1-\delta/L)-1,N}\\
        0 & 0 & \cdots & 0 & 0 & 0 & 0 & 0 & 0\\
        \vdots & \vdots & \cdots & \vdots & \vdots & \vdots & \vdots & \vdots & \vdots\\
        0 & 0 & \cdots & 0 & 0 & 0 & 0 & 0 & 0
	\end{array}\right)~,
\end{align}
where
\begin{align}
    \lambda_i\equiv -\frac{1}{2}g^2 \left(\mu^2+\frac{9}{4}\right)\left[-i \frac{e^{-\pi \omega/2}}{\Gamma(i\omega)} \frac{x_0^{i\omega}}{\omega}\left(e^{\delta }-1 \right)^{i \omega }e^{i \omega r_i}+(\omega\to -\omega)\right]~.
\end{align}
The numerical solution is thus obtained with the boundary conditions \eqref{BCatrArB} and by using the regularised quadrature matrix,
\begin{keyeqn}
    \begin{align}
    \mathbf{\widehat{F}}_{\rm sol}=\left[\mathcal{P}(\mathcal{D}-\widehat{\mathcal{Q}})+\mathcal{C}_{\rm b.c.}\right]^{-1}\left(\mathcal{P} \bm{\mathcal{S}}+\bm{\mathcal{S}}_{\rm b.c.}\right)~.\label{formalFDMSolutionwithBCbeforeReg}
\end{align}
\end{keyeqn}

\subsection{Results} 

We now present the results of our numerical bootstrap.

\begin{figure}[h!]
    \centering
    \includegraphics[width=\textwidth]{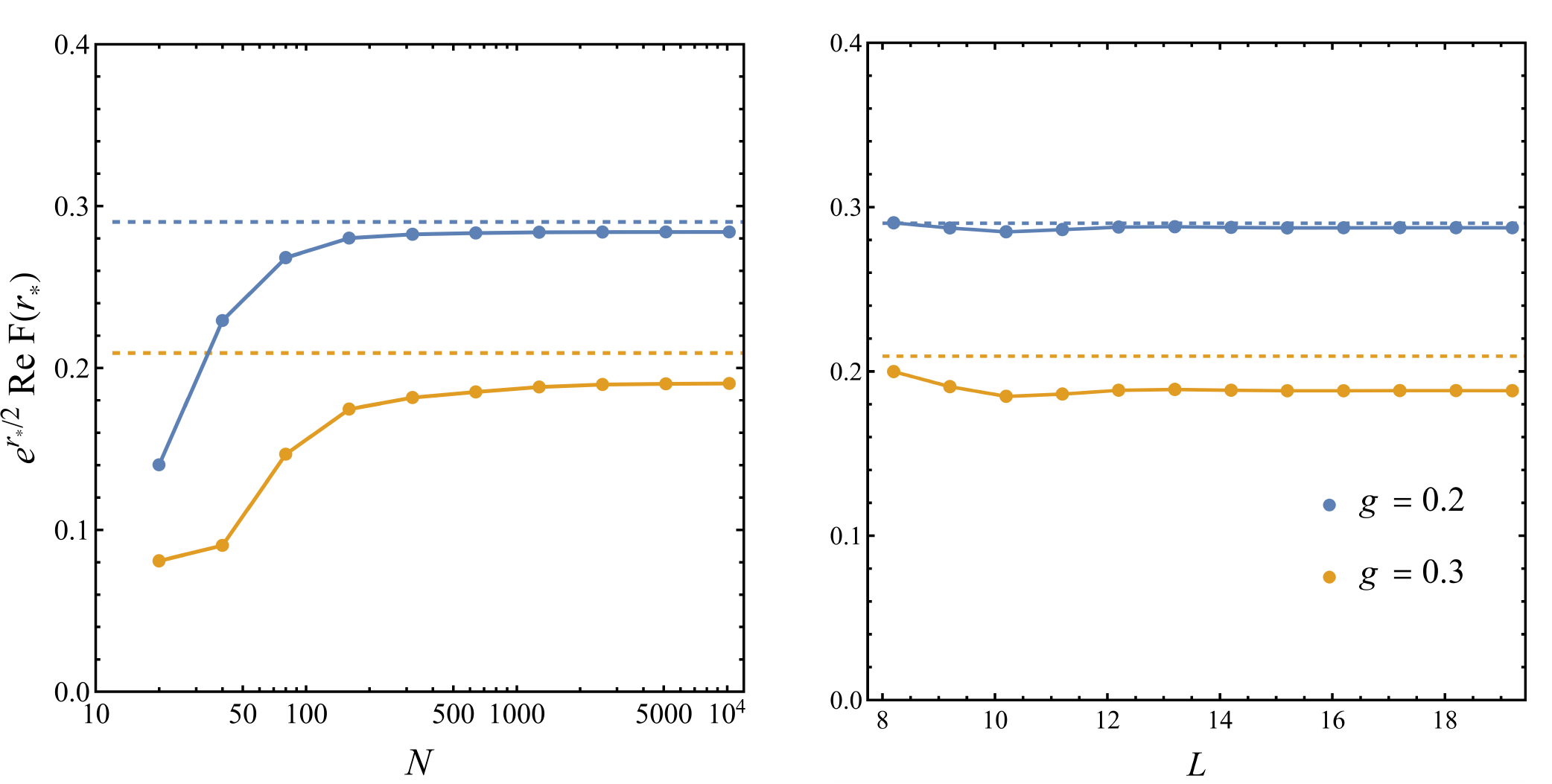}
    \caption{Convergence tests of the numerical bootstrap solution with respect to the mesh points $N$ (\textit{left panel}) and the cutoff $L=\log (k_{12}/s)_{\text{max}}$ (\textit{right panel}). The benchmark kinematics is chosen to be $r_*=\log (k_{12}/s)_*=8$ and the \textcolor{mmaBlue}{blue} and \textcolor{mmaYellow}{yellow} points correspond to $g=0.2$ and $g=0.3$, respectively. The dashed lines represent the prediction of the perturbative analytical solution. The other parameters are chosen as $\mu=1$, $\omega=1/3$, $x_0=-s\eta_0=1$. The boundary conditions are implemented at $r_{\rm I}=1$ and at $r_{\rm II}=1+2L/N$. The regularisation gap is chosen to be $\delta=10^{-1}$. In the \textit{left panel}, the kinematic cutoff is fixed to be $L=20$ and in the \textit{right panel}, the mesh points are chosen such that the step size $L/N=10^{-2}$ is fixed. We conclude that the numerical solution converges at large $N$ and $L$ towards the perturbative prediction, but with a small $\mathcal{O}(g^4)$ mismatch. The weak dependence on $L\gtrsim r_*$ also shows the quasi-locality of the bootstrap equation.}
    \label{FigCheckConvergenceConsistency}
\end{figure}

\paragraph{Convergence and consistency} To first check that our numerical algorithm converges to a result consistent with the perturbative expectation, we solve \eqref{formalFDMSolutionwithBCbeforeReg} with an increasing number of mesh points $N$ and kinematic cutoff $L$, and plot the result at a randomly chosen benchmark point ($r_*=8$ in our case) in Figure \ref{FigCheckConvergenceConsistency}. We observe that with a finer mesh resolution $N$, the numerical solution slowly converges towards the prediction of the perturbative solution (dashed lines in Figure \ref{FigCheckConvergenceConsistency}), yet not being quite the same. This is to be expected since our numerical bootstrap covers a non-perturbative resummation of higher-order effects in $g$ whereas the perturbative solution covers only up to $\mathcal{O}(g^2)$. There is therefore a natural mismatch at order $\mathcal{O}(g^4)$, which is confirmed from its increase with the coupling $g$ in Figure \ref{FigCheckConvergenceConsistency}. 

We also observe that the numerical solution at $r_*<L$ is not very sensitive to the kinematic cutoff $L$, as the result converges very quickly with an increasing cutoff $L$. This suggest that, surprisingly, although the bootstrap equation appears highly non-local as it involves an integral over the infinite kinematic domain $r_*<r<\infty$, its effective behaviour remains \textit{quasi-local}, meaning that the value of the solution at one point $F(r_*)$ only depends on its behaviour near that point approximately.\footnote{By contrast, at zero coupling $g=0$, the bootstrap equation is a differential equation and $F(r_*)$ solely depends on its properties exactly at $r=r_*$, i.e. its derivatives.}

\begin{figure}[h!]
    \centering
    \includegraphics[width=\textwidth]{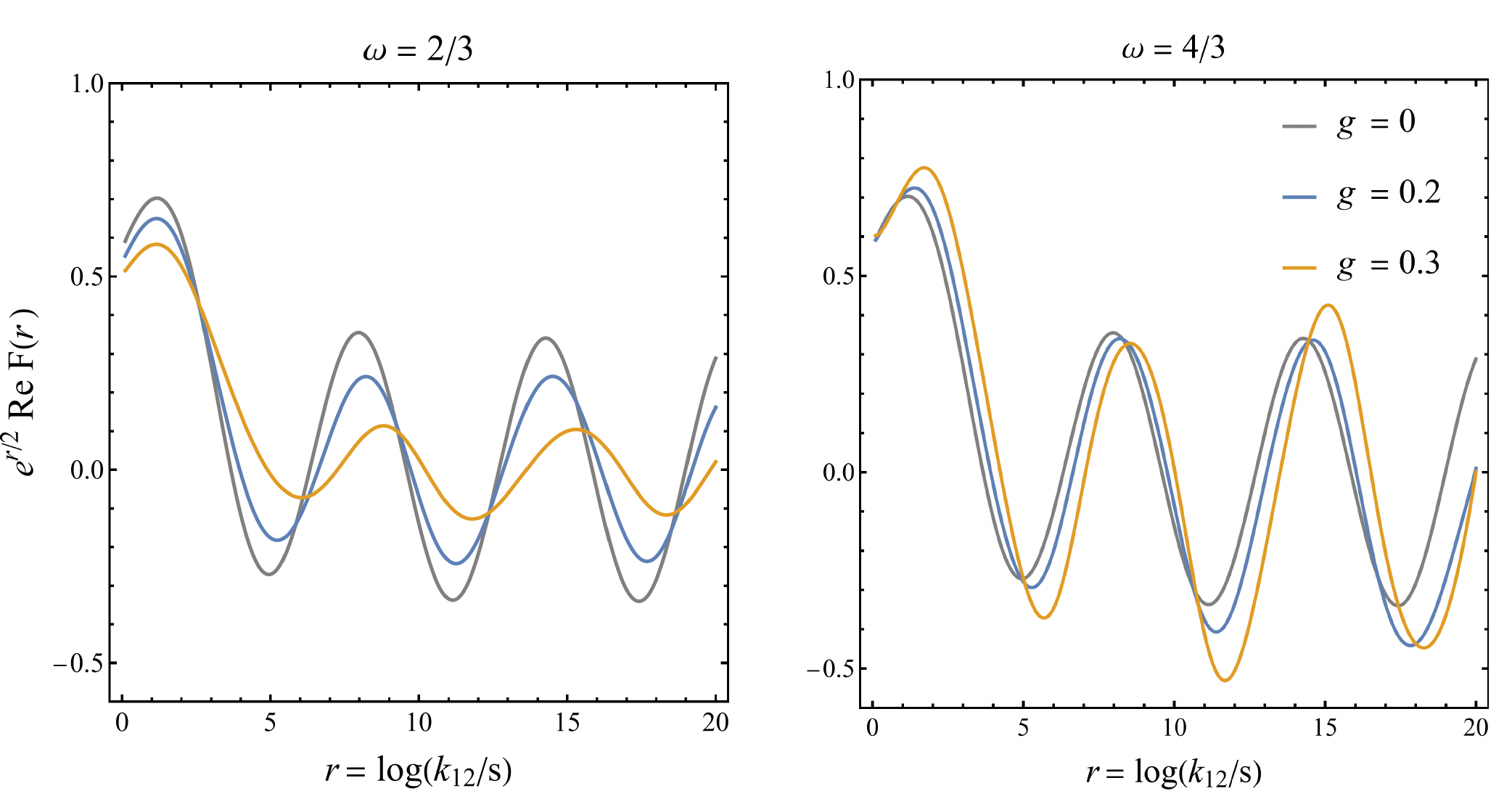}
    \caption{The dependence of the rescaled shape function on the momentum ratio $r=\log(k_{12}/s)$. The \textit{left} and \textit{right} panels correspond to frequency choices $\omega=2/3$ and $\omega=4/3$, respectively. The \textcolor{mmaGrey}{grey}, \textcolor{mmaBlue}{blue} and \textcolor{mmaYellow}{yellow} curves correspond to different sizes of the coupling $g$ as specified in the right panel. Other model parameters are chosen as $\mu=1$, $x_0=-s\eta_0=2$. The number of mesh point is $N=10^4$ with a kinematic cutoff $L=20$. The boundary conditions are implemented at $r_{\rm I}=1$ and at $r_{\rm II}=1+2L/N$ with a regularisation gap $\delta=10^{-1}$. }
    \label{FigNumericalSolution}
\end{figure}

\paragraph{Shape function off the resonances} We then plot in Figure \ref{FigNumericalSolution} the kinematical dependence of our numerical solution under different model parameter choices and fixed mesh configuration. We observe from the plot that the size of the cosmological collider signals is \textit{not} a monotonic function of the coupling $g$ and can reduce or enhance the signal strength depending on the mass-oscillation frequency $\omega$. This is in sharp contrast with the perturbative analysis as the signal always increases with the coupling at linear order in $g^2$. We also observe that the waveform of cosmological collider signals deviates from a simple sinusoidal function at larger couplings and frequencies.

\begin{figure}[h!]
    \centering
    \includegraphics[width=\textwidth]{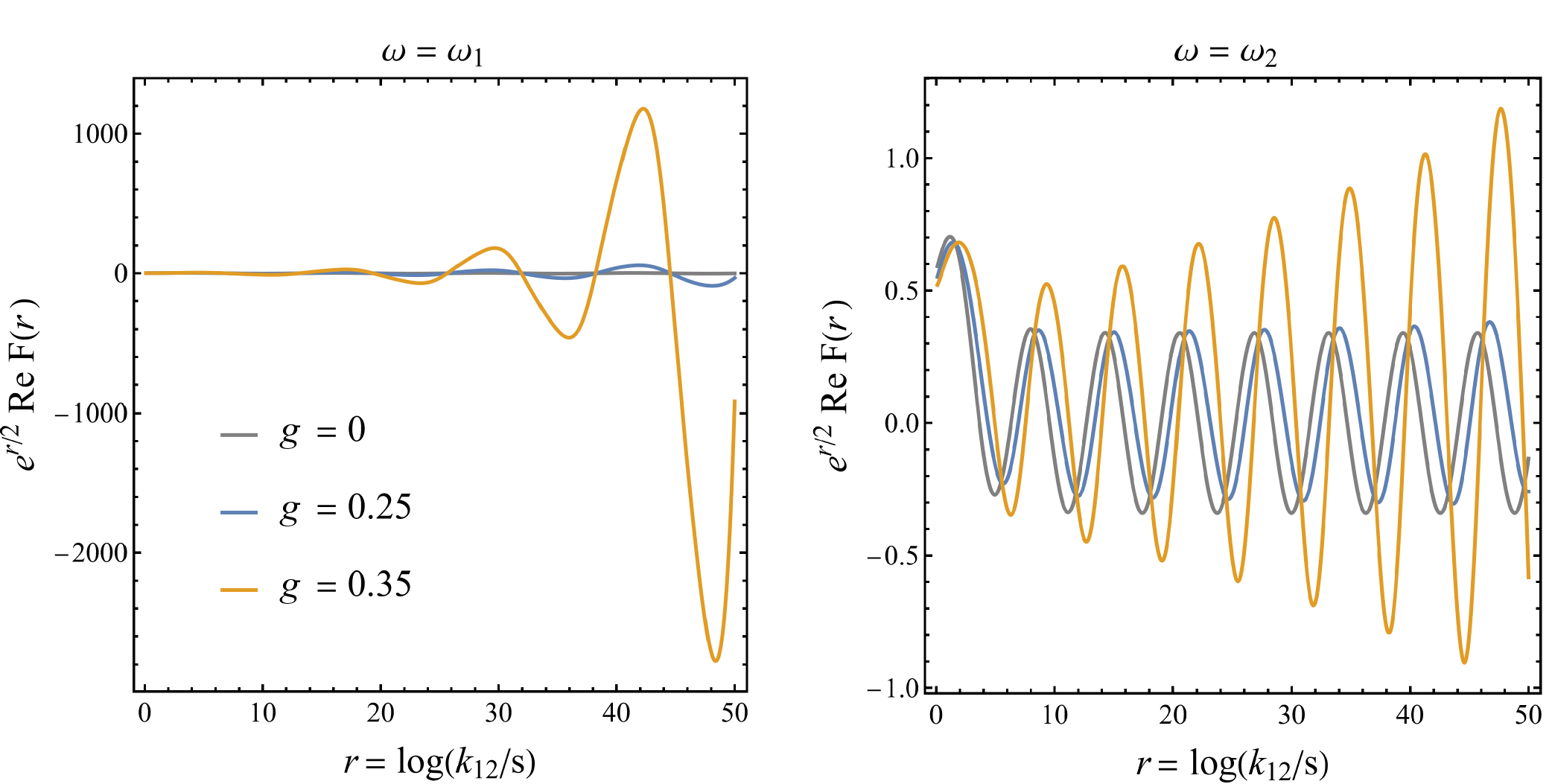}
    \caption{The rescaled shape function in the resonant regime. The \textit{left} and \textit{right} panels correspond to frequencies at the primary resonance $\omega_1\simeq 2\mu$ and the secondary resonance $\omega_2\simeq\mu$, respectively. The \textcolor{mmaGrey}{grey}, \textcolor{mmaBlue}{blue} and \textcolor{mmaYellow}{yellow} curves correspond to different sizes of the coupling $g$ as specified in the \textit{left} panel. To maintain $\omega\lesssim 1$ for algorithm stability, we choose different masses for these two resonances, i.e. $\mu=0.5$ for the primary resonance (\textit{left}) and $\mu=1$ for the secondary resonance (\textit{right}). We also choose $x_0=-s\eta_0=2$ and $N=16000$ with $L=50$. The boundary conditions are implemented at $r_{\rm I}=1$ and at $r_{\rm II}=1+2L/N$ with a regularisation gap $\delta=10^{-1}$. We observe that the cosmological collider signal oscillations are exponentially enhanced towards the squeezed limit thanks to the IR parametric resonance. The enhancement effect is most pronounced near the primary resonance and grows non-linearly with respect to the coupling strength $g$.}
    \label{figResonanceWaveform}
\end{figure}

\paragraph{Shape function at the resonances and the universality of scaling exponents} Due to the limitation of our knowledge on the precise boundary conditions, one might be led to conclude that our numerical bootstrap is not more useful than the very perturbative solution used to determine the boundary conditions, as there is a systematic $\mathcal{O}(g^4)$ error in numerics. However, we stress that this is not the case. Our numerical result is to be considered as a non-perturbative solution to the bootstrap equation that lies close to the genuine Bunch-Davies one in solution space. In other words, despite the $\mathcal{O}(g^4)$ systematic error, the numerical bootstrap automatically resums a subset of the infinite series in the $g^2$-expansion, which is e.g. crucial to capture the IR resonance effect discussed in Section \ref{IRresonanceSubSect}. In Figure \ref{figResonanceWaveform}, we plot the waveform in the presence of IR parametric resonances and find an exponentially growing enhancement to the cosmological collider signal in the squeezed limit. 

\begin{figure}[h!]
    \centering
    \includegraphics[width=\textwidth]{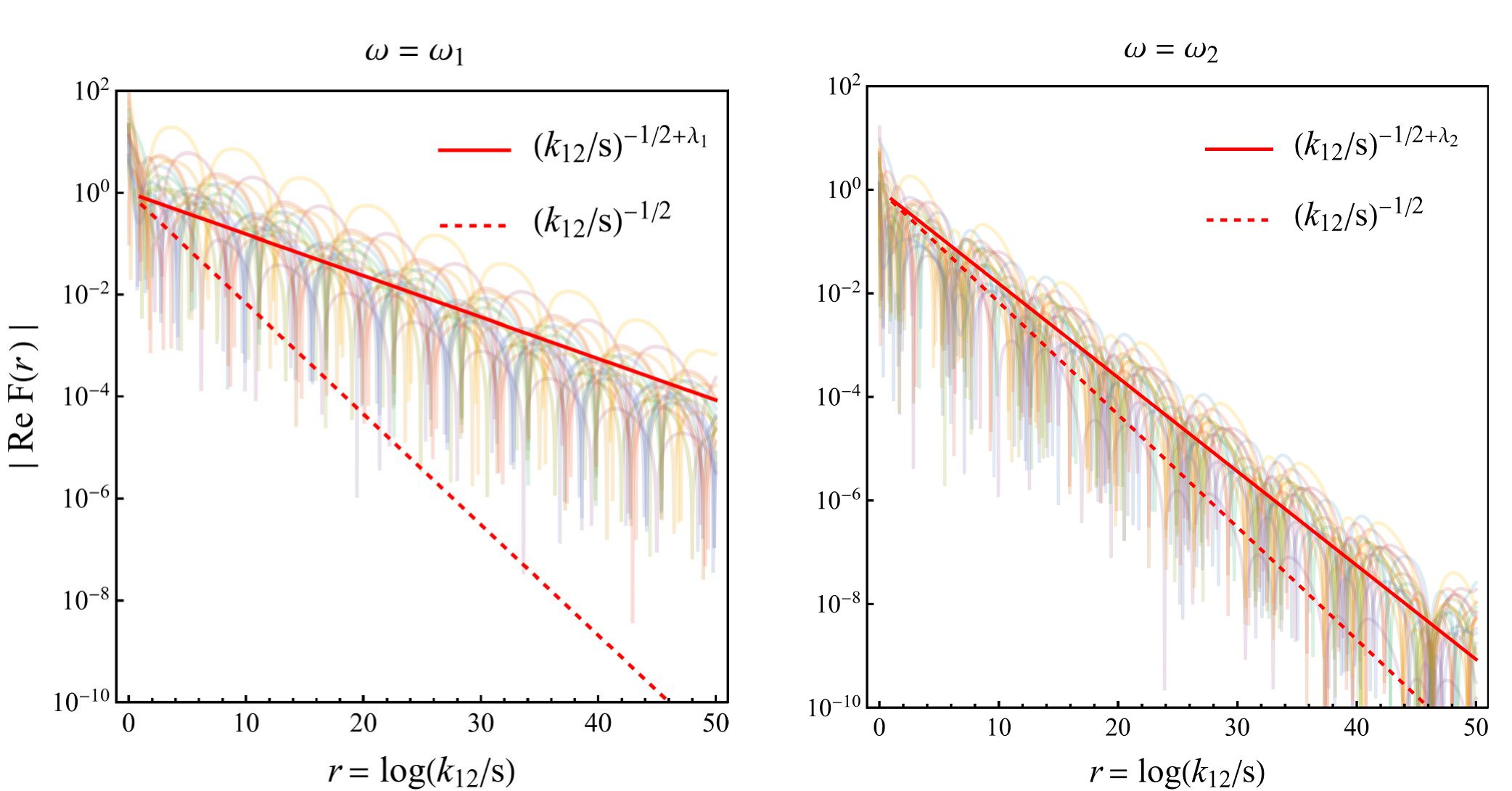}
    \caption{20 samples of numerical bootstrap with random boundary conditions near the primary (\textit{left}) and secondary (\textit{right}) resonances. The \textcolor{red}{red} solid lines are the prediction from boundary eigenfrequency analysis (see  \eqref{growthRateFromBoundaryEq} and \eqref{growthRateSecondary}) while the \textcolor{red}{red} dashed lines are the natural scaling in the absence of IR parametric resonances. Here the coupling constant is $g=0.5$ and we choose $\mu=0.5$ and $\mu=1$ for the primary and secondary resonances, respectively. Other parameters are chosen as $N=18000$, $L=50$ and $\delta=0.1$. We implement boundary conditions at $r_{\rm I}=1$ and $r_{\rm II}=1+2L/N$ by randomly deforming the $\mathcal{O}(g^2)$ perturbative solution as $F(r_{\rm I,II})=Z_{\rm I,II} F^{(g^2)}(r_{\rm I,II})$, where $Z_{\rm I,II}$ are complex numbers randomly sampled from a log-normal distribution $\exp(-\frac{1}{2}\log^2 |Z|)$. With $20$ samples of numerical bootstrap with random boundary conditions, we observe that the scaling exponents are indeed universal and match the analytical predictions $\lambda_1=\frac{g^2\mu}{4}\left(1+\frac{9}{4\mu^2}\right)$ and $\lambda_2=\frac{g^4\mu}{8} \left(1+\frac{9}{4\mu^2}\right)^2$. As expected, the primary resonance grows much faster than the secondary under the same choice of small coupling constant.}
    \label{figResonanceSampling}
\end{figure}

To test the universality of the scaling exponent and its independence on the boundary conditions, we explore the collective behaviour of solutions by randomly sampling boundary conditions in the vicinity of our perturbative Bunch-Davies solution. This can be achieved via deforming the boundary conditions $\mathcal{C}_{\rm b.c.}\mathbf{F}=Z \bm{\mathcal{S}}_{\rm b.c.}$ by a random complex matrix $Z$. We implement this by modifying $F(r_{\rm I})=Z_{\rm I} F^{(g^2)}(r_{\rm I})$ and $F(r_{\rm II})=Z_{\rm II} F^{(g^2)}(r_{\rm II})$ with $Z_{\rm I}, Z_{\rm II}$ two random complex numbers generated from a log-normal distribution $\exp(-\frac{1}{2}\log^2 |Z|)$. In Figure \ref{figResonanceSampling}, we plot the scaling behaviour of the cosmological collider signal and confirm that it is indeed independent of the boundary conditions and that it agrees with the prediction from the eigenfrequency analysis in Section \ref{IRresonanceSubSect},
\begin{subequations}
    \begin{align}
        \lambda_1&=\frac{g^2\mu}{4}\left(1+\frac{9}{4\mu^2}\right)~,\quad \text{at}\quad\omega_1\simeq 2\mu~,\\
        \lambda_2&=\frac{g^4\mu}{8} \left(1+\frac{9}{4\mu^2}\right)^2~,\quad  \text{at}\quad\omega_2 \simeq \mu\left[1-\frac{g^4}{12}\left(1+\frac{9}{4\mu^2}\right)^2\right]~.
    \end{align}
\end{subequations}

\paragraph{Limitations and potential optimisations} Note that a drawback of our boundary numerical approach is the limitation to small frequencies i.e. $\omega\lesssim 1$. This is because the kernel $\mathcal{K}(r,r')$ contains a piece that grows exponentially with $\omega$ (see \eqref{KernelFormInrSpace}),
\begin{align}
    \mathcal{K}(r,r')\supset \frac{e^{\pi\omega/2}}{\Gamma(-i\omega)}\sim e^{\pi \omega}~,\quad \omega\gg 1~,
\end{align}
leading to the superficial vanishing of the numerical solution when inverted using \eqref{formalFDMSolutionwithBCbeforeReg}. However, anticipating $F(r')\propto e^{\pm i\mu r'}$ in the squeezed limit, we expect the integral near the threshold to behave as
\begin{align}
    \int_r^\infty \d r' \mathcal{K}(r,r')F(r')\sim  \frac{e^{\pi\omega/2}}{\Gamma(-i\omega)} x_0^{-i\omega} \int_r^\infty \d r' \frac{(r'-r)^{-i\omega}}{r'-r}e^{\pm i\mu r'}+\cdots~,
\end{align}
where the saddle point of the $e^{-i\mu r'}$ component brings an extra suppression factor $e^{-\pi\omega}$, cancelling the large prefactor and rendering the result regular. The $e^{+i\mu r'}$ component, on the other hand, does not receive suppression from the saddle point and must therefore itself be suppressed in order to maintain regularity. As a result, we expect the shape function to be dominated by $F\supset e^{-i\mu r}$ at large frequencies $\omega\gg 1$. However, at the level of numerics within the real kinematic domain $r\in \mathbb{R}_+$, the information of saddle points is concealed and the large prefactor remains, causing the divergence problem. Consequently, we expect that to resolve this issue, one would need to perform numerics in the complex kinematic domain $r\in \mathbb{C}$, which we leave for future exploration.

Another limitation lies in the efficiency of numerics, in particular, the large memory consumption. This is due to the dense nature of the quadrature matrix $\widehat{\mathcal{Q}}$, with its memory consumption rising up as $N^2$ at large $N$.\footnote{As an instance, $N=10^4$ produces a quadrature matrix $\widehat{\mathcal{Q}}$ of size $1.6\,$GB in {\sf Mathematica}. By contrast, the differentiation matrix $\mathcal{D}$ is sparse and takes up only $1.0\,$MB.} To optimise the memory consumption and accelerate the convergence, one might consider adaptative mesh refinement with a recursive algorithm.

\section{Summary and outlooks}\label{conclusionSection}

The boundary approach to cosmological correlators provides new perspectives on the dynamics of the underlying bulk evolution, often in the form of differential equations in kinematics space. However, in this work, we have shown that in the absence of exact scale invariance a.k.a dilation, the boundary bootstrap equation can become integro-differential equations, which serve as non-local constraints in kinematic space that link one momentum configuration to another. More specifically, motivated by axion-monodromy-like inflation models, we considered turning on an oscillatory time dependence in the EFT of inflation involving a massive scalar field. Upon coupling to the Goldstone boson of time-diffeomorphism breaking i.e. the inflaton, these oscillations source non-trivial corrections to the power spectrum as well as the non-Gaussian bispectrum and trispectrum of curvature perturbations. The seed integral of these observables are shown to satisfy a set of integro-differential equations that generalise the conventional bootstrap equations to cases beyond scale invariance.

We then focused on the simplest scenario where the mass of the heavy field is the only source of oscillation, reducing the integration kernel of the bootstrap equation to a monochromatic function of kinematics. We were able to find analytical solutions to the integro-differential bootstrap equation perturbatively at order $\mathcal{O}(g^2)$ from a purely boundary perspective. We also implemented the first numerical approach to cosmological bootstrap beyond scale invariance, where we find consistency between the numerics and perturbative analytical solutions. Using the numerical bootstrap, we were able to non-perturbatively verify the universality of soft-limit scaling exponents near the parametric resonances and found them to match the prediction of eigenfrequency analysis on the boundary.

This work also leads to various avenues for future exploration:

\begin{itemize}
    \item \textbf{Towards better analytical solutions:} The non-local nature of the integro-differential bootstrap equation naturally implies the complexity of its solutions. Indeed, the mere fact that we are allowed to analytically solve it to the first non-trivial order in perturbation theory is already surprising, and the result turns out to be extremely complicated. This suggests that going up to higher orders in the coupling $g$ is likely no longer illuminating. Therefore, it would be interesting to investigate the bootstrap equation from an alternative perspective (perhaps with a different perturbative expansion parameter) and try to find a better way to organise the analytical solutions, if there exists an exact analytical solution at all.
    
    \item \textbf{Dispersive bootstrap:} As we have seen in Section \ref{Three-point exchange diagram}, the factorisation property of the exchange diagram imposes highly non-trivial constraints on its analytic structure. Along the lines of \cite{Meltzer:2021zin, Jazayeri:2021fvk}, this suggests an alternative bootstrap approach, in which the exchange graph---without invoking the inhomogeneous IDE---is computed directly from the knowledge of the three-point function $f$, using an appropriate dispersive representation in the $s^2$ complex plane. By following this method, it might be possible to obtain alternative analytic expressions for the exchange diagram with better convergence properties near the equilateral limit $U=1$ than our series expansion formulas.
    
    \item \textbf{One-loop diagrams:} In this paper, we concentrated on tree-level contribution from the heavy sector to the inflationary correlators. As we alluded to earlier, from the mass term $m^2(t+\pi)\sigma^2$ alone, several contributions to the power spectrum and higher order correlation functions arise at one-loop order, see e.g. \cite{Flauger:2016idt}. It would be interesting to study these loop contributions in more detail, leveraging similar bootstrap techniques such as cutting rules and dispersive representations.
    
    \item \textbf{Optimisation of numerics:} As discussed in Section \ref{numericalBootstrapSection}, our finite difference method has been proven useful when the oscillation frequency is below the Hubble scale i.e. $\omega\lesssim H$, but becomes unstable at large frequencies. One might be able to improve the stability of the algorithm by extending the kinematics to complex domain. It is also interesting to adopt adaptative mesh refinement to optimise the convergence rate and the memory consumption in the future.

    \item \textbf{Modulated external fields:} In this work, we limited ourselves to the simplest case where the mass of the heavy field is the sole source of explicit time dependence. However, in general, oscillations in one parameter of a theory are naturally accompanied by those in the rest of parameters (unless they are forbidden by symmetries). In axion monodromy inflation, for instance, the external inflatons also acquire oscillations via the kinetic term, which are conventionally treated perturbatively (see a recent endeavour to capture this effect non-perturbatively in \cite{Creminelli:2025tae}). It is therefore an interesting question if we can incorporate the oscillation effects of external fields non-perturbatively in the boundary bootstrap equation, as we did for the internal massive field here.

    \item \textbf{General time dependence:} As suggested in \eqref{IDEs}, our bootstrap equation is applicable to general time dependence other than simple oscillations. It would be interesting to explore couplings $g_\omega$ with a non-trivial support in the frequency domain, as expected from inflationary models with features such as steps or sharp turns (see e.g. \cite{Chen:2016vvw} for a review).
\end{itemize}

\paragraph{Acknowledgments} We would like to thank Ali Akbar Abolhasani, Sebasti{\'a}n C{\'e}spedes, Xingang Chen, Siyao Li, Zhehan Qin, Sébastien Renaux-Petel, Guilherme L. Pimentel, Arthur Poisson, David Stefanyszyn, Denis Werth, Dong-Gang Wang, Zhong-Zhi Xianyu and Bowei Zhang for stimulating discussions. We thank Xingang Chen, Guilherme L. Pimentel, Lucas Pinol and Sébastien Renaux-Petel for comments on the manuscript. 
SJ would like to thank the organisers of the workshop ``Constraining Effective Field Theories without Lorentz" (July 2025 at IFPU, Trieste) for their hospitality, when part of this work was under development, and the participants for lively discussions. S.J. is supported
by the STFC Consolidated Grant ST/X000575/1 and a
Simons Investigator Award 690508.
X.T. is supported by STFC consolidated grants ST/T000694/1 and ST/X000664/1.
Y.Z. is supported by the IBS under the project code, IBS-R018-D3.

\appendix

\section{An infinite set of coupled bootstrap equations}
\label{vertexbootstrap}

In this appendix, as an alternative to the integro-differential equations (IDEs) discussed in the main text, we establish a bootstrap framework entirely based on ordinary differential equations (ODEs). This framework will be specifically designed for computing the exchange diagram with the time-dependent intermediate mass $m^2(t)=m_0^2(1+g^2\cos(\omega (t-t_0))$. 

Starting with \eqref{bootstrapset}, recall that we landed on a set of integro-differential equations in \eqref{IDEs}, rather than ODEs, because we substituted an integral relation between $J_{\aa\bb}$ in \eqref{Jppmm} and the Schwinger-Keldysh components $F_{\aa\bb}$. 
To avoid this non-local form, we will introduce an infinite set of auxiliary correlators $F^{(n,l)}(k_{12},s)$, labelled by two integers $n,l$. We define the element $F^{(n,l)}$ in this series by the same exchange diagram and intermediate mass as before, except for time-dependent left/right vertices that oscillate as
\begin{align}
    \lambda^{(n)}_L(\eta)=(\eta/\eta_0)^{-in\omega}\,,\quad \lambda^{(l)}_R(\eta)=(\eta/\eta_0)^{-il\omega}\,,\qquad \text{with}\qquad n,l\in \mathbb{Z}\,.
\end{align}
We will shortly prove these correlators satisfy a recursive set of \textit{ordinary} differential equations. So, our non-local description for correlators, from this new perspective, is traded for an infinite set of ODEs involving an infinite series of exchange diagrams. Among these, only the $n=l=0$ element is of real interest, while the remaining components are auxiliary diagrams introduced only to form a closed system of equations.  

We begin by the observation that each correlator $F^{(n,l)}(k_{12},s)$ satisfies a local differential equation like \eqref{bootstrapset} in which the source $J^{n,l}_{\aa\bb}(k_{12},k_{34})$ can be locally specified in terms of $F_{\aa\bb}^{(n\pm 1,l)}(k_{12},k_{34})$. Inserting this relation into the original equation \eqref{bootstrapset}, one finds a recursive set of ordinary differential equations for  $F^{(n,l)}(k_{12},s)$: 
\begin{align}
    \hat{{\cal O}}^{(\textcolor{red}{n},\textcolor{blue}{l})}_{12}F^{(\textcolor{red}{n},\textcolor{blue}{l})}_{++}&=\Gamma\left[1-i(\textcolor{red}{n}+\textcolor{blue}{l})\omega\right]\,k_T^{-1}(-k_T\eta_0)^{+i(\textcolor{red}{n}+\textcolor{blue}{l})\omega}\,c_{++}^{\textcolor{red}{n},\textcolor{blue}{l}}-\dfrac{1}{2}\,g^2m_0^2\left(F^{(\textcolor{red}{n-1},\textcolor{blue}{l})}_{++}+F^{(\textcolor{red}{n}+1,\textcolor{blue}{l})}_{++}\right)\,,\\
    \hat{{\cal O}}^{(\textcolor{red}{n},\textcolor{blue}{l})}_{12}F^{(\textcolor{red}{n},\textcolor{blue}{l})}_{+-}&=-\dfrac{1}{2}\,g^2m_0^2\left(F^{(\textcolor{red}{n-1},\textcolor{blue}{l})}_{+-}+F^{(\textcolor{red}{n+1},\textcolor{blue}{l})}_{+-}\right)\,,\\
     \hat{{\cal O}}^{(\textcolor{red}{n},\textcolor{blue}{l})}_{34}F^{(\textcolor{red}{n},\textcolor{blue}{l})}_{++}&=\Gamma\left[1-i(\textcolor{red}{n}+\textcolor{blue}{l})\omega\right]\,k_T^{-1}(-k_T\eta_0)^{+i(\textcolor{red}{n}+\textcolor{blue}{l})\omega}\,c_{++}^{\textcolor{red}{n},\textcolor{blue}{l}}-\dfrac{1}{2}\,g^2m_0^2\left(F^{(\textcolor{red}{n},\textcolor{blue}{l-1})}_{++}+F^{(\textcolor{red}{n},\textcolor{blue}{l+1})}_{++}\right)\,,\\
    \hat{{\cal O}}^{(\textcolor{red}{n},\textcolor{blue}{l})}_{34}F^{(\textcolor{red}{n},\textcolor{blue}{l})}_{+-}&=-\dfrac{1}{2}\,g^2m_0^2\left(F^{(\textcolor{red}{n},\textcolor{blue}{l-1})}_{+-}+F^{(\textcolor{red}{n},\textcolor{blue}{l+1})}_{+-}\right)\,,
\end{align}
where the derivative operators ${\cal O}_{12,34}^{(n,l)}$ are defined by
\begin{align}
\nn
   \hat{{\cal O}}^{(n,l)}_{12}&=(k_{12}^2-s^2)\partial^2_{k_{12}}+2(1-in\omega)k_{12}\partial_{k_{12}}+\left(m_0^2-2-in\omega-n^2\omega^2\right)\,,\\ 
     \hat{{\cal O}}_{34}^{(n,l)}&=(k_{34}^2-s^2)\partial^2_{k_{34}}+2(1-il\,\omega)k_{34}\partial_{k_{34}}+\left(m_0^2-2-il\omega-l^2\omega^2\right)\,,
\end{align}
and 
\begin{align}
    c^{n,l}_{++}=\exp(-\pi\,(n+l)\omega/2)\,.
\end{align}
Indeed, $F^{(0,0)}$, which was computed via the IDEs in the main text, could be re-derived using the above bootstrap equations. At linear order in $g^2$, this requires us to replace \eqref{eq: F0g0}--\eqref{DFpm1} with
\begin{align}
\label{Fpp00}
    \hat{{\cal O}}^{(0,0)}_{12}F^{(0,0)}_{++}&=\,k_T^{-1}-\dfrac{1}{2}\,g^2m_0^2\left(F^{(-1,0)}_{++}+F^{(+1,0)}_{++}\right)\,.\\ 
    \hat{{\cal O}}^{(-1,0)}_{12}F^{(-1,0)}_{++}&=\hat{{\cal O}}^{(+1,0)}_{12}F^{(+1,0)}_{++}={\cal O}(g^2)\,.
\end{align}
Up to ${\cal O}(g^4)$ corrections to $F_{++}^{(0,0)}$, the last two equations can be solved by ignoring the would-be sources on the right-hand side. Plugging the solutions for $F^{(\pm 1,0)}_{++}$ back into the first line, and after decomposing $F^{(0,0)}_{++}$ as \eqref{Fexchange}, it is straightforward to show that the right-hand sides of \eqref{Fpp00} and \eqref{DFpm1} will match, consistent with the fact that the two sets of equations describe identical correlators. 

\section{Detailed derivations for solving the boundary IDE} \label{sec: derivations}

\subsection{Three-point contact diagram}

In this section, we spell out the technical details for finding the $\mathcal{O}(g^2)$ three-point contact function $f_{\pm}(u)$, which satisfies the boundary IDE
\begin{align}
    \hat{\Delta}_u\,f_{\pm 1}(u)=\int_0^\infty dx\,K_\pm(x)f_0\left(\dfrac{u}{1+u\,x}\right)\,.\label{eq:3ptContact_IDE}
\end{align}
Note that this three-point function can differ by an arbitrary constant phase depending on the convention of mode functions. Here we set the convention that the zeroth-order mode function of the massive field as
\begin{align}
    \sigma_{0-}(s,\eta)\equiv \frac{\sqrt{\pi}}{2}e^{\frac{\pi\mu}{2}}e^{-i\frac{\pi}{4}}(-\eta)^{\frac{3}{2}}H^{(2)}_{i\mu}(-s\eta)\,,
\end{align}
then the scale invariant component of the three-point function $f_0(u)$ is given by \eqref{eq: f0}.

\paragraph{Expanding the source} As the first step, we expand $f_0(u)$ expressed using Hypergeometric functions into an infinite series,
\begin{align}
    f_0(u)=\sum_{n=0}^{\infty} \sqrt{\frac{\pi}{2}}\text{csch}(\pi\mu)\,\left(\frac{u}{2}\right)^{\frac{1}{2}+2n-i\mu}\,\Gamma\,\left[\bgm \frac{1}{2}+2n-i\mu
    \\1+n,1+n-i\mu\edm\right]+(\mu\to-\mu)\,,
\end{align}
then the integration on the right-hand side of \eqref{eq:3ptContact_IDE} can be easily performed using \eqref{eq: momentum_int_identity}, yielding the following result,
\begin{align}
    &\int_0^\infty dx\,K_\pm(x)f_0\left(\dfrac{u}{1+u\,x}\right)\nonumber\\
    &=-\sum_{n=0}^{\infty}\frac{g^2m_0^2}{2}\sqrt{\frac{\pi}{2}}\,\csch(\pi\mu)\,2^{\mp i\omega}\,\left(\frac{u}{2}\right)^{\frac{1}{2}+2n+i\mu\mp i\omega}\times\Gamma\,\left[\bgm \frac{1}{2}+2n-i\mu\mp i\omega
    \\1+n,1+n-i\mu\edm\right]+(\mu\to-\mu)\,.
\end{align}
In this way, we have reduced the original IDE into a differential equation with sources.

\paragraph{Ansatz and recursive relations} Based on the form of the source term, we make the following ansatz,
\begin{align}
    f^{\text{ansatz}}_{\pm 1}(u)=\sum_{k,n=0}^\infty d^\pm_{k,n}(\mu,\omega)\,u^{\frac{1}{2}+2k+2n}u^{-i\mu\mp i\omega}+(\mu\leftrightarrow -\mu)\,,\label{eq: fpm_ansatz}
\end{align}
and plug it into \eqref{eq:3ptContact_IDE} to obtain the recurrence relation of coefficients as
\begin{align}
    d^{\pm}_{k,n+1}=\frac{(\frac{1}{4}+k+n-\frac{i\mu}{2}\mp\frac{i\omega}{2})(\frac{3}{4}+k+n-\frac{i\mu}{2}\mp\frac{i\omega}{2})}{(1+k+n\mp \frac{i\omega}{2})(1+k+n-i\mu\mp \frac{i\omega}{2})}\times d^{\pm}_{k,n}\,,
\end{align}
with the initial value obtained from matching the source terms:
\begin{align}
    d^{\pm}_{k,0}=-g^2m_0^2\sqrt{\pi}\csch(\pi\mu)\,\frac{2^{-4-2k+i\mu}\,e^{\mp{\frac{\pi\omega}{2}}}}{(k\mp \frac{i\omega}{2})(k-i\mu\mp \frac{i\omega}{2})}\,\Gamma\,\left[\bgm \frac{1}{2}+2k-i\mu\mp i\omega
    \\1+k,1+k-i\mu\edm\right]\,.
\end{align}
This recurrence relation can be easily solved, and the general expression reads
\begin{align}
    d^{\pm}_{k,n}=-\frac{g^2m_0^2\sqrt{\pi}\,e^{\mp \frac{\pi\omega}{2}}}{16\sinh\pi\mu}\,{2^{-2k-2n+i\mu}}\,\Gamma\,\left[\bgm k\mp \frac{i\omega}{2}, \frac{1}{2}+2k+2n-i\mu\mp i\omega,k-i\mu\mp \frac{i\omega}{2}
    \\1+k,1+k-i\mu,1+k+n\mp \frac{i\omega}{2},1+k+n-i\mu\mp \frac{i\omega}{2}\edm\right]\,.
\end{align}
Plugging this into the ansatz \eqref{eq: fpm_ansatz}, we obtain a solution involving two layers of infinite series, one of which can be summed explicitly, yielding the following particular solution:
\begin{align}
    \nn
        f^{\text{ansatz}}_{\pm 1}(u)&=g^2m_0^2\sum_{k=0}^{\infty}b_k^\pm (\mu,\omega)\,u^{\frac{1}{2}+2k-i\mu\mp i\omega}\pregFq{3}{2}{1,\tfrac{1}{4}+k-\tfrac{i\mu}{2}\mp \tfrac{i\omega}{2},\tfrac{3}{4}+k-\tfrac{i\mu}{2}\mp\tfrac{ i\omega}{2}}{1+k\mp \tfrac{ i\omega}{2},1+k- i\mu\mp \tfrac{ i\omega}{2}}{u^2}\\ 
        \label{fansatz2}
        &+(\mu\to -\mu)\,,
\end{align}
where the coefficient is given by
\begin{align}
    b_k^{\pm}(\mu,\omega)=&-\frac{\sqrt{\pi}}{16}\,e^{\mp\frac{\pi\omega}{2}}\text{csch}(\pi \mu)\,2^{-2k+i\mu}\nonumber\\
    &\times \Gamma\left[\bgm k\mp \frac{i\omega}{2},k- i\mu\mp \frac{i\omega}{2},\frac{1}{2}+2k- i\mu\mp i\omega\\
    1+k,1+k- i\mu\edm\right]\,.
\end{align}

\paragraph{Adding the homogenous solution} To get the physical solution consistent with the Bunch-Davies vacuum choice that is free from any folded singularities, we need to add additional pieces that satisfy the homogenous equation
\begin{align}
    f^{\text{hom}}_{\pm 1}(u)= \chi^\pm_1\cdot\left(\frac{u}{2}\right)^{\frac{1}{2}+  i\mu}\pFq{2}{1}{\tfrac{1}{4}+\tfrac{ i\mu}{2}, \tfrac{3}{4}+\tfrac{ i\mu}{2}}{1+ i \mu}{u^2}+\chi^\pm_2\cdot\left(\frac{u}{2}\right)^{\frac{1}{2}-  i\mu}\pFq{2}{1}{\tfrac{1}{4}-\tfrac{ i\mu}{2}, \tfrac{3}{4}-\tfrac{ i\mu}{2}}{1- i \mu}{u^2}\,.
\end{align}
To fix these undetermined parameters, we impose two conditions: (i) The cancellation of folded poles and (ii) finiteness under the total energy limit at $\mathcal{O}(g^2)$ as shown in  \eqref{eq: TotalEnergyLimit}.

\paragraph{Folded limit} In the folded limit $u\rightarrow 1$ or equivalently $k_{12}\to s$, the particular solution $f_{\pm1}^{\text{ansatz}}$ exhibits logarithmic divergence:
\begin{align}
    \lim_{u\to 1} f^{\text{ansatz}}_{\pm1}(u)
    &=\frac{g^2m_0^2}{16\sqrt{2\pi}}\csch(\pi\mu)\,e^{\mp\frac{\pi\omega}{2}}\,\Gamma\left[\bgm \mp \frac{i\omega}{2},\frac{1}{2}\pm \frac{i\omega}{2},  - i\mu\mp \frac{i\omega}{2}\\
    1- i\mu\pm\frac{i\omega}{2}\edm\right]\,\textcolor{red3}{\log(1-u)}\nonumber\\
     &+(\mu\to -\mu)\,,
\end{align}
where we have resummed the infinite series after expanding Hypergeometric function around $u=1$. Similarly, the singularity of the homogenous solution in the folded limit is given by
\begin{align}
    \lim_{u\to 1} f^{\text{hom}}_{\pm1}(u)=-\Bigg\{\chi_1^{\pm}\,2^{-\frac{1}{2}+i\mu}\,\Gamma\left[\bgm 1-i\mu\\
    \frac{1}{4}-\frac{i\mu}{2},\frac{3}{4}-\frac{i\mu}{2}\edm\right]+\chi_2^{\pm}\,2^{-\frac{1}{2}-i\mu}\,\Gamma\left[\bgm 1+i\mu\\
    \frac{1}{4}+\frac{i\mu}{2},\frac{3}{4}+\frac{i\mu}{2}\edm\right]\Bigg\}\textcolor{red3}{\log(1-u)}\,,
\end{align}
then we get the first constraint by requiring 
\begin{align}
    \lim_{u\to 1}\Big[ f^{\text{hom}}_{\pm1}(u)+f^{\text{ansatz}}_{\pm1}(u)\Big]=\text{finite}\,,
\end{align}
i.e. the logarithmic divergences in the above equations should cancel each other out.

\paragraph{Total-energy limit} Under this limit, $u\rightarrow -1$ or equivalently $k_{12}+s\to 0$, and the particular solution $f_{\pm1}^{\text{ansatz}}$ has a divergence,
\begin{align}
    \lim_{k_{12}+s\to 0} f^{\text{ansatz}}_{\pm1}
    &=\frac{ig^2m_0^2}{16\sqrt{2\pi}}\left(1+\coth \pi\mu\right)\,e^{\pm\frac{\pi\omega}{2}}\,\Gamma\left[\bgm \mp \frac{i\omega}{2},\frac{1}{2}\pm \frac{i\omega}{2},  - i\mu\mp \frac{i\omega}{2}\\
    1- i\mu\pm\frac{i\omega}{2}\edm\right]\,\textcolor{blue3}{\log(1+u)}\nonumber\\
     &+(\mu\to -\mu)\,.
\end{align}
The divergence in the homogenous part is 
\begin{align}
    \lim_{k_{12}+s\to0} f^{\text{hom}}_{\pm1}
    &=\frac{\mu\cosh(\pi\mu)}{2\pi^{3/2}}\left(\chi_2^{\pm}e^{-\pi\mu}\,\Gamma\left[\frac{1}{2}-i\mu,i\mu\right]-\chi_1^{\pm}e^{\pi\mu}\,\Gamma\left[\frac{1}{2}+i\mu,-i\mu\right]\right)\textcolor{blue3}{\log(1+u)}\,.
\end{align}
The second constraint is thus from the cancellation of $\textcolor{blue3}{\log(1+u)}$, such that 
\begin{align}
    \lim_{k_{12}+s\to 0}\Big[ f^{\text{hom}}_{\pm1}(u)+f^{\text{ansatz}}_{\pm1}(u)\Big]=\text{finite}\,.
\end{align}

\paragraph{Final expression} Combine those two constraints, we finally arrive at the solution:
\begin{align}
    f_{\pm1}(u)=f^{\text{hom}}_{\pm1}(u)+f^{\text{ansatz}}_{\pm1}(u)\,,
\end{align}
with the coefficients in the homogenous solution given by 
\begin{align}
    \chi_1^{\pm}(\mu,\omega)=\chi_2^{\pm}(-\mu,\omega)=\frac{g^2m_0^2\pi}{4\sqrt{2}}\,\frac{1-\tanh(\pi\mu)}{e^{\pm\pi\omega}-e^{-2\pi\mu}}\,\Gamma\left[\bgm  -i\mu,\,\frac 1 2\pm\frac{i\omega}{2},\,\mp \frac{i\omega}{2}\\
    \frac{1}{2}-i\mu,\,1- i\mu\pm \frac{i\omega}{2},\,1+i\mu\pm \frac{i\omega}{2}\edm\right].
\end{align}

\subsection{Three-point exchange diagram}

In this subsection, we present the detailed derivation of the scale-breaking three-point exchange diagram $F_{\pm1}$ that satisfies the IDE \eqref{DFpm1}. The procedure largely parallels that of the simpler example discussed in the previous subsection.

\paragraph{Expanding the source} Firstly, we expand the closed form of $F_0(U)$ as an infinite series:
\begin{align}
    F_0(U)=&\sum_{k=0}^{\infty}\frac{\pi}{2 \cosh \pi\mu}\,\Gamma\left[\bgm  1+k,\,1+k\\
    \frac{3}{2}+k+i\mu,\,\frac{3}{2}+k-i\mu\edm\right]\times U^{1+k}\nonumber\\
    &+\left\{\frac{\pi\,e^{-\pi\mu}}{2 \sinh (2\pi\mu)}\,\Gamma\left[\bgm  \frac{1}{2}+k-i\mu,\,\frac{1}{2}+k-i\mu\\
    1+k,\,1+k-2i\mu\edm\right]\times U^{\frac{1}{2}+k-i\mu}+(\mu\to-\mu)\right\}\,,
\end{align}
where the first line is the EFT contribution and the second line corresponds to cosmological collider signals associated with particle production. Then we plug the $\mathcal{O}(g^0)$ solution into the integral \eqref{eq: Source_int} and obtain
\begin{align}
    &\int_0^\infty dx\,K_\pm(\omega,x)F_0\left(\dfrac{U}{1+U\,x/2}\right)\nonumber\\
    &=-\frac{g^2 m_0^2\pi\,}{4}\frac{e^{\mp \frac{\omega}{2}}}{2^{\mp i\omega}}\sum_{k=0}^{\infty}\Bigg\{\frac{e^{-\pi\mu}}{\sinh(2\pi\mu)}\,\Gamma\left[\bgm  \frac{1}{2}+k-i\mu,\,\frac{1}{2}+k-i\mu\mp i\omega\\
    1+k,\,1+k-2i\mu\edm\right]\times U^{\frac{1}{2}+k-i\mu\mp i\omega}+(\mu\to-\mu)\nonumber\\
    &\qquad\qquad\qquad\
    \qquad~~~~~+\,\frac{1}{\cosh \pi\mu}\,\Gamma\left[\bgm  1+k,\,1+k\mp i\omega\\
    \frac{3}{2}+k+i\mu,\,\frac{3}{2}+k-i\mu\edm\right]\times U^{1+k\mp i\omega}\Bigg\}\,.
\end{align}
This reduces the problem to solving standard differential equations with sources.

\paragraph{Ansatz and recursive relations} Based on the structure of the series of sources, we adopt the ansatz 
\begin{align}
        F^{\text{ansatz}}_{\pm1}= \sum_{k,n=0}^{\infty} d^{\,\text{part}}_{k,n}(\mu,\pm \omega)\, U^{1+k+n\mp i\omega}+\left[d^{\,\text{hom}}_{k,n}(\mu,\pm\omega)\, U^{\frac{1}{2}+k+n+i\mu\mp i\omega}+(\mu\to-\mu)\right]\,,
\end{align}
where the coefficients satisfy the following recurrence relation,
\begin{align}
    &d^{\,\text{part}}_{k,n+1}(\mu, \omega)=\frac{(1+k+n-i\omega)^2}{(\frac{3}{2}+k+n-i\mu-i\omega)(\frac{3}{2}+k+n+i\mu-i\omega)}d^{\,\text{part}}_{k,n}(\mu, \omega)\,,\\
    &d^{\,\text{hom}}_{k,n+1}(\mu, \omega)=\frac{(\frac{1}{2}+k+n-i\omega)^2}{(1+k+n-i\omega)(1+k+n+2i\mu-i\omega)}d^{\,\text{hom}}_{k,n}(\mu, \omega)\,.
\end{align}
The initial values are determined from matching with the sources,
\begin{align}
    &d^{\,\text{part}}_{k,0}(\mu, \omega)=-\frac{g^2m_0^2\pi}{4\cosh \pi\mu}\,\frac{2^{i\omega}\,e^{-{\frac{\pi\omega}{2}}}}{(\frac{1}{2}+k+i\mu-i\omega)(\frac{1}{2}+k-i\mu-i\omega)}\,\Gamma\,\left[\bgm 1+k,\,1+k-i\omega
    \\
    \frac{3}{2}+k-i\mu,\,\frac{3}{2}+k+i\mu
    \edm\right]\,,\nonumber\\
    &d^{\,\text{hom}}_{k,0}(\mu, \omega)=\frac{g^2m_0^2\pi}{4\sinh (2\pi\mu)}\,\frac{2^{i\omega}\,e^{\pi\mu-{\frac{\pi\omega}{2}}}}{(k+2i\mu-i\omega)(k-i\omega)}\,\Gamma\,\left[\bgm \frac{1}{2}+k+i\mu,\,\frac{1}{2}+k+i\mu-i\omega
    \\
    1+k,\,1+k+2i\mu
    \edm\right]\,.
\end{align}
The general expressions for the coefficients can then be solved straightforwardly as
\begin{align}
    &d^{\,\text{part}}_{k,n}(\mu, \omega)\nonumber\\
    &=-\frac{e^{-\frac{\pi\omega}{2}}g^2m_0^2\pi }{2^{2-i\omega}\cosh \pi\mu}\,\Gamma\,\left[\bgm 1+k,1+k+n-i\omega,1+k+n-i\omega,\frac{1}{2}+k-i\mu-i\omega,\frac{1}{2}+k+i\mu-i\omega
    \\
    \frac{3}{2}+k-i\mu,\frac{3}{2}+k+i\mu,1+k-i\omega,\frac{3}{2}+n+k-i\mu-i\omega,\frac{3}{2}+n+k+i\mu-i\omega
    \edm\right]\,,\\
    &d^{\,\text{hom}}_{k,n}(\mu, \omega)\nonumber\\
    &=\frac{e^{\pi\mu-\frac{\pi\omega}{2}}g^2m_0^2\pi }{2^{2-i\omega}\sinh( 2\pi\mu)}\,\Gamma\,\left[\bgm \frac{1}{2}+k-i\mu,\,k-i\omega,\frac{1}{2}+k+n+i\mu-i\omega,\frac{1}{2}+k+n+i\mu-i\omega,k+2i\mu-i\omega
    \\
    1+k,1+k+2i\mu,1+k+n-i\omega,\frac{1}{2}+k+i\mu-i\omega,1+k+n+2i\mu-i\omega
    \edm\right]\,.
\end{align}
We have now expressed $F^{\text{ansatz}}_{\pm 1}$ as a double-layered summation over both the $n$ and $k$ directions, where one of the layers can be explicitly performed to yield 
\begin{align}
        F^{\text{ansatz}}_{\pm1}(U)=&\sum_{n=0}^{\infty}\,\mathcal{A}^{\pm}_n(\mu,\omega)\times U^{1+n\mp i\omega}\,\pFq{3}{2}{1,1+n\mp i\omega,1+n\mp i\omega}{\frac{3}{2}+n-i\mu\mp i\omega,\frac{3}{2}+n+i\mu\mp i\omega}{U}\nonumber\\
        +&\sum_{n=0}^{\infty}\Big(\mathcal{B}_n^{\pm}(\mu,\omega)\times U^{\frac{1}{2}+n+i\mu\mp i\omega}\,\pFq{3}{2}{1,\frac{1}{2}+n+i\mu\mp i\omega,\frac{1}{2}+n+i\mu\mp i\omega}{1+n\mp i\omega,1+n+2i\mu\mp i\omega}{U}\nonumber\\
        &\qquad+(\mu\to-\mu)\Big)\,, \label{eq: Fpm_part2}
\end{align}
where the coefficients are given by 
\begin{align}
    &\mathcal{A}^{\pm}_n(\mu,\omega)\equiv-\frac{m_0^2g^2\pi}{4}\frac{2^{\pm i\omega}e^{\mp\frac{\pi\omega}{2}}\,{\rm{sech}}(\pi\mu)}{(\frac{1}{2}+n+i\mu\mp i\omega)(\frac{1}{2}+n-i\mu\mp i\omega)}\Gamma\left[\bgm 1+n,1+n\mp i
    \omega\\\frac{3}{2}+n-i\mu,\frac{3}{2}+n+i\mu\edm\right],\\
    &\mathcal{B}^{\pm}_n(\mu,\omega)\equiv\frac{m_0^2g^2\pi}{4}\frac{2^{\pm i\omega}e^{\mp\frac{\pi\omega}{2}+\pi\mu}\,{\rm{csch}}(2\pi\mu)}{(n\mp i\omega)(n+2i\mu\mp i\omega)}\Gamma\,\left[\bgm \frac{1}{2}+n+i\mu,\frac{1}{2}+n+i\mu\mp i
    \omega\\1+n,1+n+2i\mu\edm\right]\,.
\end{align}

\paragraph{Adding the homogenous solution} The particular solution $F^{\text{ansatz}}_{\pm 1}$ we found above possesses a folded divergence at $U\rightarrow1$. To cancel this divergence, we need to include homogeneous solutions that satisfy the associated homogeneous differential equation:
 \begin{align}
        F_{\pm1}^{\text{hom}}(U)= \xi_1^{\pm} (\mu,\omega)\cdot\mathcal{Y}_{1}(U)+\xi_2^{\pm} (\mu,\omega)\cdot\mathcal{Y}_{2}(U)\,.
    \end{align}
There are two coefficients $\xi_1$ and $\xi_2$ to be determined from two boundary conditions. These may come from physical requirements such as the cancellation of the folded pole. However, as previously emphasized in this paper, we find \textit{microcausality} to be a very powerful tool for fixing these coefficients. As shown explicitly in \eqref{eq: soft_Fac}, the exchange diagram should factorise into two parts in the soft limit:
\begin{align}
   \lim_{\boldsymbol{s}\to 0} F_{++}(k_{12},k_{34},s)= -\left[f(-k_{12}-i\epsilon,s)\right]^*\times f(k_{34}-i\epsilon,s)+\text{analytic}\,,
\end{align}
this relation should hold non-perturbatively in $g$ and should remain valid in the three-point limit $k_4\to 0$. For our purpose, only its perturbative version is needed to fix the coefficients here. 

In the soft limit, there are two distinct types of non-analyticities, namely $U^{\pm i\omega \pm i\mu}$ and $U^{\pm i\mu}$. They originate from the particular solution $F^{\text{ansatz}}_{\pm}$ and the homogeneous solution $F^{\text{hom}}_{\pm}$, respectively. The first type can be directly confirmed to satisfy the relation using the analytical expression of $f_{\pm}$ we derived before. Thus here we will concentrate on $U^{i\mu}$ (or equivalently $u^{i\mu}$, since under the soft limit $U\sim 2 u$), which should allow us to fix the coefficients of the homogeneous solution at order $\mathcal{O}(g^2)$. 
Recalling the definition of the three-point-contact $f(k_{12},s)$ in (\ref{finalfk12s}) and the three-point-exchange $F_{++}$ in (\ref{Fexchange}), the soft limit factorisation \eqref{eq: soft_Fac} yields
\begin{align}
    \lim_{\boldsymbol{s}\to 0} F^{\text{hom}}_{\pm 1}(U)=-f^{\text{hom}}_{\mp 1}\left[(-u-i\epsilon)\right]^*\times f_0(1)-\left[f_0(-u-i\epsilon)\right]^*\times f_{\pm1}(1)\,, \label{eq: fact_u^imu}
\end{align}
where we have truncated at $\mathcal{O}(g^2)$, retaining only the contribution that sources the non-analytic behaviour $u^{\pm i \mu}$. Note that taking the complex conjugate also flips the $x_0^{\pm i \omega }$ factor, which is why the first term on the left-hand side carries an opposite label from the others. The left-hand side gives 
\begin{align}
   \lim_{\boldsymbol{s}\to 0} F^{\text{hom}}_{\pm 1}(U)=\beta^{\pm}_1(\mu,\omega)\cdot\,2^{-i\mu}\,\Gamma\left[\frac{1}{2}+i\mu,\,-i\mu\right]\,\textcolor{red3}{u^{\frac{1}{2}+i\mu}}+\xi_2^{\pm}(\mu,\omega)\cdot2^{i\mu}\,\Gamma\left[\frac{1}{2}-i\mu,\, i\mu\right]\,\textcolor{red3}{u^{\frac{1}{2}-i\mu}}\,,
\end{align}
where we have used $U=2u$ in the soft limit. The zeroth-order terms are easily calculated as 
\begin{align}
    f_0(1)&=\frac{i\pi}{\sqrt{2}\cosh\pi\mu}\,,\\
    \lim_{\boldsymbol{s}\to 0}  f^{*}_{0}(-u)&=-\,\frac{2^{-1-i\mu}e^{\pi\mu}}{\sqrt{\pi}}\,\Gamma\left[\frac{1}{2}+i\mu,\,-i\mu\right]\textcolor{red3}{u^{\frac{1}{2}+i\mu}}-\,\frac{2^{-1+i\mu}e^{-\pi\mu}}{\sqrt{\pi}}\,\Gamma\left[\frac{1}{2}-i\mu,\,i\mu\right]\,\textcolor{red3}{u^{\frac{1}{2}-i\mu}}\,,
\end{align}
and using the expression for three-point contact diagram previously derived in \eqref{eq: fpm1_hom} along with coefficients in \eqref{eq: f0_final_coeffs}, we obtain
\begin{align}
    \lim_{\boldsymbol{s}\to 0}  \left[f^{\text{hom}}_{\mp 1}(-u-i\epsilon)\right]^*=\frac{i\,2^{-3-i\mu}\,g^2m_0^2}{1-e^{-2\pi\mu\mp\pi\omega}} \,\,\Gamma\left[\bgm \frac{1}{2}+i\mu,\, -i\mu,\,\frac 1 2\pm\frac{i\omega}{2},\,\mp \frac{i\omega}{2}\\
    1- i\mu\pm \frac{i\omega}{2},\,1+i\mu\pm \frac{i\omega}{2}\edm\right]\,\textcolor{red3}{u^{\frac{1}{2}+i\mu}}+(u\to-u)\,,
\end{align}
The final piece is the three-point function in the folded limit $f_{\pm 1}(1)$ , which is complicated but just yields a number ultimately. We will leave it for now, and later present its explicit form by taking the limit of the three-point function. Finally, from the constraint (\ref{eq: fact_u^imu}), we get the coefficients
\begin{align}
    &\xi_2^{\pm}(\mu,\omega)=\xi_1^{\pm}(-\mu,\omega)\,,\\
    &\xi_1^{\pm}(\mu,\omega)=\frac{g^2m_0^2\pi}{16\sqrt{2}}\frac{1+\coth(\pi(\mu\pm \frac{\omega}{2}))}{\cosh \pi\mu}\Gamma\left[\bgm \frac 1 2\pm\frac{i\omega}{2},\,\mp \frac{i\omega}{2}\\
    1- i\mu\pm \frac{i\omega}{2},\,1+i\mu\pm \frac{i\omega}{2}\edm\right]+\frac{e^{\pi\mu}}{2\sqrt{\pi}}\,f_{\pm1}(1)\,. 
\end{align}

\paragraph{Two-point function} To obtain the two-point function, which is required for the coefficients, we can take the limit of the three-point function $f_{\pm}(u)$ by $u\to 1$. Both the $f_{\pm 1}^{\text{hom}}$ and $f_{\pm 1}^{\text{ansatz}}$ parts have logarithmic divergences under this limit, yet they cancel each other and yield a finite expression:
\begin{align}
    &f_{\pm1}(1)\nonumber\\
    &=-\frac{ig^2m_0^2\sqrt{\pi}}{16\sqrt{2}}\frac{(\gamma-\log4+2\psi\left(\frac{1}{2}+i\mu)\right)e^{\mp\frac{\pi\omega}{2}}}{\sinh(\pi\mu)\,\sinh(\pi\mu\pm\frac{\pi\omega}{2})}\times\Gamma\left[\bgm \mp\frac{i\omega}{2},\,\frac 1 2\pm \frac{i\omega}{2}\\
    1- i\mu\pm \frac{i\omega}{2},\,1+i\mu\pm \frac{i\omega}{2}\edm\right]\nonumber\\
    &-\sum_{m=0}^{\infty}\frac{g^2 m_0^2}{\sinh \pi\mu}\,e^{\mp \frac{\pi\omega}{2}}2^{-\frac{9}{2}\mp i\omega}\psi\left(\frac{1}{4}+m+\frac{i\mu}{2}\mp \frac{i\omega}{2}\right)\times\Gamma\left[\bgm m\mp\frac{i\omega}{2},\,m+i\mu\mp \frac{i\omega}{2}\\
    1+m,\,1+m+i\mu\edm\right]\nonumber\\
    &+\sum_{m=0}^{\infty}\frac{g^2 m_0^2}{\sinh \pi\mu}\frac{2^{-\frac{13}{2}\mp i\omega}e^{\mp \frac{\pi\omega}{2}}(1+4\mu^2)}{1+4m+2i\mu\mp2i\omega}\Gamma\left[\bgm m\mp\frac{i\omega}{2},\,m+i\mu\mp \frac{i\omega}{2}\\
    1+m,\,1+m+i\mu\edm\right]\pFq{4}{3}{1,1,\frac{5}{4}-\frac{i\mu}{2},\frac{5}{4}+\frac{i\mu}{2}}{2,2,\frac{5}{4}+m+\frac{i\mu}{2}-\frac{i\omega}{2}}{1}\nonumber\\
    &+(\mu\to-\mu)\,,
\end{align}
here $\psi(x)\equiv\Gamma'(x)/\Gamma(x)$ denotes the digamma function, and the final expression involves a single-layer summation. This number can also be written in terms of higher-order hypergeometric functions without summation, allowing efficient evaluation:
\begin{align}
     &f_{\pm1}(1)=- \frac{i\,\pi g^2m_0^2\,e^{\mp\frac{\pi\omega}{2}} }{2^{\frac{3}{2}\mp i\omega}\cosh(\pi\mu)} \Gamma\left[\frac{1}{2}-i\mu\mp i\omega,\frac{1}{2}+i\mu\mp i\omega\right]\,\pregFq{4}{3}{1,1,\frac{1}{2}-i\mu\mp i\omega,\frac{1}{2}+i\mu\mp i\omega}{\frac{3}{2}-i\mu,\frac{3}{2}+i\mu,1\mp i\omega}{1}\,.
\end{align}

\paragraph{Another contribution} The non-time-ordered exchange diagram can be directly expressed in terms of the three-point contact one $f(k,s)$,
\begin{align}
    F_{+-}(k_{12},k_{34},s)=\frac{1}{s}f(k_{12},s)f^*(k_{34},s)\,,
\end{align}
and this relation holds non-perturbatively. If we expand to $\mathcal{O}(g^2)$, it gives
\begin{align}
    s F_{+-}(u)=f_0(u)f_0^*(1)+\big[f_{\pm1}(u)f_0^*(1)+f_0(u)f_{\mp 1}^*(1)\big]x_0^{\pm i\omega}\,.
\end{align}
Building on previous efforts, all necessary ingredients above are now available.

\section{Light fields and the anomalous scaling}\label{LightFieldSection}

We have mostly been concerned with heavy field exchanges in the main body of this work. However, the analytical and numerical techniques readily extend to the case of light fields. First, we note that the boundary bootstrap equation retains its form \eqref{IDEintro1}, albeit with a light mass $m_0$ lying in the complementary series with $\nu=(9/4-m_0^2/H^2)^{1/2}>0$. To avoid IR divergences in the $\dot \pi_c \sigma$ and $\dot \pi_c^2 \sigma$ vertices. In the absence of parametric resonances this implies
\begin{align}
    \int \d\eta\, a^3(\eta) \pi_c'(\eta)\sigma(\eta)\sim \int \d\eta \, (-\eta)^{-1/2-\nu}\quad \Rightarrow \quad 0<\nu<1/2~,
\end{align}
meaning that the $\sigma$ field must be heavier than the conformally coupled scalar.\footnote{In the scale-invariant case $g=0$, the IR divergences in the Schwinger-Keldysh diagrams $F_{\sf ab}$ are known to cancel out in correlator observables \cite{Chen:2009zp}. In the scale-breaking case $g\neq 0$, it remains unclear if the cancellation persists, we therefore choose to assume a sufficiently large mass to ensure IR convergence.} In this regime, the perturbative solution in \eqref{ThreeptsSeed} and \eqref{BispectrumFinal} directly translates to the light-mass case after replacing $\mu\to -i\nu$, resulting in an analytical solution accurate at the order of $g^2$. One might therefore expect the translation to the light-mass case to be trivial at each order. Yet surprisingly, this turns out not to be the case. In fact, we discover an \textit{entirely new} effect emerging at the order of $g^4$, which will first manifest itself in the numerics below.

\begin{figure}[h!]
    \centering
    \includegraphics[width=\textwidth]{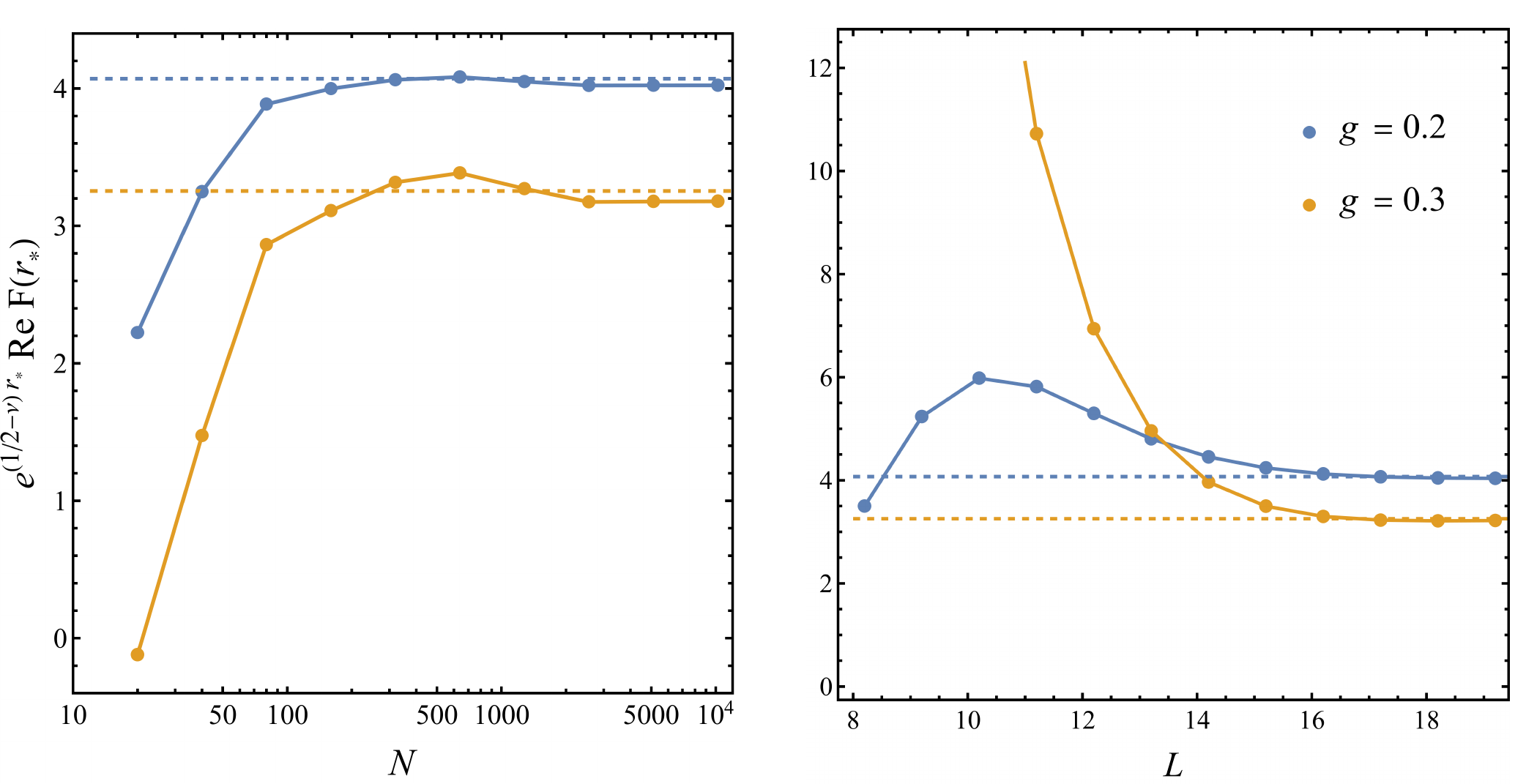}
    \caption{Convergence tests of the numerical bootstrap solution for the light-field exchange with respect to the mesh points $N$ (\textit{left panel}) and the cutoff $L=\log (k_{12}/s)_{\text{max}}$ (\textit{right panel}). The benchmark kinematics is chosen to be $r_*=\log (k_{12}/s)_*=8$ and the \textcolor{mmaBlue}{blue} and \textcolor{mmaYellow}{yellow} points correspond to $g=0.2$ and $g=0.3$, respectively. The dashed lines represent the prediction of the perturbative analytical solution. The other parameters are chosen as $\nu=0.2$, $\omega=1/3$, $x_0=-s\eta_0=1$. The boundary conditions are implemented at $r_{\rm I}=1$ and at $r_{\rm II}=1+2L/N$. The regularisation gap is chosen to be $\delta=10^{-1}$. In the \textit{left panel}, the kinematic cutoff is fixed to be $L=20$ and in the \textit{right panel}, the mesh points are chosen such that the step size $L/N=10^{-2}$ is fixed. Thus the numerical solution again converges at large $N$ and $L$ towards the perturbative prediction. Comparison to the heavy field case in Figure \ref{FigCheckConvergenceConsistency}, we observe a much stronger non-locality of the bootstrap equation.}
    \label{FigCheckConvergenceConsistencyLight}
\end{figure}

\paragraph{Numerical bootstrap} The numerical bootstrap can be performed in the same fashion as in the heavy-mass case. We confirm the numerical convergence in Figure \ref{FigCheckConvergenceConsistencyLight}. From the plots, we also observe that the non-locality in the light-field-exchange case appears to be much stronger than in the heavy one (c.f. Figure \ref{FigCheckConvergenceConsistency}), as the numerics only converges for $L-r_*=\mathcal{O}(10)$. This matches the intuition that light fields propagate a longer distance than heavy fields and are therefore more non-local.

\begin{figure}[h!]
    \centering
    \includegraphics[width=\textwidth]{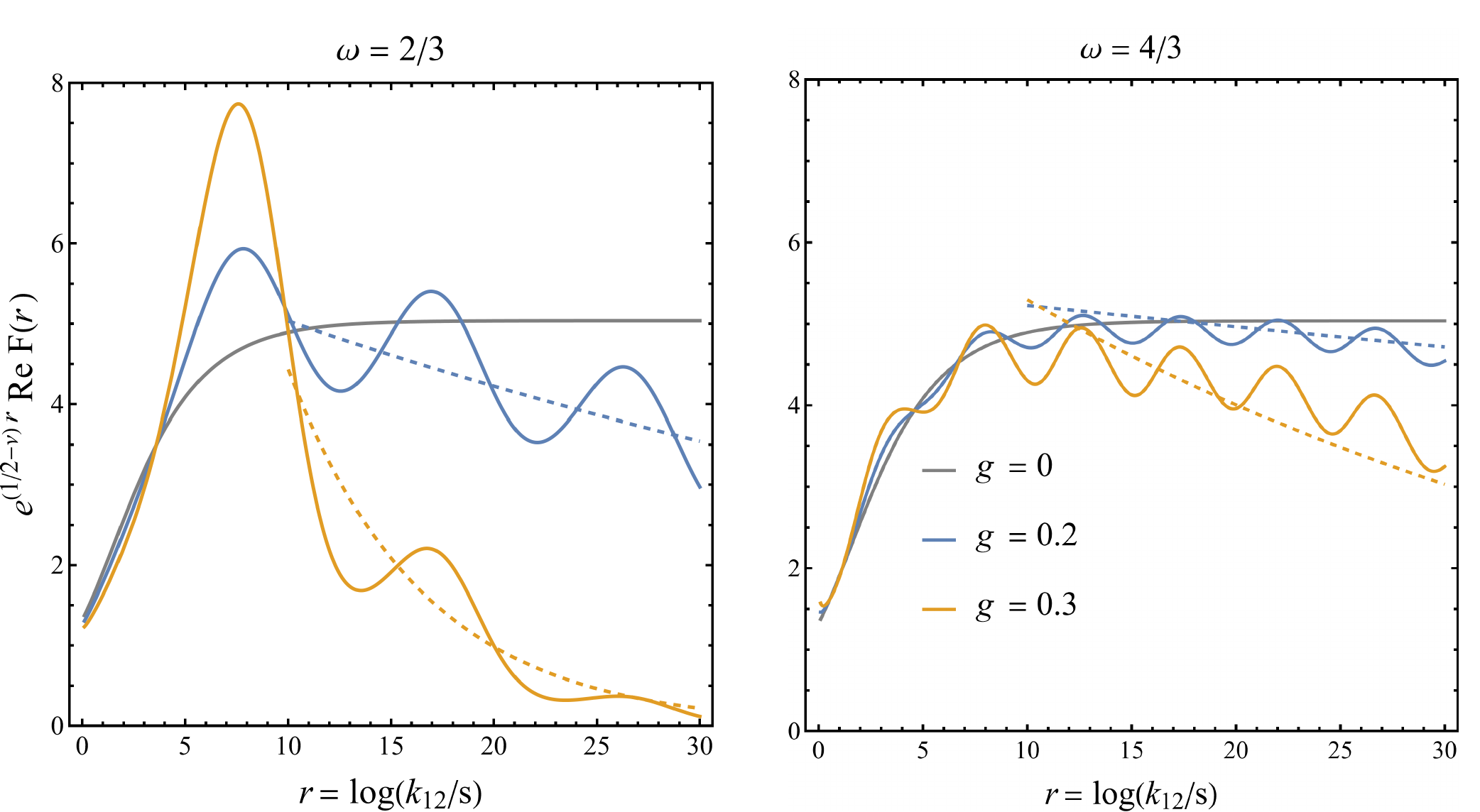}
    \caption{The numerical solution (solid curves) as a function of the momentum ratio $r=\log(k_{12}/s)$ normalised by the naive scaling exponent. The dotted curves indicates the anomalous scaling predicted from the boundary \eqref{lightAnomalousScalingExponent}. The \textit{left} and \textit{right} panels correspond to frequency choices $\omega=2/3$ and $\omega=4/3$, respectively. The \textcolor{mmaGrey}{grey}, \textcolor{mmaBlue}{blue} and \textcolor{mmaYellow}{yellow} curves correspond to different sizes of the coupling $g$ as specified in the right panel. Other model parameters are chosen as $\nu=0.2$, $x_0=-s\eta_0=2$. The number of mesh point is $N=10^4$ with a kinematic cutoff $L=40$. The boundary conditions are implemented at $r_{\rm I}=1$ and at $r_{\rm II}=1+2L/N$ with a regularisation gap $\delta=10^{-1}$. }
    \label{FigNumericalSolutionLight}
\end{figure}

The waveforms normalised according to the naive scaling exponent $-1/2+\nu$ are shown in Figure \ref{FigNumericalSolutionLight}. Apparently, the numerical solutions start to deviate from the naive scaling at large couplings $g$ and in the squeezed limit $k_{12}/s\gg 1$, signalling the breakdown of the $\mathcal{O}(g^2)$ perturbative solution. Carefully examining the plots indeed points to the emergence of a new expansion parameter $g^4\log(k_{12}/s)$ which is absent in our analytical solution at order $\mathcal{O}(g^2)$ due to its high order in $g$. A resummation of this new expansion parameter would then lead to an anomalous correction to the scaling exponent in the squeezed limit:
\begin{align}
    F(k_{12},s)\sim \left(\frac{k_{12}}{s}\right)^{-1/2+\nu}\left(1+\mathcal{O}(g^4)\times \log\frac{k_{12}}{s}+\cdots\right)\to\left(\frac{k_{12}}{s}\right)^{-1/2+\nu+\mathcal{O}(g^4)}~.
\end{align}

\paragraph{Anomalous scaling from the boundary} Interestingly, the emergence of a new scaling can be very easily predicted from the boundary. In fact, the anomalous scaling exponent can be analytically computed via the same eigenfrequency analysis as in the IR resonances. We make the same ansatz as in \eqref{squeezedAnsatz},
\begin{align}
    F(k_{12},s)=\sum_{n=-\infty}^{\infty} \xi_n \left(\frac{k_{12}}{s}\right)^{\alpha+i n \omega}~,
\end{align}
and truncate the resultant recurrence relation at $\xi_{\pm 2}=\xi_{\pm 3}=\cdots=0$. The constraint equation is identical to \eqref{boundaryResExponentConstraintEq} but with $\mu\to -i\nu$. Now, instead of expanding around $\omega=2\mu$, we directly solve the constraint near $\alpha=-1/2+\nu$, yielding
\begin{keyeqn}
    \begin{align}
        \alpha=-1/2+\nu-\frac{g^4 \left(9/4-\nu ^2\right)^2}{4 \nu  \left(4 \nu^2+\omega ^2\right)}+\mathcal{O}(g^8)~,\label{lightAnomalousScalingExponent}
    \end{align}
\end{keyeqn}
which produces exactly the expected scaling exponent (see the dotted lines in Figure \ref{FigNumericalSolutionLight}). This eigenfrequency analysis therefore non-perturbatively resums large logarithms of the momentum ratio from a pure boundary perspective. 

\paragraph{Effective mass and an analogy to Kapitza's pendulum} 

\begin{wrapfigure}{r}{0.3\textwidth}
    \centering
    \begin{tikzpicture}[>=stealth, thick, scale=0.9]
        \def\L{2.8}      
        \def\B{0.4}      
        \def\Theta{18}   

        \coordinate (P) at (0, \B); 
        \coordinate (M) at ({\L*sin(\Theta)}, {\B + \L*cos(\Theta)});
        
        \draw[dashed, gray] (0, -0.5) -- (0, 4.0);

        \draw[<->, blue, thick] (0, -\B-0.1) -- (0, \B-0.1) 
            node[midway, left, font=\small] {$\textsl{b} \cos(\omega t)$};

        \draw[very thick] (P) -- (M) 
            node[midway, right, xshift=2pt] {$\ell$};

        \filldraw[fill=white, draw=black] (P) circle (2.5pt);

        \filldraw[fill=black] (M) circle (4pt) 
            node[above right] {$m$};

        \draw[->, red, thick] ([xshift=10pt]M) -- ++(0, -0.8) 
            node[below, font=\small] {$\textsl{g}$};

        \draw[->] (0, {\B + 1.2}) arc (90:{90-\Theta}:1.2);
        \node at (0.2, {\B + 1.5}) {$\theta$};
    \end{tikzpicture}
    \caption{Kapitza's pendulum of mass $m$, length $\ell$, frequency $\omega$, amplitude $\textsl{b}$ and gravity $\textsl{g}$.}\label{KapitzaPendulumIllustration}
\end{wrapfigure}

The anomalous scaling can be understood through the analogy to a classical mechanics system known as the \textit{Kapitza pendulum} \cite{1572824500035291648}. The Kapitza pendulum consists of an inverted pendulum with a vertically oscillating pivot point (see Figure \ref{KapitzaPendulumIllustration} for an illustration). The nature of the inverted position implies an instability for the pendulum, yet the fast-oscillating pivot point can distort its motion and effectively stabilise it. Up to total derivatives, its Lagrangian
\begin{align}
    L=\frac{1}{2}m \ell^2\dot{\theta}^2-m\ell \left[\textsl{g}-\textsl{b} \,\omega^2 \cos (\omega t)\right]\cos\theta
\end{align}
leads to an equation of motion of the form
\begin{align}
    \left[\frac{\partial^2}{\partial t^2}-\frac{\textsl{g}}{\ell}\left(1-\frac{\textsl{b} \, \omega^2}{\textsl{g}}\cos(\omega t)\right)\right]\hat{\theta}+\mathcal{O}(\hat{\theta}^3)=0~,\label{KapitzaPendulumEoM}
\end{align}
where we have performed a linear expansion around $\theta=0$ and $\hat{\theta}\equiv \sqrt{m}\ell \theta$ is the canonical variable. In the double-scaling limit $\textsl{b}\to 0$, $\omega\to \infty$ with $\textsl{b}\,\omega$ fixed, perturbative analysis using the separation of scales shows that the pendulum oscillates with an effective potential \cite{wiki:kapitza_pendulum}
\begin{align}
    \frac{\partial^2\hat{\theta}}{\partial t^2}=-\frac{\partial V_{\rm eff}(\hat{\theta})}{\partial\hat\theta}~,\quad V_{\rm eff}(\hat{\theta})=\frac{1}{2}\left(-\frac{\textsl{g}}{\ell}+\frac{\textsl{b}^2\omega^2}{2\ell^2}\right)\hat{\theta}^2+\mathcal{O}(\hat\theta^4)~.
\end{align}
The second term in the last bracket provides a positive correction to the effective frequency of the pendulum which tends to stabilise the system. Comparing \eqref{KapitzaPendulumEoM} to the equation of motion of the $\sigma$ field with light masses,
\begin{align}
    \left[\frac{\partial^2}{\partial t^2}-\nu^2\left(1-\tilde{g}^2 \cos(\omega t)\right)\right](a^{3/2}\sigma(s,t))\approx 0~,\quad \tilde{g}^2\equiv g^2 \left(\frac{9}{4\nu^2}-1\right)~,
\end{align}
we make the identification
\begin{align}
    \nu^2=\frac{\textsl{g}}{\ell}\quad,\quad \tilde{g}^2=\frac{\textsl{b} \, \omega^2}{\textsl{g}}~.
\end{align}
Thus the corrected effective mass reads
\begin{align}
    \nu_{\rm eff}^2=\nu^2-\frac{\tilde{g}^4\nu^4}{2\omega^2}~,
\end{align}
leading to the effective scaling exponent
\begin{align}
    (\alpha)_{\rm Kapitza}=-1/2+\nu_{\rm eff}=-1/2+\nu-\frac{g^4 \left(9/4-\nu ^2\right)^2}{4 \nu  \omega ^2}~.
\end{align}
This matches nicely with our result from the boundary bootstrap \eqref{lightAnomalousScalingExponent} in the double-scaling limit with $\omega\gg \nu$. Taking the analogy, we therefore conclude that time-dependent oscillations provide an (necessarily positive) effective correction to the mass of the light field, resulting in the anomalous scaling in the squeezed limit of boundary correlators. It is now transparent why this effect only appears at order $\mathcal{O}(g^4)$ since one needs at least \textit{two} insertions of cosine oscillations to generate a non-zero effective mean i.e. $\overline{\cos^2(\omega t)}=1/2$. This also suggests that such corrections exist for heavy fields as well. A complete understanding of this phenomenon would require an analytical investigation at order $\mathcal{O}(g^4)$. We leave such a more thorough analysis for future works.

\newpage

\bibliographystyle{JHEP}
\bibliography{Refs}

\providecommand{\href}[2]{#2}\begingroup\raggedright\begin{thebibliography}{100}

\bibitem{Arkani-Hamed:2018kmz}
N.~Arkani-Hamed, D.~Baumann, H.~Lee and G.L.~Pimentel, \emph{{The Cosmological
  Bootstrap: Inflationary Correlators from Symmetries and Singularities}},
  \href{https://doi.org/10.1007/JHEP04(2020)105}{\emph{JHEP} {\bfseries 04}
  (2020) 105} [\href{https://arxiv.org/abs/1811.00024}{{\ttfamily
  1811.00024}}].

\bibitem{Baumann:2020dch}
D.~Baumann, C.~Duaso~Pueyo, A.~Joyce, H.~Lee and G.L.~Pimentel, \emph{{The
  Cosmological Bootstrap: Spinning Correlators from Symmetries and
  Factorization}},
  \href{https://doi.org/10.21468/SciPostPhys.11.3.071}{\emph{SciPost Phys.}
  {\bfseries 11} (2021) 071}
  [\href{https://arxiv.org/abs/2005.04234}{{\ttfamily 2005.04234}}].

\bibitem{Hogervorst:2021uvp}
M.~Hogervorst, J.a.~Penedones and K.S.~Vaziri, \emph{{Towards the
  non-perturbative cosmological bootstrap}},
  \href{https://doi.org/10.1007/JHEP02(2023)162}{\emph{JHEP} {\bfseries 02}
  (2023) 162} [\href{https://arxiv.org/abs/2107.13871}{{\ttfamily
  2107.13871}}].

\bibitem{Baumann:2022jpr}
D.~Baumann, D.~Green, A.~Joyce, E.~Pajer, G.L.~Pimentel, C.~Sleight et~al.,
  \emph{{Snowmass White Paper: The Cosmological Bootstrap}},  in \emph{{2022
  Snowmass Summer Study}}, 3, 2022
  [\href{https://arxiv.org/abs/2203.08121}{{\ttfamily 2203.08121}}].

\bibitem{deRham:2025mjh}
C.~de~Rham, S.~Jazayeri and A.J.~Tolley, \emph{{Bispectrum islands: Bootstrap
  bounds on cosmological correlators}},
  \href{https://doi.org/10.1103/q6rq-sj9t}{\emph{Phys. Rev. D} {\bfseries 112}
  (2025) 083531}.

\bibitem{Bzowski:2013sza}
A.~Bzowski, P.~McFadden and K.~Skenderis, \emph{{Implications of conformal
  invariance in momentum space}},
  \href{https://doi.org/10.1007/JHEP03(2014)111}{\emph{JHEP} {\bfseries 03}
  (2014) 111} [\href{https://arxiv.org/abs/1304.7760}{{\ttfamily 1304.7760}}].

\bibitem{Sleight:2019hfp}
C.~Sleight and M.~Taronna, \emph{{Bootstrapping Inflationary Correlators in
  Mellin Space}}, \href{https://doi.org/10.1007/JHEP02(2020)098}{\emph{JHEP}
  {\bfseries 02} (2020) 098}
  [\href{https://arxiv.org/abs/1907.01143}{{\ttfamily 1907.01143}}].

\bibitem{Sleight:2019mgd}
C.~Sleight, \emph{{A Mellin Space Approach to Cosmological Correlators}},
  \href{https://doi.org/10.1007/JHEP01(2020)090}{\emph{JHEP} {\bfseries 01}
  (2020) 090} [\href{https://arxiv.org/abs/1906.12302}{{\ttfamily
  1906.12302}}].

\bibitem{Pajer:2020wnj}
E.~Pajer, D.~Stefanyszyn and J.~Supe\l{}, \emph{{The Boostless Bootstrap:
  Amplitudes without Lorentz boosts}},
  \href{https://doi.org/10.1007/JHEP12(2020)198}{\emph{JHEP} {\bfseries 12}
  (2020) 198} [\href{https://arxiv.org/abs/2007.00027}{{\ttfamily
  2007.00027}}].

\bibitem{Qin:2022lva}
Z.~Qin and Z.-Z.~Xianyu, \emph{{Phase information in cosmological collider
  signals}}, \href{https://doi.org/10.1007/JHEP10(2022)192}{\emph{JHEP}
  {\bfseries 10} (2022) 192}
  [\href{https://arxiv.org/abs/2205.01692}{{\ttfamily 2205.01692}}].

\bibitem{Qin:2022fbv}
Z.~Qin and Z.-Z.~Xianyu, \emph{{Helical inflation correlators: partial
  Mellin-Barnes and bootstrap equations}},
  \href{https://doi.org/10.1007/JHEP04(2023)059}{\emph{JHEP} {\bfseries 04}
  (2023) 059} [\href{https://arxiv.org/abs/2208.13790}{{\ttfamily
  2208.13790}}].

\bibitem{Wang:2022eop}
D.-G.~Wang, G.L.~Pimentel and A.~Ach\'ucarro, \emph{{Bootstrapping multi-field
  inflation: non-Gaussianities from light scalars revisited}},
  \href{https://doi.org/10.1088/1475-7516/2023/05/043}{\emph{JCAP} {\bfseries
  05} (2023) 043} [\href{https://arxiv.org/abs/2212.14035}{{\ttfamily
  2212.14035}}].

\bibitem{Salcedo:2022aal}
S.A.~Salcedo, M.H.G.~Lee, S.~Melville and E.~Pajer, \emph{{The Analytic
  Wavefunction}}, \href{https://doi.org/10.1007/JHEP06(2023)020}{\emph{JHEP}
  {\bfseries 06} (2023) 020}
  [\href{https://arxiv.org/abs/2212.08009}{{\ttfamily 2212.08009}}].

\bibitem{Jazayeri:2022kjy}
S.~Jazayeri and S.~Renaux-Petel, \emph{{Cosmological bootstrap in slow
  motion}}, \href{https://doi.org/10.1007/JHEP12(2022)137}{\emph{JHEP}
  {\bfseries 12} (2022) 137}
  [\href{https://arxiv.org/abs/2205.10340}{{\ttfamily 2205.10340}}].

\bibitem{Pimentel:2022fsc}
G.L.~Pimentel and D.-G.~Wang, \emph{{Boostless cosmological collider
  bootstrap}}, \href{https://doi.org/10.1007/JHEP10(2022)177}{\emph{JHEP}
  {\bfseries 10} (2022) 177}
  [\href{https://arxiv.org/abs/2205.00013}{{\ttfamily 2205.00013}}].

\bibitem{Melville:2023kgd}
S.~Melville and G.L.~Pimentel, \emph{{de Sitter S matrix for the masses}},
  \href{https://doi.org/10.1103/PhysRevD.110.103530}{\emph{Phys. Rev. D}
  {\bfseries 110} (2024) 103530}
  [\href{https://arxiv.org/abs/2309.07092}{{\ttfamily 2309.07092}}].

\bibitem{Armstrong:2023phb}
C.~Armstrong, H.~Goodhew, A.~Lipstein and J.~Mei, \emph{{Graviton trispectrum
  from gluons}}, \href{https://doi.org/10.1007/JHEP08(2023)206}{\emph{JHEP}
  {\bfseries 08} (2023) 206}
  [\href{https://arxiv.org/abs/2304.07206}{{\ttfamily 2304.07206}}].

\bibitem{Qin:2023nhv}
Z.~Qin and Z.-Z.~Xianyu, \emph{{Nonanalyticity and on-shell factorization of
  inflation correlators at all loop orders}},
  \href{https://doi.org/10.1007/JHEP01(2024)168}{\emph{JHEP} {\bfseries 01}
  (2024) 168} [\href{https://arxiv.org/abs/2308.14802}{{\ttfamily
  2308.14802}}].

\bibitem{Qin:2023bjk}
Z.~Qin and Z.-Z.~Xianyu, \emph{{Inflation correlators at the one-loop order:
  nonanalyticity, factorization, cutting rule, and OPE}},
  \href{https://doi.org/10.1007/JHEP09(2023)116}{\emph{JHEP} {\bfseries 09}
  (2023) 116} [\href{https://arxiv.org/abs/2304.13295}{{\ttfamily
  2304.13295}}].

\bibitem{Qin:2023ejc}
Z.~Qin and Z.-Z.~Xianyu, \emph{{Closed-form formulae for inflation
  correlators}}, \href{https://doi.org/10.1007/JHEP07(2023)001}{\emph{JHEP}
  {\bfseries 07} (2023) 001}
  [\href{https://arxiv.org/abs/2301.07047}{{\ttfamily 2301.07047}}].

\bibitem{Chowdhury:2023arc}
C.~Chowdhury, A.~Lipstein, J.~Mei, I.~Sachs and P.~Vanhove, \emph{{The Subtle
  Simplicity of Cosmological Correlators}},
  \href{https://arxiv.org/abs/2312.13803}{{\ttfamily 2312.13803}}.

\bibitem{Fan:2024iek}
B.~Fan and Z.-Z.~Xianyu, \emph{{Cosmological amplitudes in power-law FRW
  universe}}, \href{https://doi.org/10.1007/JHEP12(2024)042}{\emph{JHEP}
  {\bfseries 12} (2024) 042}
  [\href{https://arxiv.org/abs/2403.07050}{{\ttfamily 2403.07050}}].

\bibitem{Aoki:2024uyi}
S.~Aoki, L.~Pinol, F.~Sano, M.~Yamaguchi and Y.~Zhu, \emph{{Cosmological
  correlators with double massive exchanges: bootstrap equation and
  phenomenology}}, \href{https://doi.org/10.1007/JHEP09(2024)176}{\emph{JHEP}
  {\bfseries 09} (2024) 176}
  [\href{https://arxiv.org/abs/2404.09547}{{\ttfamily 2404.09547}}].

\bibitem{Goodhew:2024eup}
H.~Goodhew, A.~Thavanesan and A.C.~Wall, \emph{{The Cosmological CPT Theorem}},
   \href{https://arxiv.org/abs/2408.17406}{{\ttfamily 2408.17406}}.

\bibitem{Werth:2024mjg}
D.~Werth, \emph{{Spectral representation of cosmological correlators}},
  \href{https://doi.org/10.1007/JHEP12(2024)017}{\emph{JHEP} {\bfseries 12}
  (2024) 017} [\href{https://arxiv.org/abs/2409.02072}{{\ttfamily
  2409.02072}}].

\bibitem{Melville:2024ove}
S.~Melville and G.L.~Pimentel, \emph{{A de Sitter S-matrix from amputated
  cosmological correlators}},
  \href{https://doi.org/10.1007/JHEP08(2024)211}{\emph{JHEP} {\bfseries 08}
  (2024) 211} [\href{https://arxiv.org/abs/2404.05712}{{\ttfamily
  2404.05712}}].

\bibitem{Qin:2025xct}
Z.~Qin, S.~Renaux-Petel, X.~Tong, D.~Werth and Y.~Zhu, \emph{{The exact and
  approximate tales of boost-breaking cosmological correlators}},
  \href{https://doi.org/10.1088/1475-7516/2025/09/058}{\emph{JCAP} {\bfseries
  09} (2025) 058} [\href{https://arxiv.org/abs/2506.01555}{{\ttfamily
  2506.01555}}].

\bibitem{Wang:2025qww}
D.-G.~Wang and B.~Zhang, \emph{{Bootstrapping the cosmological collider with
  resonant features}},
  \href{https://doi.org/10.1007/JHEP09(2025)122}{\emph{JHEP} {\bfseries 09}
  (2025) 122} [\href{https://arxiv.org/abs/2505.19066}{{\ttfamily
  2505.19066}}].

\bibitem{Cespedes:2025dnq}
S.~Cespedes and S.~Jazayeri, \emph{{The Massive Flat Space Limit of
  Cosmological Correlators}},
  \href{https://arxiv.org/abs/2501.02119}{{\ttfamily 2501.02119}}.

\bibitem{Chowdhury:2025ohm}
C.~Chowdhury, A.~Lipstein, J.~Marshall, J.~Mei and I.~Sachs,
  \emph{{Cosmological Dressing Rules}},
  \href{https://arxiv.org/abs/2503.10598}{{\ttfamily 2503.10598}}.

\bibitem{Fan:2025scu}
B.~Fan and Z.-Z.~Xianyu, \emph{{Anatomy of Family Trees in Cosmological
  Correlators}},  \href{https://arxiv.org/abs/2509.02684}{{\ttfamily
  2509.02684}}.

\bibitem{CarrilloGonzalez:2025qjk}
M.~Carrillo~Gonz{\'a}lez and T.~Keseman, \emph{{Spinning Boundary Correlators
  from (A)dS$_4$ Twistors}},
  \href{https://arxiv.org/abs/2510.00096}{{\ttfamily 2510.00096}}.

\bibitem{Jazayeri:2021fvk}
S.~Jazayeri, E.~Pajer and D.~Stefanyszyn, \emph{{From locality and unitarity to
  cosmological correlators}},
  \href{https://doi.org/10.1007/JHEP10(2021)065}{\emph{JHEP} {\bfseries 10}
  (2021) 065} [\href{https://arxiv.org/abs/2103.08649}{{\ttfamily
  2103.08649}}].

\bibitem{Arkani-Hamed:2023kig}
N.~Arkani-Hamed, D.~Baumann, A.~Hillman, A.~Joyce, H.~Lee and G.L.~Pimentel,
  \emph{{Differential Equations for Cosmological Correlators}},
  \href{https://arxiv.org/abs/2312.05303}{{\ttfamily 2312.05303}}.

\bibitem{Grimm:2024tbg}
T.W.~Grimm and A.~Hoefnagels, \emph{{Reductions of GKZ systems and applications
  to cosmological correlators}},
  \href{https://doi.org/10.1007/JHEP04(2025)196}{\emph{JHEP} {\bfseries 04}
  (2025) 196} [\href{https://arxiv.org/abs/2409.13815}{{\ttfamily
  2409.13815}}].

\bibitem{Chen:2024glu}
J.~Chen, B.~Feng and Y.-X.~Tao, \emph{{Multivariate hypergeometric solutions of
  cosmological (dS) correlators by $\text{d} \log$-form differential
  equations}},  \href{https://arxiv.org/abs/2411.03088}{{\ttfamily
  2411.03088}}.

\bibitem{Liu:2024str}
H.~Liu and Z.-Z.~Xianyu, \emph{{Massive inflationary amplitudes: differential
  equations and complete solutions for general trees}},
  \href{https://doi.org/10.1007/JHEP09(2025)183}{\emph{JHEP} {\bfseries 09}
  (2025) 183} [\href{https://arxiv.org/abs/2412.07843}{{\ttfamily
  2412.07843}}].

\bibitem{Cheung:2007st}
C.~Cheung, P.~Creminelli, A.L.~Fitzpatrick, J.~Kaplan and L.~Senatore,
  \emph{{The Effective Field Theory of Inflation}},
  \href{https://doi.org/10.1088/1126-6708/2008/03/014}{\emph{JHEP} {\bfseries
  03} (2008) 014} [\href{https://arxiv.org/abs/0709.0293}{{\ttfamily
  0709.0293}}].

\bibitem{Senatore:2010wk}
L.~Senatore and M.~Zaldarriaga, \emph{{The Effective Field Theory of Multifield
  Inflation}}, \href{https://doi.org/10.1007/JHEP04(2012)024}{\emph{JHEP}
  {\bfseries 04} (2012) 024} [\href{https://arxiv.org/abs/1009.2093}{{\ttfamily
  1009.2093}}].

\bibitem{Noumi:2012vr}
T.~Noumi, M.~Yamaguchi and D.~Yokoyama, \emph{{Effective field theory approach
  to quasi-single field inflation and effects of heavy fields}},
  \href{https://doi.org/10.1007/JHEP06(2013)051}{\emph{JHEP} {\bfseries 06}
  (2013) 051} [\href{https://arxiv.org/abs/1211.1624}{{\ttfamily 1211.1624}}].

\bibitem{Cabass:2022jda}
G.~Cabass, D.~Stefanyszyn, J.~Supe\l{} and A.~Thavanesan, \emph{{On graviton
  non-Gaussianities in the Effective Field Theory of Inflation}},
  \href{https://doi.org/10.1007/JHEP10(2022)154}{\emph{JHEP} {\bfseries 10}
  (2022) 154} [\href{https://arxiv.org/abs/2209.00677}{{\ttfamily
  2209.00677}}].

\bibitem{Salcedo:2024smn}
S.A.~Salcedo, T.~Colas and E.~Pajer, \emph{{The open effective field theory of
  inflation}}, \href{https://doi.org/10.1007/JHEP10(2024)248}{\emph{JHEP}
  {\bfseries 10} (2024) 248}
  [\href{https://arxiv.org/abs/2404.15416}{{\ttfamily 2404.15416}}].

\bibitem{Pinol:2024arz}
L.~Pinol, \emph{{Effective field theory of multifield inflationary
  fluctuations}},
  \href{https://doi.org/10.1103/PhysRevD.110.L041302}{\emph{Phys. Rev. D}
  {\bfseries 110} (2024) L041302}
  [\href{https://arxiv.org/abs/2405.02190}{{\ttfamily 2405.02190}}].

\bibitem{EnricoDan}
D.~Green and E.~Pajer, \emph{{On the Symmetries of Cosmological
  Perturbations}},  \href{https://arxiv.org/abs/2004.09587}{{\ttfamily
  2004.09587}}.

\bibitem{Chen:2009zp}
X.~Chen and Y.~Wang, \emph{{Quasi-Single Field Inflation and
  Non-Gaussianities}},
  \href{https://doi.org/10.1088/1475-7516/2010/04/027}{\emph{JCAP} {\bfseries
  04} (2010) 027} [\href{https://arxiv.org/abs/0911.3380}{{\ttfamily
  0911.3380}}].

\bibitem{Baumann:2011nk}
D.~Baumann and D.~Green, \emph{{Signatures of Supersymmetry from the Early
  Universe}}, \href{https://doi.org/10.1103/PhysRevD.85.103520}{\emph{Phys.
  Rev. D} {\bfseries 85} (2012) 103520}
  [\href{https://arxiv.org/abs/1109.0292}{{\ttfamily 1109.0292}}].

\bibitem{Arkani-Hamed:2015bza}
N.~Arkani-Hamed and J.~Maldacena, \emph{{Cosmological Collider Physics}},
  \href{https://arxiv.org/abs/1503.08043}{{\ttfamily 1503.08043}}.

\bibitem{Lee:2016vti}
H.~Lee, D.~Baumann and G.L.~Pimentel, \emph{{Non-Gaussianity as a Particle
  Detector}}, \href{https://doi.org/10.1007/JHEP12(2016)040}{\emph{JHEP}
  {\bfseries 12} (2016) 040}
  [\href{https://arxiv.org/abs/1607.03735}{{\ttfamily 1607.03735}}].

\bibitem{Flauger:2016idt}
R.~Flauger, M.~Mirbabayi, L.~Senatore and E.~Silverstein, \emph{{Productive
  Interactions: heavy particles and non-Gaussianity}},
  \href{https://doi.org/10.1088/1475-7516/2017/10/058}{\emph{JCAP} {\bfseries
  1710} (2017) 058} [\href{https://arxiv.org/abs/1606.00513}{{\ttfamily
  1606.00513}}].

\bibitem{Chen:2018xck}
X.~Chen, Y.~Wang and Z.-Z.~Xianyu, \emph{{Neutrino Signatures in Primordial
  Non-Gaussianities}},
  \href{https://doi.org/10.1007/JHEP09(2018)022}{\emph{JHEP} {\bfseries 09}
  (2018) 022} [\href{https://arxiv.org/abs/1805.02656}{{\ttfamily
  1805.02656}}].

\bibitem{Hook:2019zxa}
A.~Hook, J.~Huang and D.~Racco, \emph{{Searches for other vacua. Part II. A new
  Higgstory at the cosmological collider}},
  \href{https://doi.org/10.1007/JHEP01(2020)105}{\emph{JHEP} {\bfseries 01}
  (2020) 105} [\href{https://arxiv.org/abs/1907.10624}{{\ttfamily
  1907.10624}}].

\bibitem{Wang:2020ioa}
L.-T.~Wang and Z.-Z.~Xianyu, \emph{{Gauge Boson Signals at the Cosmological
  Collider}}, \href{https://doi.org/10.1007/JHEP11(2020)082}{\emph{JHEP}
  {\bfseries 11} (2020) 082}
  [\href{https://arxiv.org/abs/2004.02887}{{\ttfamily 2004.02887}}].

\bibitem{Bodas:2020yho}
A.~Bodas, S.~Kumar and R.~Sundrum, \emph{{The Scalar Chemical Potential in
  Cosmological Collider Physics}},
  \href{https://doi.org/10.1007/JHEP02(2021)079}{\emph{JHEP} {\bfseries 02}
  (2021) 079} [\href{https://arxiv.org/abs/2010.04727}{{\ttfamily
  2010.04727}}].

\bibitem{Sou:2021juh}
C.M.~Sou, X.~Tong and Y.~Wang, \emph{{Chemical-potential-assisted particle
  production in FRW spacetimes}},
  \href{https://doi.org/10.1007/JHEP06(2021)129}{\emph{JHEP} {\bfseries 06}
  (2021) 129} [\href{https://arxiv.org/abs/2104.08772}{{\ttfamily
  2104.08772}}].

\bibitem{Tong:2022cdz}
X.~Tong and Z.-Z.~Xianyu, \emph{{Large spin-2 signals at the cosmological
  collider}}, \href{https://doi.org/10.1007/JHEP10(2022)194}{\emph{JHEP}
  {\bfseries 10} (2022) 194}
  [\href{https://arxiv.org/abs/2203.06349}{{\ttfamily 2203.06349}}].

\bibitem{Chen:2022vzh}
X.~Chen, R.~Ebadi and S.~Kumar, \emph{{Classical cosmological collider physics
  and primordial features}},
  \href{https://doi.org/10.1088/1475-7516/2022/08/083}{\emph{JCAP} {\bfseries
  08} (2022) 083} [\href{https://arxiv.org/abs/2205.01107}{{\ttfamily
  2205.01107}}].

\bibitem{Chen:2023txq}
X.~Chen, J.~Fan and L.~Li, \emph{{New inflationary probes of axion dark
  matter}}, \href{https://doi.org/10.1007/JHEP12(2023)197}{\emph{JHEP}
  {\bfseries 12} (2023) 197}
  [\href{https://arxiv.org/abs/2303.03406}{{\ttfamily 2303.03406}}].

\bibitem{Stefanyszyn:2023qov}
D.~Stefanyszyn, X.~Tong and Y.~Zhu, \emph{{Cosmological correlators through the
  looking glass: reality, parity, and factorisation}},
  \href{https://doi.org/10.1007/JHEP05(2024)196}{\emph{JHEP} {\bfseries 05}
  (2024) 196} [\href{https://arxiv.org/abs/2309.07769}{{\ttfamily
  2309.07769}}].

\bibitem{Bodas:2024hih}
A.~Bodas, E.~Broadberry and R.~Sundrum, \emph{{Grand unification at the
  cosmological collider with chemical potential}},
  \href{https://doi.org/10.1007/JHEP01(2025)115}{\emph{JHEP} {\bfseries 01}
  (2025) 115} [\href{https://arxiv.org/abs/2409.07524}{{\ttfamily
  2409.07524}}].

\bibitem{An:2025mdb}
H.~An, Z.~Qin, Z.-Z.~Xianyu and B.~Zhang, \emph{{Primordial stochastic
  gravitational waves from massive higher-spin bosons}},
  \href{https://doi.org/10.1007/JHEP09(2025)203}{\emph{JHEP} {\bfseries 09}
  (2025) 203} [\href{https://arxiv.org/abs/2504.05389}{{\ttfamily
  2504.05389}}].

\bibitem{Werth:2023pfl}
D.~Werth, L.~Pinol and S.~Renaux-Petel, \emph{{Cosmological Flow of Primordial
  Correlators}},
  \href{https://doi.org/10.1103/PhysRevLett.133.141002}{\emph{Phys. Rev. Lett.}
  {\bfseries 133} (2024) 141002}
  [\href{https://arxiv.org/abs/2302.00655}{{\ttfamily 2302.00655}}].

\bibitem{Pinol:2023oux}
L.~Pinol, S.~Renaux-Petel and D.~Werth, \emph{{The cosmological flow: a
  systematic approach to primordial correlators}},
  \href{https://doi.org/10.1088/1475-7516/2025/02/019}{\emph{JCAP} {\bfseries
  02} (2025) 019} [\href{https://arxiv.org/abs/2312.06559}{{\ttfamily
  2312.06559}}].

\bibitem{Jazayeri:2023xcj}
S.~Jazayeri, S.~Renaux-Petel and D.~Werth, \emph{{Shapes of the cosmological
  low-speed collider}},
  \href{https://doi.org/10.1088/1475-7516/2023/12/035}{\emph{JCAP} {\bfseries
  12} (2023) 035} [\href{https://arxiv.org/abs/2307.01751}{{\ttfamily
  2307.01751}}].

\bibitem{Meltzer:2020qbr}
D.~Meltzer and A.~Sivaramakrishnan, \emph{{CFT unitarity and the AdS Cutkosky
  rules}}, \href{https://doi.org/10.1007/JHEP11(2020)073}{\emph{JHEP}
  {\bfseries 11} (2020) 073}
  [\href{https://arxiv.org/abs/2008.11730}{{\ttfamily 2008.11730}}].

\bibitem{Goodhew:2020hob}
H.~Goodhew, S.~Jazayeri and E.~Pajer, \emph{{The Cosmological Optical
  Theorem}}, \href{https://doi.org/10.1088/1475-7516/2021/04/021}{\emph{JCAP}
  {\bfseries 04} (2021) 021}
  [\href{https://arxiv.org/abs/2009.02898}{{\ttfamily 2009.02898}}].

\bibitem{Melville:2021lst}
S.~Melville and E.~Pajer, \emph{{Cosmological Cutting Rules}},
  \href{https://doi.org/10.1007/JHEP05(2021)249}{\emph{JHEP} {\bfseries 05}
  (2021) 249} [\href{https://arxiv.org/abs/2103.09832}{{\ttfamily
  2103.09832}}].

\bibitem{Goodhew:2021oqg}
H.~Goodhew, S.~Jazayeri, M.H.~Gordon~Lee and E.~Pajer, \emph{{Cutting
  cosmological correlators}},
  \href{https://doi.org/10.1088/1475-7516/2021/08/003}{\emph{JCAP} {\bfseries
  08} (2021) 003} [\href{https://arxiv.org/abs/2104.06587}{{\ttfamily
  2104.06587}}].

\bibitem{Stefanyszyn:2024msm}
D.~Stefanyszyn, X.~Tong and Y.~Zhu, \emph{{There and Back Again: Mapping and
  Factorizing Cosmological Observables}},
  \href{https://doi.org/10.1103/PhysRevLett.133.221501}{\emph{Phys. Rev. Lett.}
  {\bfseries 133} (2024) 221501}
  [\href{https://arxiv.org/abs/2406.00099}{{\ttfamily 2406.00099}}].

\bibitem{Stefanyszyn:2025yhq}
D.~Stefanyszyn, X.~Tong and Y.~Zhu, \emph{{A Match Made in Heaven: Linking
  Observables in Inflationary Cosmology}},
  \href{https://arxiv.org/abs/2505.16071}{{\ttfamily 2505.16071}}.

\bibitem{Tong:2021wai}
X.~Tong, Y.~Wang and Y.~Zhu, \emph{{Cutting rule for cosmological collider
  signals: a bulk evolution perspective}},
  \href{https://doi.org/10.1007/JHEP03(2022)181}{\emph{JHEP} {\bfseries 03}
  (2022) 181} [\href{https://arxiv.org/abs/2112.03448}{{\ttfamily
  2112.03448}}].

\bibitem{AguiSalcedo:2023nds}
S.~Agui~Salcedo and S.~Melville, \emph{{The cosmological tree theorem}},
  \href{https://doi.org/10.1007/JHEP12(2023)076}{\emph{JHEP} {\bfseries 12}
  (2023) 076} [\href{https://arxiv.org/abs/2308.00680}{{\ttfamily
  2308.00680}}].

\bibitem{Ema:2024hkj}
Y.~Ema and K.~Mukaida, \emph{{Cutting rule for in-in correlators and
  cosmological collider}},
  \href{https://doi.org/10.1007/JHEP12(2024)194}{\emph{JHEP} {\bfseries 12}
  (2024) 194} [\href{https://arxiv.org/abs/2409.07521}{{\ttfamily
  2409.07521}}].

\bibitem{Liu:2024xyi}
H.~Liu, Z.~Qin and Z.-Z.~Xianyu, \emph{{Dispersive bootstrap of massive
  inflation correlators}},
  \href{https://doi.org/10.1007/JHEP02(2025)101}{\emph{JHEP} {\bfseries 02}
  (2025) 101} [\href{https://arxiv.org/abs/2407.12299}{{\ttfamily
  2407.12299}}].

\bibitem{Stefanyszyn:2020kay}
D.~Stefanyszyn and J.~Supe{\l}, \emph{{The Boostless Bootstrap and BCFW
  Momentum Shifts}}, \href{https://doi.org/10.1007/JHEP03(2021)091}{\emph{JHEP}
  {\bfseries 03} (2021) 091}
  [\href{https://arxiv.org/abs/2009.14289}{{\ttfamily 2009.14289}}].

\bibitem{Pajer:2020wxk}
E.~Pajer, \emph{{Building a Boostless Bootstrap for the Bispectrum}},
  \href{https://doi.org/10.1088/1475-7516/2021/01/023}{\emph{JCAP} {\bfseries
  01} (2021) 023} [\href{https://arxiv.org/abs/2010.12818}{{\ttfamily
  2010.12818}}].

\bibitem{Meltzer:2021zin}
D.~Meltzer, \emph{{The inflationary wavefunction from analyticity and
  factorization}},
  \href{https://doi.org/10.1088/1475-7516/2021/12/018}{\emph{JCAP} {\bfseries
  12} (2021) 018} [\href{https://arxiv.org/abs/2107.10266}{{\ttfamily
  2107.10266}}].

\bibitem{Meltzer:2021bmb}
D.~Meltzer, \emph{{Dispersion Formulas in QFTs, CFTs, and Holography}},
  \href{https://doi.org/10.1007/JHEP05(2021)098}{\emph{JHEP} {\bfseries 05}
  (2021) 098} [\href{https://arxiv.org/abs/2103.15839}{{\ttfamily
  2103.15839}}].

\bibitem{Cabass:2021fnw}
G.~Cabass, E.~Pajer, D.~Stefanyszyn and J.~Supe\l{}, \emph{{Bootstrapping large
  graviton non-Gaussianities}},
  \href{https://doi.org/10.1007/JHEP05(2022)077}{\emph{JHEP} {\bfseries 05}
  (2022) 077} [\href{https://arxiv.org/abs/2109.10189}{{\ttfamily
  2109.10189}}].

\bibitem{Achucarro:2022qrl}
A.~Ach\'ucarro et~al., \emph{{Inflation: Theory and Observations}},
  \href{https://arxiv.org/abs/2203.08128}{{\ttfamily 2203.08128}}.

\bibitem{Braglia:2021ckn}
M.~Braglia, X.~Chen and D.K.~Hazra, \emph{{Comparing multi-field primordial
  feature models with the Planck data}},
  \href{https://doi.org/10.1088/1475-7516/2021/06/005}{\emph{JCAP} {\bfseries
  06} (2021) 005} [\href{https://arxiv.org/abs/2103.03025}{{\ttfamily
  2103.03025}}].

\bibitem{Braglia:2022ftm}
M.~Braglia, X.~Chen, D.K.~Hazra and L.~Pinol, \emph{{Back to the features:
  assessing the discriminating power of future CMB missions on inflationary
  models}}, \href{https://doi.org/10.1088/1475-7516/2023/03/014}{\emph{JCAP}
  {\bfseries 03} (2023) 014}
  [\href{https://arxiv.org/abs/2210.07028}{{\ttfamily 2210.07028}}].

\bibitem{Chen:2016vvw}
X.~Chen, C.~Dvorkin, Z.~Huang, M.H.~Namjoo and L.~Verde, \emph{{The Future of
  Primordial Features with Large-Scale Structure Surveys}},
  \href{https://doi.org/10.1088/1475-7516/2016/11/014}{\emph{JCAP} {\bfseries
  11} (2016) 014} [\href{https://arxiv.org/abs/1605.09365}{{\ttfamily
  1605.09365}}].

\bibitem{Slosar:2019gvt}
A.~Slosar et~al., \emph{{Scratches from the Past: Inflationary Archaeology
  through Features in the Power Spectrum of Primordial Fluctuations}},
  {\emph{Bull. Am. Astron. Soc.} {\bfseries 51} (2019) 98}
  [\href{https://arxiv.org/abs/1903.09883}{{\ttfamily 1903.09883}}].

\bibitem{Beutler:2019ojk}
F.~Beutler, M.~Biagetti, D.~Green, A.~Slosar and B.~Wallisch, \emph{{Primordial
  Features from Linear to Nonlinear Scales}},
  \href{https://doi.org/10.1103/PhysRevResearch.1.033209}{\emph{Phys. Rev.
  Res.} {\bfseries 1} (2019) 033209}
  [\href{https://arxiv.org/abs/1906.08758}{{\ttfamily 1906.08758}}].

\bibitem{Euclid:2023shr}
{\scshape Euclid} collaboration, \emph{{Euclid: The search for primordial
  features}}, \href{https://doi.org/10.1051/0004-6361/202348162}{\emph{Astron.
  Astrophys.} {\bfseries 683} (2024) A220}
  [\href{https://arxiv.org/abs/2309.17287}{{\ttfamily 2309.17287}}].

\bibitem{Chen:2011zf}
X.~Chen, \emph{{Primordial Features as Evidence for Inflation}},
  \href{https://doi.org/10.1088/1475-7516/2012/01/038}{\emph{JCAP} {\bfseries
  01} (2012) 038} [\href{https://arxiv.org/abs/1104.1323}{{\ttfamily
  1104.1323}}].

\bibitem{Chen:2012ja}
X.~Chen and C.~Ringeval, \emph{{Searching for Standard Clocks in the Primordial
  Universe}}, \href{https://doi.org/10.1088/1475-7516/2012/08/014}{\emph{JCAP}
  {\bfseries 08} (2012) 014} [\href{https://arxiv.org/abs/1205.6085}{{\ttfamily
  1205.6085}}].

\bibitem{Abolhasani:2019wri}
A.A.~Abolhasani and S.~Jazayeri, \emph{{Cherenkov radiation in the sky:
  Measuring the sound speed of primordial scalar fluctuations}},
  \href{https://doi.org/10.1103/PhysRevD.100.023520}{\emph{Phys. Rev. D}
  {\bfseries 100} (2019) 023520}
  [\href{https://arxiv.org/abs/1904.05589}{{\ttfamily 1904.05589}}].

\bibitem{Abolhasani:2019lwu}
A.A.~Abolhasani and M.M.~Sheikh-Jabbari, \emph{{Resonant reconciliation of
  convex-potential inflation models and the Planck data}},
  \href{https://doi.org/10.1103/PhysRevD.100.103505}{\emph{Phys. Rev. D}
  {\bfseries 100} (2019) 103505}
  [\href{https://arxiv.org/abs/1903.05120}{{\ttfamily 1903.05120}}].

\bibitem{Abolhasani:2020xcg}
A.A.~Abolhasani and M.M.~Sheikh-Jabbari, \emph{{Observable Quantum Loop Effects
  in the Sky}},
  \href{https://doi.org/10.1088/1475-7516/2020/06/031}{\emph{JCAP} {\bfseries
  06} (2020) 031} [\href{https://arxiv.org/abs/2003.09640}{{\ttfamily
  2003.09640}}].

\bibitem{Behbahani:2011it}
S.R.~Behbahani, A.~Dymarsky, M.~Mirbabayi and L.~Senatore, \emph{{(Small)
  Resonant non-Gaussianities: Signatures of a Discrete Shift Symmetry in the
  Effective Field Theory of Inflation}},
  \href{https://doi.org/10.1088/1475-7516/2012/12/036}{\emph{JCAP} {\bfseries
  12} (2012) 036} [\href{https://arxiv.org/abs/1111.3373}{{\ttfamily
  1111.3373}}].

\bibitem{DuasoPueyo:2023kyh}
C.~Duaso~Pueyo and E.~Pajer, \emph{{A cosmological bootstrap for resonant
  non-Gaussianity}}, \href{https://doi.org/10.1007/JHEP03(2024)098}{\emph{JHEP}
  {\bfseries 03} (2024) 098}
  [\href{https://arxiv.org/abs/2311.01395}{{\ttfamily 2311.01395}}].

\bibitem{Pajer:2024ckd}
E.~Pajer, D.-G.~Wang and B.~Zhang, \emph{{The UV Sensitivity of Axion Monodromy
  Inflation}},  \href{https://arxiv.org/abs/2412.05762}{{\ttfamily
  2412.05762}}.

\bibitem{Aoki:2023wdc}
S.~Aoki, T.~Noumi, F.~Sano and M.~Yamaguchi, \emph{{Analytic formulae for
  inflationary correlators with dynamical mass}},
  \href{https://doi.org/10.1007/JHEP03(2024)073}{\emph{JHEP} {\bfseries 03}
  (2024) 073} [\href{https://arxiv.org/abs/2312.09642}{{\ttfamily
  2312.09642}}].

\bibitem{Hui:2025aja}
L.~Hui, A.~Nicolis, A.~Podo and S.~Zhou, \emph{{Microcausality without Lorentz
  invariance}}, \href{https://doi.org/10.1007/JHEP07(2025)188}{\emph{JHEP}
  {\bfseries 07} (2025) 188}
  [\href{https://arxiv.org/abs/2502.04215}{{\ttfamily 2502.04215}}].

\bibitem{deRham:2020zyh}
C.~de~Rham and A.J.~Tolley, \emph{{Causality in curved spacetimes: The speed of
  light and gravity}},
  \href{https://doi.org/10.1103/PhysRevD.102.084048}{\emph{Phys. Rev. D}
  {\bfseries 102} (2020) 084048}
  [\href{https://arxiv.org/abs/2007.01847}{{\ttfamily 2007.01847}}].

\bibitem{Creminelli:2022onn}
P.~Creminelli, O.~Janssen and L.~Senatore, \emph{{Positivity bounds on
  effective field theories with spontaneously broken Lorentz invariance}},
  \href{https://doi.org/10.1007/JHEP09(2022)201}{\emph{JHEP} {\bfseries 09}
  (2022) 201} [\href{https://arxiv.org/abs/2207.14224}{{\ttfamily
  2207.14224}}].

\bibitem{CarrilloGonzalez:2022fwg}
M.~Carrillo~Gonzalez, C.~de~Rham, V.~Pozsgay and A.J.~Tolley, \emph{{Causal
  effective field theories}},
  \href{https://doi.org/10.1103/PhysRevD.106.105018}{\emph{Phys. Rev. D}
  {\bfseries 106} (2022) 105018}
  [\href{https://arxiv.org/abs/2207.03491}{{\ttfamily 2207.03491}}].

\bibitem{CarrilloGonzalez:2023emp}
M.~Carrillo~Gonz{\'a}lez, \emph{{Bounds on EFT{\textquoteright}s in an
  expanding universe}},
  \href{https://doi.org/10.1103/PhysRevD.109.085008}{\emph{Phys. Rev. D}
  {\bfseries 109} (2024) 085008}
  [\href{https://arxiv.org/abs/2312.07651}{{\ttfamily 2312.07651}}].

\bibitem{Creminelli:2024lhd}
P.~Creminelli, O.~Janssen, B.~Salehian and L.~Senatore, \emph{{Positivity
  bounds on electromagnetic properties of media}},
  \href{https://doi.org/10.1007/JHEP08(2024)066}{\emph{JHEP} {\bfseries 08}
  (2024) 066} [\href{https://arxiv.org/abs/2405.09614}{{\ttfamily
  2405.09614}}].

\bibitem{CarrilloGonzalez:2025fqq}
M.~Carrillo~Gonz{\'a}lez and S.~C{\'e}spedes, \emph{{Causality bounds on the
  primordial power spectrum}},
  \href{https://doi.org/10.1088/1475-7516/2025/08/071}{\emph{JCAP} {\bfseries
  08} (2025) 071} [\href{https://arxiv.org/abs/2502.19477}{{\ttfamily
  2502.19477}}].

\bibitem{Chen:2008wn}
X.~Chen, R.~Easther and E.A.~Lim, \emph{{Generation and Characterization of
  Large Non-Gaussianities in Single Field Inflation}},
  \href{https://doi.org/10.1088/1475-7516/2008/04/010}{\emph{JCAP} {\bfseries
  04} (2008) 010} [\href{https://arxiv.org/abs/0801.3295}{{\ttfamily
  0801.3295}}].

\bibitem{Flauger:2010ja}
R.~Flauger and E.~Pajer, \emph{{Resonant Non-Gaussianity}},
  \href{https://doi.org/10.1088/1475-7516/2011/01/017}{\emph{JCAP} {\bfseries
  01} (2011) 017} [\href{https://arxiv.org/abs/1002.0833}{{\ttfamily
  1002.0833}}].

\bibitem{Leblond:2010yq}
L.~Leblond and E.~Pajer, \emph{{Resonant Trispectrum and a Dozen More
  Primordial N-point functions}},
  \href{https://doi.org/10.1088/1475-7516/2011/01/035}{\emph{JCAP} {\bfseries
  01} (2011) 035} [\href{https://arxiv.org/abs/1010.4565}{{\ttfamily
  1010.4565}}].

\bibitem{Creminelli:2024cge}
P.~Creminelli, S.~Renaux-Petel, G.~Tambalo and V.~Yingcharoenrat,
  \emph{{Non-perturbative wavefunction of the universe in inflation with
  (resonant) features}},
  \href{https://doi.org/10.1007/JHEP03(2024)010}{\emph{JHEP} {\bfseries 03}
  (2024) 010} [\href{https://arxiv.org/abs/2401.10212}{{\ttfamily
  2401.10212}}].

\bibitem{Creminelli:2025tae}
P.~Creminelli, S.~Renaux-Petel, G.~Tambalo and V.~Yingcharoenrat, \emph{{Large
  $n$-point Functions in Resonant Inflation}},
  \href{https://arxiv.org/abs/2508.19240}{{\ttfamily 2508.19240}}.

\bibitem{Baumann:2019oyu}
D.~Baumann, C.~Duaso~Pueyo, A.~Joyce, H.~Lee and G.L.~Pimentel, \emph{{The
  cosmological bootstrap: weight-shifting operators and scalar seeds}},
  \href{https://doi.org/10.1007/JHEP12(2020)204}{\emph{JHEP} {\bfseries 12}
  (2020) 204} [\href{https://arxiv.org/abs/1910.14051}{{\ttfamily
  1910.14051}}].

\bibitem{Flauger:2014ana}
R.~Flauger, L.~McAllister, E.~Silverstein and A.~Westphal, \emph{{Drifting
  Oscillations in Axion Monodromy}},
  \href{https://doi.org/10.1088/1475-7516/2017/10/055}{\emph{JCAP} {\bfseries
  10} (2017) 055} [\href{https://arxiv.org/abs/1412.1814}{{\ttfamily
  1412.1814}}].

\bibitem{Silverstein:2008sg}
E.~Silverstein and A.~Westphal, \emph{{Monodromy in the CMB: Gravity Waves and
  String Inflation}},
  \href{https://doi.org/10.1103/PhysRevD.78.106003}{\emph{Phys. Rev. D}
  {\bfseries 78} (2008) 106003}
  [\href{https://arxiv.org/abs/0803.3085}{{\ttfamily 0803.3085}}].

\bibitem{Flauger:2009ab}
R.~Flauger, L.~McAllister, E.~Pajer, A.~Westphal and G.~Xu, \emph{{Oscillations
  in the CMB from Axion Monodromy Inflation}},
  \href{https://doi.org/10.1088/1475-7516/2010/06/009}{\emph{JCAP} {\bfseries
  06} (2010) 009} [\href{https://arxiv.org/abs/0907.2916}{{\ttfamily
  0907.2916}}].

\bibitem{McAllister:2008hb}
L.~McAllister, E.~Silverstein and A.~Westphal, \emph{{Gravity Waves and Linear
  Inflation from Axion Monodromy}},
  \href{https://doi.org/10.1103/PhysRevD.82.046003}{\emph{Phys. Rev. D}
  {\bfseries 82} (2010) 046003}
  [\href{https://arxiv.org/abs/0808.0706}{{\ttfamily 0808.0706}}].

\bibitem{Colas:2025ind}
T.~Colas, Z.~Qin and X.~Tong, \emph{{Open Effective Field Theory and the
  Physics of Cosmological Collider Signals}},
  \href{https://arxiv.org/abs/2512.07941}{{\ttfamily 2512.07941}}.

\bibitem{Hook:2023pba}
A.~Hook and R.~Rattazzi, \emph{{Softening the UV without new particles}},
  \href{https://doi.org/10.1103/PhysRevD.108.115019}{\emph{Phys. Rev. D}
  {\bfseries 108} (2023) 115019}
  [\href{https://arxiv.org/abs/2306.12489}{{\ttfamily 2306.12489}}].

\bibitem{Akrami:2018odb}
{\scshape Planck} collaboration, \emph{{Planck 2018 results. X. Constraints on
  inflation}}, \href{https://doi.org/10.1051/0004-6361/201833887}{\emph{Astron.
  Astrophys.} {\bfseries 641} (2020) A10}
  [\href{https://arxiv.org/abs/1807.06211}{{\ttfamily 1807.06211}}].

\bibitem{Kofman:1994rk}
L.~Kofman, A.D.~Linde and A.A.~Starobinsky, \emph{{Reheating after inflation}},
  \href{https://doi.org/10.1103/PhysRevLett.73.3195}{\emph{Phys. Rev. Lett.}
  {\bfseries 73} (1994) 3195}
  [\href{https://arxiv.org/abs/hep-th/9405187}{{\ttfamily hep-th/9405187}}].

\bibitem{Kofman:1997yn}
L.~Kofman, A.D.~Linde and A.A.~Starobinsky, \emph{{Towards the theory of
  reheating after inflation}},
  \href{https://doi.org/10.1103/PhysRevD.56.3258}{\emph{Phys. Rev. D}
  {\bfseries 56} (1997) 3258}
  [\href{https://arxiv.org/abs/hep-ph/9704452}{{\ttfamily hep-ph/9704452}}].

\bibitem{Greene:1997fu}
P.B.~Greene, L.~Kofman, A.D.~Linde and A.A.~Starobinsky, \emph{{Structure of
  resonance in preheating after inflation}},
  \href{https://doi.org/10.1103/PhysRevD.56.6175}{\emph{Phys. Rev. D}
  {\bfseries 56} (1997) 6175}
  [\href{https://arxiv.org/abs/hep-ph/9705347}{{\ttfamily hep-ph/9705347}}].

\bibitem{Bezrukov:2008ut}
F.~Bezrukov, D.~Gorbunov and M.~Shaposhnikov, \emph{{On initial conditions for
  the Hot Big Bang}},
  \href{https://doi.org/10.1088/1475-7516/2009/06/029}{\emph{JCAP} {\bfseries
  06} (2009) 029} [\href{https://arxiv.org/abs/0812.3622}{{\ttfamily
  0812.3622}}].

\bibitem{Garcia-Bellido:2008ycs}
J.~Garcia-Bellido, D.G.~Figueroa and J.~Rubio, \emph{{Preheating in the
  Standard Model with the Higgs-Inflaton coupled to gravity}},
  \href{https://doi.org/10.1103/PhysRevD.79.063531}{\emph{Phys. Rev. D}
  {\bfseries 79} (2009) 063531}
  [\href{https://arxiv.org/abs/0812.4624}{{\ttfamily 0812.4624}}].

\bibitem{Lozanov:2017hjm}
K.D.~Lozanov and M.A.~Amin, \emph{{Self-resonance after inflation: oscillons,
  transients and radiation domination}},
  \href{https://doi.org/10.1103/PhysRevD.97.023533}{\emph{Phys. Rev. D}
  {\bfseries 97} (2018) 023533}
  [\href{https://arxiv.org/abs/1710.06851}{{\ttfamily 1710.06851}}].

\bibitem{Cai:2020ovp}
Y.-F.~Cai, C.~Lin, B.~Wang and S.-F.~Yan, \emph{{Sound speed resonance of the
  stochastic gravitational wave background}},
  \href{https://doi.org/10.1103/PhysRevLett.126.071303}{\emph{Phys. Rev. Lett.}
  {\bfseries 126} (2021) 071303}
  [\href{https://arxiv.org/abs/2009.09833}{{\ttfamily 2009.09833}}].

\bibitem{Cai:2021yvq}
Y.-F.~Cai, J.~Jiang, M.~Sasaki, V.~Vardanyan and Z.~Zhou, \emph{{Beating the
  Lyth Bound by Parametric Resonance during Inflation}},
  \href{https://doi.org/10.1103/PhysRevLett.127.251301}{\emph{Phys. Rev. Lett.}
  {\bfseries 127} (2021) 251301}
  [\href{https://arxiv.org/abs/2105.12554}{{\ttfamily 2105.12554}}].

\bibitem{Brandenberger:2022xbu}
R.~Brandenberger, P.C.M.~Delgado, A.~Ganz and C.~Lin, \emph{{Graviton to photon
  conversion via parametric resonance}},
  \href{https://doi.org/10.1016/j.dark.2023.101202}{\emph{Phys. Dark Univ.}
  {\bfseries 40} (2023) 101202}
  [\href{https://arxiv.org/abs/2205.08767}{{\ttfamily 2205.08767}}].

\bibitem{Hertzberg:2018zte}
M.P.~Hertzberg and E.D.~Schiappacasse, \emph{{Dark Matter Axion Clump Resonance
  of Photons}},
  \href{https://doi.org/10.1088/1475-7516/2018/11/004}{\emph{JCAP} {\bfseries
  11} (2018) 004} [\href{https://arxiv.org/abs/1805.00430}{{\ttfamily
  1805.00430}}].

\bibitem{Dror:2018pdh}
J.A.~Dror, K.~Harigaya and V.~Narayan, \emph{{Parametric Resonance Production
  of Ultralight Vector Dark Matter}},
  \href{https://doi.org/10.1103/PhysRevD.99.035036}{\emph{Phys. Rev. D}
  {\bfseries 99} (2019) 035036}
  [\href{https://arxiv.org/abs/1810.07195}{{\ttfamily 1810.07195}}].

\bibitem{Brandenberger:2023idg}
R.~Brandenberger, V.~Kamali and R.~O.~Ramos, \emph{{Decay of ALP condensates
  via gravitation-induced resonance}},
  \href{https://doi.org/10.1088/1475-7516/2023/11/009}{\emph{JCAP} {\bfseries
  11} (2023) 009} [\href{https://arxiv.org/abs/2303.14800}{{\ttfamily
  2303.14800}}].

\bibitem{Cai:2018tuh}
Y.-F.~Cai, X.~Tong, D.-G.~Wang and S.-F.~Yan, \emph{{Primordial Black Holes
  from Sound Speed Resonance during Inflation}},
  \href{https://doi.org/10.1103/PhysRevLett.121.081306}{\emph{Phys. Rev. Lett.}
  {\bfseries 121} (2018) 081306}
  [\href{https://arxiv.org/abs/1805.03639}{{\ttfamily 1805.03639}}].

\bibitem{Chen:2020uhe}
C.~Chen, X.-H.~Ma and Y.-F.~Cai, \emph{{Dirac-Born-Infeld realization of sound
  speed resonance mechanism for primordial black holes}},
  \href{https://doi.org/10.1103/PhysRevD.102.063526}{\emph{Phys. Rev. D}
  {\bfseries 102} (2020) 063526}
  [\href{https://arxiv.org/abs/2003.03821}{{\ttfamily 2003.03821}}].

\bibitem{Zhou:2020kkf}
Z.~Zhou, J.~Jiang, Y.-F.~Cai, M.~Sasaki and S.~Pi, \emph{{Primordial black
  holes and gravitational waves from resonant amplification during inflation}},
  \href{https://doi.org/10.1103/PhysRevD.102.103527}{\emph{Phys. Rev. D}
  {\bfseries 102} (2020) 103527}
  [\href{https://arxiv.org/abs/2010.03537}{{\ttfamily 2010.03537}}].

\bibitem{mclachlan1951theory}
N.~McLachlan, \emph{Theory and Application of Mathieu Functions}, Clarendon
  Press (1951).

\bibitem{1572824500035291648}
K.P.~L., \emph{Dynamical stability of a pendulum when its point of suspension
  vibrates, and pendulum with a vibrating suspension}, {\emph{Collected Papers
  of P. L. Kapitza} {\bfseries 2} (1965) 714}.

\bibitem{wiki:kapitza_pendulum}
{Wikipedia contributors}, ``Kapitza's pendulum --- {W}ikipedia{,} the free
  encyclopedia.'' \url{https://en.wikipedia.org/wiki/Kapitza%27s_pendulum},
  2024.

\end{thebibliography}\endgroup
\end{document}